\documentclass[useAMS,usenatbib,onecolumn]{mn2e}
\usepackage{epsfig}

\newcommand{\li}{\rm{Li}}
\newcommand{\Limi}{\rm{Li^{-}}}
\newcommand{\lip}{\rm Li^{+}}
\newcommand{\lih}{\rm{LiH}}
\newcommand{\liD}{\rm{LiD}}
\newcommand{\Hp}{\rm{H}^{+}}
\newcommand{\me}{\rm{e^{-}}}
\newcommand{\mH}{\rm{H}}
\newcommand{\He}{\rm{He}}
\newcommand{\Hep}{\rm{He}^{+}}
\newcommand{\mHt}{\rm{H}_{2}}
\newcommand{\hd}{\rm{HD}}
\newcommand{\DD}{\rm{D_{2}}}
\newcommand{\Hm}{\rm{H}^{-}}
\newcommand{\Dm}{\rm{D}^{-}}
\newcommand{\Dp}{\rm{D}^{+}}
\newcommand{\mD}{\rm{D}}
\newcommand{\mHtp}{\rm{H}_{2}^{+}}
\newcommand{\ddp}{{\rm D_{2}^{+}}}
\newcommand{\htp}{\rm{H}_{3}^{+}}
\newcommand{\hdp}{{\rm HD^{+}}}
\newcommand{\hhdp}{{\rm H_{2}D^{+}}}
\newcommand{\hddp}{{\rm HD_{2}^{+}}}
\newcommand{\dtp}{{\rm D_{3}^{+}}}
\newcommand{\hehp}{{\rm HeH^{+}}}
\newcommand{\hedp}{{\rm HeD^{+}}}
\newcommand{\hehep}{{\rm He_{2}^{+}}}
\newcommand{\lihp}{{\rm LiH^{+}}}
\newcommand{\lidp}{{\rm LiD^{+}}}
\newcommand{\lihhp}{{\rm LiH_{2}^{+}}}
\newcommand{\msun}{{\rm M}_{\odot}}
\newcommand{\expf}[3]{\exp \left(#1\frac{#2}{#3}\right)}

\def\simless{\mathbin{\lower 3pt\hbox
   {$\rlap{\raise 5pt\hbox{$\char'074$}}\mathchar"7218$}}}   
\def\simgreat{\mathbin{\lower 3pt\hbox  
   {$\rlap{\raise 5pt\hbox{$\char'076$}}\mathchar"7218$}}}

\title[Is $H_{3}^{+}$ cooling important?]{Is ${\mathbf H_{3}^{+}}$ cooling ever important in primordial gas?}

\author[S.~C.~O. Glover \& D.~W. Savin]{S. C. O. Glover$^{1,2}$
 \& D. W. Savin$^{3}$ \\
$^{1}$Astrophysikalisches Institut Potsdam, An der Sternwarte 16,
D-14482 Potsdam, Germany \\
$^{2}$Institut f\"ur Theoretische Astrophysik, Albert-Ueberle-Str.\ 2, 69120 Heidelberg, Germany \\
$^{3}$Columbia Astrophysics Laboratory, Columbia University,
550 West 120th Street, New York, NY 10027-6601}

\begin{document}
\maketitle

\begin{abstract}
Studies of the formation of metal-free Population III stars usually focus
primarily on the role played by $\mHt$ cooling, on account of its large 
chemical abundance relative to other possible molecular or ionic
coolants. However, while $\mHt$ is generally the most important
coolant at low gas densities, it is not an effective coolant at high gas
densities, owing to the low critical density at which it reaches local 
thermodynamic equilibrium (LTE) and to the large opacities that develop 
in its emission lines. It is therefore possible that emission from other
chemical species may play an important role in cooling high density
primordial gas. 

A particularly interesting candidate is the $\htp$ molecular ion. This
ion has an LTE cooling rate that is roughly a billion times larger than
that of $\mHt$, and unlike other primordial molecular ions such
as $\mHtp$ or ${\rm HeH^{+}}$, it is not easily removed from the gas
by collisions with $\mH$ or $\mHt$. It is already known to be an
important coolant in at least one astrophysical context -- the upper
atmospheres of gas giants -- but its role in the cooling of primordial 
gas has received little previous study.

In this paper, we investigate the potential importance of $\htp$ cooling
in primordial gas using a newly-developed $\htp$ cooling function
and the most detailed model of primordial chemistry published
to date. We show that although $\htp$ is, in most circumstances,
the third most important coolant in dense primordial gas 
(after $\mHt$ and HD), it is nevertheless unimportant, as it 
contributes no more than a few percent of the total cooling. 
We also show that in gas irradiated by a sufficiently strong flux of 
cosmic rays or X-rays, $\htp$ can become the dominant coolant in 
the gas, although the size of the flux required renders this scenario
unlikely to occur.
\end{abstract}

\section{Introduction}
Over the past decade, we have made substantial progress in understanding
how the very first stars in the universe formed.  We know that in cosmological
models based on cold dark matter (CDM), the first stars will form in small
protogalaxies, with total masses of the order of $10^{5}$ -- $10^{6} \: \msun$,
and that by a redshift $z \sim 30$ we expect to find at least one such 
star-forming system per comoving ${\rm Mpc}^{3}$ \citep{yahs03}. We also know 
that although molecular hydrogen formation is inefficient, owing to the absence 
of dust, it is nevertheless the most abundant molecule in primordial gas, and
is the main source of cooling at low densities, for temperatures between $\sim 200$~K 
and  $8000$~K. 

Furthermore,  simple semi-analytical estimates \citep[e.g.,][]{teg97}, later confirmed 
by detailed simulations \citep[e.g.,][]{yahs03}, demonstrate that $\mHt$ provides 
enough cooling in these small protogalaxies to allow the gas to collapse under 
the influence of gravity on a timescale comparable to its gravitational free-fall 
timescale, thereby allowing star formation to occur. High-resolution, adaptive 
mesh simulations performed by \citet{abn00, abn02} have taught us much about 
the dynamics of the gas in these first protogalaxies. They consider gas cooled 
only by $\mHt$, and find that gravitational fragmentation of the collapsing gas is 
inefficient and that therefore only a single, massive star will form during the collapse.
This will then suppress further star formation 
through a variety of feedbacks \cite[see the recent review by][]{cf05}. Other 
simulations, making simplifying assumptions such as the adoption of spherical 
symmetry, have allowed us to examine the importance of various aspects of the 
physics of the gas which are currently difficult to treat in the high-resolution 
adaptive mesh simulations (for instance, the development of large optical depths 
in the rotational and  vibrational lines of $\mHt$ at high densities; see e.g.\
\citealt{on98,ripa02}).

Much of the theoretical uncertainty that remains concerns the behaviour of
dense gas during the later stages of gravitational collapse, and during the
period of accretion that follows the formation of the first protostar. There is much
that is still unknown here -- for instance, a complete understanding of the mechanism by
which the collapsing gas loses much of its initial angular momentum still eludes
us -- but in this paper we intend to focus on one relatively simple aspect: the 
identification of the dominant coolant(s) in the dense gas.

As previously noted, $\mHt$ cooling has long been known to dominate at low
densities, and it is frequently assumed that it also dominates at high densities.
However, it is not at all obvious that this is actually the case.  Two factors 
dramatically reduce the effectiveness of $\mHt$ as a coolant in high density
gas. The first is the fact that the excited rotational and vibrational levels of 
$\mHt$ have small radiative transition probabilities, and hence long radiative 
lifetimes ($\tau \simgreat 10^{6} \: {\rm s}$; \citealt{wsd98}). This means that 
collisional de-excitation of excited $\mHt$ becomes competitive with radiative 
de-excitation at fairly low gas densities ($n \sim 10^{4} \: {\rm cm^{-3}}$),
and so as the number density $n$ increases, the cooling rate of $\mHt$ quickly 
reaches its local thermodynamic equilibrium (LTE) value, given by
\begin{equation}
 \Lambda_{\rm \mHt, LTE} = \sum_{i, j>i} A_{ji} E_{ji} n_{j},
\end{equation}
where $A_{ji}$ is the transition probability for a transition from $j \rightarrow i$,
$E_{ji}$ is the energy of this transition, $n_{j}$ is the number density of 
$\mHt$ molecules in level $j$, computed assuming LTE, and where we sum 
over all bound levels $i$ and over all bound levels $j$ with energies greater than 
$i$. In the LTE limit, the cooling rate per $\mHt$ molecule is independent of the
gas density, and is largely determined by the size of the transition probabilities.
Since these are small, the LTE cooling rate is also small. Consequently, at high
gas densities, a molecular species whose excited states have much shorter radiative 
lifetimes than those of $\mHt$ will provide far more cooling {\em per molecule}
than $\mHt$. 

The second factor reducing the effectiveness of $\mHt$ cooling at very large
$n$ is the fact that the gas eventually becomes optically thick in the cores of
the main rovibrational lines of $\mHt$. The effects of this cannot currently be
treated fully in three-dimensional simulations, due to the high computational
cost of solving the resulting radiative transfer problem, but it has been 
modeled accurately in simple spherically symmetric, one-dimensional 
simulations \citep{on98,ripa02,ra04} and in an approximate fashion in
three dimensions \citep{yoha06}. These studies confirm that at densities
$n \simgreat 10^{10} \: {\rm cm^{-3}}$, optical depth effects significantly
suppress $\mHt$ cooling.

Together, these factors combine to render $\mHt$ a fairly ineffective coolant
in high density gas, despite the fact that at $n > 10^{10} \: {\rm cm^{-3}}$,
several three-body $\mHt$ formation reactions
\begin{eqnarray}
\mH + \mH + \mH  & \rightarrow & \mHt + \mH, \\
\mH + \mH + \He  & \rightarrow & \mHt + \He, \\
\mH + \mH + \mHt & \rightarrow & \mHt + \mHt,
\end{eqnarray}
rapidly convert almost all of the hydrogen in the gas to $\mHt$ \citep{pss83}.
It is therefore reasonable to ask whether cooling from any of the other molecular 
species present in the gas will become competitive with $\mHt$ cooling at 
these densities. 

One obvious possibility is deuterated hydrogen, $\hd$. Its excited rotational 
and vibrational levels have radiative lifetimes that are about a factor of 100
shorter than those of $\mHt$,  and so the $\hd$ cooling rate does not reach its
LTE limit until $n \sim 10^{6} \: {\rm cm^{-3}}$. It is also a far more effective
coolant than $\mHt$ at low temperatures ($T \simless 200 \: {\rm K}$;
see e.g., \citealt{flpr00}). This is due primarily to the fact that radiative 
transitions can occur between rotational levels with odd and even values of 
$J$, allowing cooling to occur through the $J=1 \rightarrow 0$ transition. 
The corresponding odd $\leftrightarrow$ even 
transitions in the case of $\mHt$ represent conversions
from ortho-$\mHt$ to para-$\mHt$ or vice versa, and are highly forbidden. 
Furthermore, at low temperatures the ratio of $\hd$ to $\mHt$ can be 
significantly enhanced with respect to the cosmological D:H ratio by chemical 
fractionation \citep[see e.g.,][]{glo08}. 

The role of $\hd$ cooling in early protogalaxies has been investigated 
by a number of authors. In the case of the earliest generation of 
protogalaxies, which form from very cold neutral gas that is never heated to
more than a few thousand K during the course of the galaxy formation
process, $\hd$ cooling appears to be unimportant, as the collapsed gas
does not become cold enough for sufficient fractionation to occur to make
$\hd$ cooling dominant \citep{bcl02}. 

The situation is rather different, however, in primordial gas cooling from an 
initially ionized state. In that case more $\mHt$ is formed, allowing the 
gas to cool to a lower temperature, at which point fractionation becomes 
effective and $\hd$ cooling rapidly becomes dominant \citep[see e.g.,][]{nu02,no05,jb06,sv06}. 
However, the initial ionization required is much larger than that expected
to be present in the earliest protogalaxies.

Another molecule to have attracted considerable attention is lithium hydride, LiH.
This molecule has a very large dipole moment, $\mu = 5.888 \: {\rm debyes}$
\citep{zs80}, and consequently its excited levels have very short radiative lifetimes. 
Therefore, despite the very low lithium abundance in primordial gas ($x_{\rm Li} = 4.3 
\times 10^{-10}$, by number; see \citealt{cy04}), it was thought for a time  that
LiH would dominate the cooling at very high densities \citep[see e.g.,][]{ls84}.
However, accurate quantal calculations of the rate of formation of $\lih$ by
radiative association \citep{dks96,ggg96,bdlg03}
\begin{equation}
 \li + \mH \rightarrow \lih + \gamma,
\end{equation}
have shown that the rate is much smaller than was initially assumed, 
while recent work by \citet{dpgg05} has shown that the reaction
\begin{equation}
\lih + \mH \rightarrow \li + \mHt,
\end{equation}
has no activation energy and so will be an efficient destruction mechanism for 
$\lih$ for as long as some atomic hydrogen remains in the gas. Consequently,
the amount of lithium hydride present in the gas is predicted to be very small,
even at very high densities, and so $\lih$ cooling is no longer believed to be
important \citep{mon05}.

In contrast to $\hd$ or $\lih$, the various molecular ions present in the gas,
such as $\mHtp$, $\htp$ or ${\rm HeH^{+}}$, have attracted little attention.
Some early work on $\mHtp$ cooling in ionized primordial gas can be found
in \citet{ss77,ss78}, and its possible importance in hot, highly ionized conditions 
has recently been re-emphasized by \citet{yokh07}, but aside from this, there has
been little exploration of the role that cooling from these species might play in 
the evolution of primordial gas, presumably because the abundances of these 
species are expected to be small. It is this absence that the current paper 
attempts to rectify. 

We present here the results of a set of simulations of the chemical and thermal
evolution of gravitationally collapsing primordial gas. These simulations use a
very simple one-zone dynamical model for the gas, but couple this with a detailed
chemical network and a comprehensive model of the various heating and cooling
processes at work. Besides the coolants considered above, we include 
the effects of cooling from (in no particular order)
$\mHtp$, $\hdp$, $\ddp$, $\htp$, $\hhdp$, $\hddp$, $\dtp$,  
$\hehp$, $\hedp$, $\hehep$, $\lihp$, $\lidp$, $\liD$ and $\lihhp$. We focus in particular on the 
possible role of $\htp$. This ion has a very large number of excited rotational and 
vibrational levels that are energetically accessible at the temperatures of interest 
in primordial gas, and its vibrational levels have much shorter radiative lifetimes 
than those of $\mHt$ or $\hd$. In LTE, its cooling rate per molecule is roughly $10^{9}$
times larger than that of $\mHt$. It is known to be an important coolant in planetary 
atmospheres \citep{miller00} and may also be an effective coolant in high density 
primordial gas \citep{gs06}. Unlike ions such as $\mHtp$ or ${\rm HeH^{+}}$, 
it does not react with  $\mHt$, and is not easily destroyed by collisions with $\mH$, 
and so its  abundance in high density gas should be large compared to the other 
molecular ions included in our model.

The layout of this paper is as follows. In \S2 we outline the numerical method 
used to simulate the thermal and chemical evolution of the protostellar gas.
The chemical reactions included in the model are discussed in \S3, and the
thermal processes, in particular $\htp$ cooling, are discussed in \S4. We present
the results of our simulations in \S5 and close with a brief discussion in \S6.

\section{Numerical method}
We treat the thermal and chemical evolution of the gas using a 
one-zone model, in which the density is assumed to evolve as
\begin{equation}
 \frac{{\rm d} \rho}{{\rm d} t} = \eta \frac{\rho}{t_{\rm ff}} 
\end{equation}
where $t_{\rm ff} = \sqrt{3\pi / 32 G \rho}$ is the free-fall 
timescale of the gas and where $\eta$ is an adjustable constant.
In most of our simulations, we set $\eta = 1$, corresponding to 
free-fall collapse, but in a few runs we examine the effect of
slowing down the collapse by setting $\eta < 1$ (see 
\S\ref{res:dyn} for details).

To follow the temperature evolution, we solve the energy equation
\begin{equation}
 \frac{{\rm d} e}{{\rm d} t} = \frac{p}{\rho^{2}} \frac{{\rm d}\rho}{{\rm d}t} -
 \Lambda + \Gamma, \label{en_eq} 
\end{equation}
where $e$ is the internal energy density, $p$ is the thermal
pressure, $\Lambda$ is the total cooling rate (which includes
contributions from both radiative and chemical cooling, as
outlined in \S\ref{cool}) and $\Gamma$ is the total heating rate.
Since the temperature evolution is strongly 
coupled to the chemical evolution, we solve Equation~\ref{en_eq} 
simultaneously with the chemical rate equations using the DVODE 
implicit ordinary differential equation solver \citep{bbh89}.  To model 
the chemistry we use an extensive chemical network consisting of
392 reactions among 30 atomic and molecular species.
Table~\ref{abtab} lists all 30 species considered.
Tables~\ref{chemtab_ci}--\ref{cr_pd_tab} in Appendix A list the reactions 
included in this network, broken down by the type of process involved. 
These tables also give details of the rate coefficient or rate adopted for each 
reaction and the source of the data. In these tables and elsewhere in
the paper, $T$ is the gas 
temperature in K, $T_{3} = T / 1000 \: {\rm K}$, and $T_{\rm e}$ is the
temperature in units of eV. (Note that we assume, both here and 
throughout, that the kinetic temperature of the electrons is the same 
as that of the atoms and molecules).

For many of these species, we 
followed the full time-dependent, non-equilibrium chemistry, but in 
some cases -- the ions $\Hm$, $\Dm$, $\Limi$, $\mHtp$, $\hdp$, 
$\ddp$, $\hehp$, $\hedp$, $\hehep$, $\lidp$ and $\lihhp$ -- chemical equilibrium 
is reached very rapidly, on a timescale of the order of $t_{\rm eq} = 1 / (k_{\rm dest} n)
\sim 10^{9} \: n^{-1} \: {\rm s}$, where $k_{\rm dest} \sim 10^{-9} \: {\rm cm^{3}} \: {\rm s^{-1}}$
is a characteristic destruction rate coefficient and $n$ is the number density of hydrogen nuclei.
At the gas densities considered in this work ($1 < n < 3 \times
10^{13} \: {\rm cm^{-3}}$ in the majority of our simulations), this
chemical equilibrium timescale for these rapidly reacting ions is orders 
of magnitude smaller than the free-fall timescale of the gas, which is 
approximately $1.5 \times 10^{15} n^{-1/2} \: {\rm s^{-1}}$. It is therefore 
sufficient to use the equilibrium abundances for these ions. Some of
the species remaining in our non-equilibrium model may also be close 
to chemical equilibrium during a large portion of the collapse, but are
included in the non-equilibrium treatment because we cannot be sure
that they always remain in equilibrium. Further details of our chemical 
network are given in \S\ref{chem}.

\begin{table}
\caption{Initial fractional abundances in our reference calculation \label{abtab}}
\begin{tabular}{lclclc}
\hline
\multicolumn{1}{c}{Species} & 
\multicolumn{1}{c}{Initial abundance} & \multicolumn{1}{c}{Species} & 
\multicolumn{1}{c}{Initial abundance}  & \multicolumn{1}{c}{Species} & 
\multicolumn{1}{c}{Initial abundance} \\
\hline
${\rm e^{-}}$ & $2.2 \times 10^{-4}$ & $\hdp$ & --- & $\hedp$ & --- \\
$\Hp$  & $2.2 \times 10^{-4}$ & $\hd$ & $6.2 \times 10^{-11}$ & $\hehep$ & --- \\
$\mH$ & 0.99978 & $\ddp$ & --- & $\lip$ & $2.2 \times 10^{-10}$ \\
$\Hm$ & --- &  $\DD$ &$1.6 \times 10^{-15}$ & $\li$ & $2.1 \times 10^{-10}$ \\
$\mHtp$ & --- & $\hhdp$ & 0.0 & $\Limi$ & --- \\
$\mHt$ & $2.4 \times 10^{-6}$ & $\hddp$ & 0.0 & $\lihp$ & 0.0  \\
$\htp$ & 0.0 & $\dtp$ & 0.0 & $\lih$ & 0.0 \\
$\Dp$ & $5.7 \times 10^{-9}$ & $\Hep$ & $2.8 \times 10^{-26}$ &  $\lidp$ & --- \\
$\mD$ & $2.6 \times 10^{-5}$ & $\He$ & $8.3 \times 10^{-2}$  & $\liD$ & 0.0 \\
$\Dm$ & --- & $\hehp$ & --- & $\lihhp$ & --- \\
\hline
\hline
\end{tabular}
\medskip
\\
{\bf Notes}:  The quoted values are fractional abundances relative to the
number density of hydrogen nuclei. \\
Chemical species listed without initial
abundances are assumed to be in chemical equilibrium, as described in
the text.
 \\
\end{table}

We assume elemental abundances relative to hydrogen of 0.083, $2.6 \times 10^{-5}$ and
$4.3 \times 10^{-10}$ for helium, deuterium and lithium, respectively
\citep{cy04}. In \S\ref{res:IC} we explore the effects of reducing the deuterium 
and/or the lithium abundance to zero, in order to asses the impact of the deuterium 
and lithium chemistry on the evolution of the gas. The initial abundances of the
various molecular and ionic species in our standard model are summarized in
Table~\ref{abtab}. The values used for the initial $\Hp$, 
$\Hep$ and $\mHt$ abundances, and the ratio of ionized to neutral lithium are taken from the 
calculations of  \citet{sld98} -- specifically their model V. Deuterated species are 
assumed to have abundances that are a factor $(2.6 \times 10^{-5})^{N_{\rm D}}$ 
smaller than the abundances of the undeuterated equivalents, where
$N_{\rm D}$ is the number of deuterium nuclei in the species in question.
The electron abundance is computed assuming charge conservation.
To assess our sensitivity to these initial values, we have also run several
simulations with different initial abundances; the results of these runs are
discussed in \S\ref{res:IC}.

For most simulations, we adopt an initial density $n_{\rm i} = 1 \: {\rm cm^{-3}}$
and an initial temperature $T_{\rm i} = 1000 \: {\rm K}$. The effects of altering
$n_{\rm i}$ and $T_{\rm i}$ are examined in \S\ref{res:IC}. All of the simulations 
are run until the density exceeds $n_{\rm f} = 3 \times 10^{13} \: {\rm cm^{-3}}$.
At higher densities, collision-induced emission from $\mHt$ quickly comes to
dominate the cooling, and the minor species considered here are unlikely
to be important coolants in this very high density regime.

Finally, we ran all of our simulations starting at a redshift $z=20$. However, 
the main influence of the redshift is to set a minimum temperature for the gas
(since the gas cannot cool radiatively to below the CMB temperature, 
$T_{\rm CMB}$). As the gas temperature $T \gg T_{\rm CMB}$ at the 
densities of interest in this paper, we do not anticipate that changing
$z$ by a moderate amount will significantly affect our results.

\section{Chemistry}
\label{chem}
The chemical evolution of the gas is modeled using a chemical
network consisting of 392 reactions amongst 30 neutral and  
ionized species. A list of all of the reactions included can be found 
in Tables~\ref{chemtab_ci}--\ref{cr_pd_tab}. This network, which 
to the best of our knowledge is 
the largest used to date for the study of primordial gas chemistry, is 
based in part on 
previous compilations by \citet{a97}, \citet{sld98}, \citet{gp98,gp02},
\citet{lsd02}, \citet{ws02}, and \citet{wfp04}, supplemented with additional
reactions drawn directly from the chemical literature, as well as
some whose rates have not (to our knowledge) been previously
discussed. These latter are generally rates involving one or 
more deuterium nuclei in place of hydrogen nuclei; and in 
estimating rates for these reactions, we have generally followed
the same procedure as in \citet{sld98}: for a non-deuterated reaction 
with a reaction rate coefficient that has a power-law temperature dependence 
$k \propto T^{m}$, we have estimated a rate coefficient for the deuterated
reaction by multiplying this rate coefficient by a scaling factor
$(\mu_{\rm H} / \mu_{\rm D})^{m}$, where $\mu_{\rm H}$ and 
$\mu_{\rm D}$ are the reduced masses of the reactants in the
non-deuterated and deuterated reactions, respectively. Some
notable exceptions to this strategy are discussed in more detail
in section \S\ref{react-discuss} below.

For reactions where the presence of a deuteron increases the 
number of distinguishable outcomes  and 
where no good information exists on the branching ratio of the
reaction, we have assumed that each outcome is equally likely.
An example of this is the dissociative 
attachment of $\hd$ with $\me$ (reactions AD6--AD7), which can 
produce either $\mH$ and $\Dm$ or $\Hm$ and $\mD$, 
in contrast to the dissociative attachment of $\mHt$ with $\me$ 
(reaction AD5) which can only produce $\Hm$ and $\mH$.
For this particular example, this assumption gives branching 
ratios of 50\% for reactions AD6 and AD7, respectively.

In spite of the size of our chemical network, there remain a number
of chemical processes that are not included. These are discussed
in \S\ref{neglect}, along with our justifications for omitting them.

\subsection{Discussion of selected reactions}
\label{react-discuss}

\subsubsection{Photorecombination of H and He
(reactions PR1 \& PR3, Table~\ref{tab:PR})}
We assume that the ionizing photons produced by the recombination of
$\Hp$ to ground state $\mH$ are immediately consumed by the ionization
of atomic hydrogen (the on-the-spot approximation), and so we use the
case B rate coefficient for hydrogen recombination \citep{fer92}. We note
that although the fractional abundance of atomic hydrogen becomes
small at densities greater than $10^{10} \: {\rm cm^{-3}}$, the number
density of atomic hydrogen remains considerable, and so case B remains
a good approximation.

We also use the on-the-spot approximation to treat $\Hep$ recombination, but
in this case the net recombination rate is larger than the case B rate, as
some of the photons produced by recombination directly into the $n=1$
ground state are lost through photoionization of $\mH$ rather than $\He$.
For $\Hep$ recombinations directly into the ground state, occurring in primordial 
gas with a low fractional ionization and low molecular abundance, approximately 
68\% of  the resulting photons are absorbed by $\mH$, with the
remaining 32\% being absorbed by $\He$  \citep{ost89}. Therefore, the effective 
$\Hep$ recombination rate coefficient in these conditions is given by
\begin{equation}
k_{\rm PR3} = 0.68 k_{\rm PR3, rr, A} + 0.32 k_{\rm PR3, rr, B} 
+ k_{\rm PR3, di} \: {\rm cm^{3}} \: {\rm s^{-1}},
\end{equation}
where $k_{\rm PR3, rr, A}$ and $k_{\rm PR3, rr, B}$ are the case A 
and case B rate coefficients, and $k_{\rm PR3, di}$ is the dielectronic
recombination rate coefficient. This formula becomes incorrect once the
$\mHt$ fraction of the gas becomes large, but as this occurs only at
densities $n \simgreat 10^{10} \: {\rm cm^{-3}}$ at which the $\Hep$ 
fractional abundance is negligible, the error that is introduced by
using this prescription for $k_{\rm PR3}$ throughout the simulation
is unimportant.

It is also necessary to take account of the photoionization of
$\mH$ caused by the $\Hep$ recombination emission. In addition to
the contribution coming from $\Hep$ recombination into the $n=1$
ground state, there is an additional contribution made by photons
produced during transitions from $n=2$ to $n=1$ in $\He$; in other
words, even pure case B recombination of $\Hep$ produces H-ionizing
photons. The proportion of case B recombinations that yield photons 
capable of ionizing hydrogen depends upon the relative populations 
of the $n=2$ singlet and triplet states, and hence upon the electron 
density, but in the low density limit, 96\% of all recombinations to 
excited states produce photons that will ionize hydrogen \citep{ost89}. 
To model the effects of these photons, along with those produced by 
recombination direct to the $n=1$ ground-state and by dielectronic
recombination, we include in our chemical network a local $\mH$
ionization rate per unit volume $R_{\rm pi}$, with a value
\begin{eqnarray}
R_{\rm pi} & = &  \left[0.68 (k_{\rm PR3, rr, A} - k_{\rm PR3, rr, B}) + 0.96 k_{\rm PR3, rr, B} 
+ 2 k_{\rm PR3, di} \right] n_{\rm e} n_{\Hep} \: {\rm cm^{-3}} \: {\rm s^{-1}},  \nonumber \\
 & = & \left[0.68 k_{\rm PR3, rr, A} + 0.28 k_{\rm PR3, rr, B} + 2 k_{\rm PR3, di} \right] n_{\rm e} n_{\Hep}
 \: {\rm cm^{-3}} \: {\rm s^{-1}},
\end{eqnarray}
where the three terms in brackets on the first line correspond to the
contributions from recombination direct to the $n=1$ ground state, 
pure case B recombination (i.e., recombination
to all states $n \geq 2$), and dielectronic recombination, respectively. 
(Note that every dielectric 
recombination produces two photons capable of ionizing hydrogen: one due
to the radiative stabilization of the process, and one as the captured
electron cascades to the 1s level). 
If the electron density is large ($n_{\rm e} \simgreat 10^{3} \: {\rm cm^{-3}}$;
\citealt{clegg87}), then more helium recombinations will result in
two-photon transitions from $2^1$S--$1^1$S, reducing the number of 
photons produced that are capable of ionizing hydrogen \citep{ost89}. However, 
the effect on $R_{\rm pi}$ is relatively small, and in any case, we
do not expect to encounter large abundances of $\Hep$ in dense gas in the particular 
scenario that we are investigating. Therefore, adopting this simplified treatment 
at all $n$ should be sufficient for our purposes. 

\subsubsection{Dissociative recombination of $H_{3}^{+}$ (reactions DR4 \& DR5, Table~\ref{tab:DR})}
For a long time, considerable disagreement has existed on the subject
of the $\htp$ dissociative recombination rate. Measurements of the rate
in merged beam experiments \citep[e.g.,][]{sund94} typically give values 
of the order of $10^{-7} \: {\rm cm^{3}} \: {\rm s^{-1}}$ for the rate coefficient 
at temperatures near room temperature, while measurements made in
flowing afterglow experiments (e.g., \citeauthor{ss93a}~1993a,b) often give 
values of the order of $10^{-8} \: {\rm cm^{3}} \: {\rm s^{-1}}$, an order of 
magnitude smaller. At the same time, most theoretical calculations 
have indicated a smaller rate still, of the order of $10^{-11} \: 
{\rm cm^{3}} \: {\rm s^{-1}}$ \citep[see][and references therein]{oss00}, 
which is in complete disagreement with the experimental measurements.

However, it has recently become clear that three-body effects play a
highly important role in the recombination of $\htp$ in flowing afterglow
experiments \citep[see e.g.,][]{glo05}. When proper allowance is made
for these effects, the inferred two-body recombination rate is in good
agreement with the results of the merged beam experiments
\citep[see the discussion in][]{glos07}. Moreover, recent theoretical 
calculations of the rate coefficient by \citet{kg03}
that account for the effects of Jahn-Teller coupling between the 
electronic and vibrational degrees of freedom produce a result that 
is in good agreement with the experimental measurements, although
disagreement at the level of a factor of two or so is still present at some
energies.

In our calculations, we therefore take our value for the total $\htp$ 
dissociative recombination rate coefficient from the recent ion storage ring
measurements of \citet{mac04}. To convert 
this total rate coefficient -- the sum of the rate coefficients for reactions DR4 and DR5 
-- into a rate coefficient for each individual reaction, we adopt a branching ratio
of 0.25 for reaction DR4 and 0.75 for reaction DR5, based on the 
measurements of \citet{datz95}. Strictly speaking,
these values are  only appropriate for
temperatures $T < 3000 \: {\rm K}$, but in practice the behaviour of the
gas is not particularly sensitive to the values chosen.

\subsubsection{$H_{3}^{+}$ formation by radiative association (reaction RA18, Table~\ref{tab:RA})}
The rate coefficient we quote for reaction RA18, the radiative association of $\mHt$ and
$\Hp$ to form $\htp$, was taken from the study of \citet{gh92}, and 
was the rate coefficient quoted by \citet{gp98} for this reaction. However, \citet{sld98}
prefer a much smaller rate coefficient of $10^{-20} \: {\rm cm^{3}} \: {\rm s^{-1}}$ for
this reaction. In \S\ref{res:chem}, we examine the effect of  adopting 
this smaller rate coefficient. 

\subsubsection{$H_{2}$ formation by associative detachment of 
$H^{-}$ (reaction AD1, Table~\ref{tab:AD})}
The rate of this reaction is quite uncertain, and we have shown in previous 
work that this uncertainty can lead in some cases to a substantial uncertainty in the 
amount of $\mHt$ formed in the gas \citep{gsj06}. However, we do not expect this 
uncertainty to significantly affect the results in this paper. At the high densities at 
which $\htp$ cooling is potentially important, the dominant $\mHt$ formation pathway 
is three-body formation. This can produce a much larger molecular fraction than is 
possible via two-body reactions, and so uncertainty in the amount of $\mHt$ produced 
via the $\Hm$ ion has a negligible impact on the evolution of the gas at high
densities. In our simulations, we adopt a default value for the rate coefficient for 
reaction AD1 of $k_{\rm AD1} = 1.5 \times 10^{-9} T_{3}^{-0.1} \: {\rm cm^{3}} 
\: {\rm s^{-1}}$, where $T_{3} = T / 300 \: {\rm K}$, taken from \citet{llz91}.
In \S\ref{res:chem} we demonstrate that our results are insensitive to  this choice. 

We note also that the other reaction discussed in \citet{gsj06}, the mutual
neutralization of $\Hm$ by $\Hp$ (reaction MN1, Table~\ref{tab:MN}), is no longer a source of
significant uncertainty in chemical models of primordial gas. Recent 
measurements of the cross-section for this reaction at astrophysically-relevant
energies made by X. Urbain (private communication) yield values in good agreement with 
those obtained by \citet{fk86} and used as a basis for the rate coefficient of 
\citet{cdg99}. These measurements strongly suggest that the earlier measurements
of the mutual neutralization cross-section made by \citet{map70} were somehow in 
error, and that rate coefficients based on them \citep[see e.g.][]{dl87,gp98} are
incorrect. The error in the rate coefficient for this reaction has thus been reduced
from the order of magnitude discussed in \citet{gsj06} to an uncertainty of about
50\% (X. Urbain, private communication).

\subsubsection{Collisional dissociation of $H_{2}$ (reactions CD9--CD12, Table~\ref{tab:CD})}
\label{h2cd}
The rate coefficients for the collisional dissociation of $\mHt$ by $\mH$
(CD9), $\mHt$ (CD10), $\He$ (CD11) and $\me$ (CD12) are represented 
by functions of the form
\begin{equation}
\log k_{\rm i}  = \left( \frac{n/n_{\rm cr}}{1 + n/n_{\rm cr}} \right)
\log k_{\rm i, LTE} + \left(\frac{1}{1 + n/n_{\rm cr}} \right) \log k_{\rm i, v=0},
\end{equation}
where $n$ is the number density of hydrogen nuclei,
$k_{\rm i}$ is the collisional dissociation rate for collisions with species
$i$, and $k_{\rm v=0, i}$ and $k_{\rm LTE, i}$ are the rate coefficients 
for this reaction in the limits in which all of the $\mHt$ molecules are in the vibrational 
ground-state (appropriate in low-density gas), or have their LTE level
populations (appropriate for high density gas). The critical density, $n_{\rm cr}$, 
for $\mHt$ collisional dissociation in a gas containing a mix of $\mH$, 
$\mHt$, $\He$ and electrons is not well determined. For simplicity, we
therefore assume that it is given by a weighted harmonic mean of the 
(better known) critical densities corresponding to reactions CD9, 
CD10, and CD11 considered individually, i.e.,
\begin{equation}
\frac{1}{n_{\rm cr}} = \frac{1}{1 + x_{\He}} \left[ \frac{x_{\mH}}{n_{\rm cr, \mH}} + 
\frac{2x_{\mHt}}{n_{\rm cr, \mHt}} + \frac{x_{\He}}{n_{\rm cr, \He}} \right],
\end{equation}
where $x_{\mH}$, $x_{\mHt}$, and $x_{\He}$ are the fractional abundances
of H, $\mHt$ and He relative to the total number of hydrogen nuclei, we use
the approximation that $x_{\rm H} + 2 x_{\rm H_{2}} = 1$,  and where
\begin{eqnarray}
n_{\rm cr, \mH} & = &  {\rm dex} \left[3.0 - 0.416 \log{T_{4}} - 0.327 \left(\log{T_{4}} \right)^{2} \right] 
\: {\rm cm^{-3}}, \\
n_{\rm cr, \mHt} & = &  {\rm dex} \left[4.845 - 1.3 \log{T_{4}} + 1.62 \left(\log{T_{4}} \right)^{2} \right]
\: {\rm cm^{-3}}, \\
n_{\rm cr, \He} & = & {\rm dex} \left[5.0792 \left\{1.0 - 1.23 \times 10^{-5} (T - 2000) \right\} \right]
\: {\rm cm^{-3}},
\end{eqnarray}
with $T_{4} = T / 10000 \: {\rm K}$. The expression for $n_{\rm cr, \mH}$ is from 
\citet{ls83}, but has been decreased by an order of magnitude, as recommended 
by \citet{msm96}. The expression for $n_{\rm cr, \mHt}$ comes from \citet{sk87},
and the expression for $n_{\rm cr, \He}$ comes from \citet{drcm87}. Note that this 
expression for the critical density assumes that in high density gas, 
$n_{\rm e} \ll n_{\mH}$, so that electron excitation of $\mHt$ does not
significantly affect the value of $n_{\rm cr}$. Other forms of averaging to
obtain $n_{\rm cr}$ are possible, of course, but we would not expect our
results to be sensitive to our particular choice here, as any differences will
only be seen for densities $n \sim n_{\rm cr}$, and in our simulations, gas at
these densities is always far too cold for collisional dissociation of $\mHt$ to be
important.

To ensure that our adopted collisional dissociation rate coefficients 
and three-body $\mHt$ formation rate coefficients are consistent, we used 
the fact that in LTE, the equilibrium constant $K$ obeys
\begin{equation}
K = k_{\rm TB1} / k_{\rm CD9} = k_{\rm TB2} / k_{\rm CD10} = k_{\rm TB3} / k_{\rm CD11}
\end{equation}
and varies with temperature as  \citep{fh07}
\begin{equation}
K = 1.05 \times 10^{-22} T^{-0.515} \exp \left(\frac{52000}{T} \right),
\end{equation}
to determine values for the rate coefficients of reactions CD9, CD10, and 
CD11 in the LTE limit. However, we also ran some test simulations where we used 
rate coefficients from \citet{ls83} and \citet{sk87} for reactions CD9 and CD10, regardless
of the value of $k_{\rm TB1}$ or $k_{\rm TB2}$. These simulations produced
almost identical results, demonstrating that our results here are insensitive
to our treatment of $\mHt$ collisional dissociation.

\subsubsection{Collisional dissociation of HD and $D_{2}$ (reactions CD13--CD20, 
Table~\ref{tab:CD})}
\label{cdhd}
For collisions with electrons, accurate rate coefficients are available from \citet{tt02a}
and \citet{tt02b}. For collisions with H, $\mHt$, or He, however, we are 
unaware of a treatment in the literature. We have therefore assumed that the
rate coefficients of these reactions in the $v=0$ and LTE limits are the same as for the
corresponding H reactions (nos.\ CD9--CD11). For ${\rm D_{2}}$, we have 
also adopted the same value for the critical density, while for HD, we have
increased it by a factor of 100 to account for the larger radiative transition probabilities. 
Note that although these rate coefficients are highly approximate, this probably does not 
introduce much uncertainty into the chemical model, as reactions IX18 and 
IX20 (Table~\ref{tab:IX}) become effective at much lower temperatures and therefore
dominate the destruction of HD and ${\rm D_{2}}$ in warm gas.

\subsubsection{Three-body $H_{2}$ formation (reactions TB1--TB3, Table~\ref{tab:TB})}
\label{tbhf}
Although unimportant at low densities, three-body reactions are the 
dominant source of $\mHt$ in high density primordial gas, and so
these reactions represent an important part of our chemical network. 
Unfortunately, rate coefficients for these reactions are, in general,
not known to a high degree of accuracy. For three-body collisions in
which atomic hydrogen is the third body (reaction TB1), the situation 
is particularly bad. One commonly adopted rate coefficient is that
of \citet{pss83}, who quote a rate coefficient 
\begin{equation}
k_{\rm TB1, PSS} = 5.5 \times 10^{-29} T^{-1} \: {\rm cm^{6}} \: {\rm s^{-1}} 
\end{equation}
for this reaction, based on experimental work by \citet{jgc67}. Also in common 
usage is the rate coefficient adopted by \citet{abn02}, which is
\begin{eqnarray}
k_{\rm TB1, ABN} & = & 1.14 \times 10^{-31} T^{-0.38} \: {\rm cm^{6}} \: {\rm s^{-1}}
\hspace{.2in} T \leq 300 \: {\rm K} \nonumber \\
 & = & 3.90 \times 10^{-30} T^{-1.00} \: {\rm cm^{6}} \: {\rm s^{-1}}
\hspace{.2in} T > 300 \: {\rm K}.
\end{eqnarray}
The low temperature portion of this rate coefficient is based on \citet{or87}, while the high 
temperature portion is an extrapolation by \citet{abn02}.  This reaction is also 
discussed by \citet{cw83} in their large compilation and review of chemical 
kinetic data. They summarize a large number of different experimental 
measurements and argue that the precision of the data does not justify 
anything more elaborate than a constant rate coefficient
\begin{equation}
k_{\rm TB1, CW} = 8.8 \times 10^{-33} \: {\rm cm^{6}} \: {\rm s^{-1}}.
\end{equation}
Another possibility is found in \citet{schw90}, who gives calculated 
values at $T = 3000 \: {\rm K}$ and $T = 5000 \: {\rm K}$ of  $1.4 \times 10^{-32} 
\: {\rm cm^{6}} \: {\rm s^{-1}}$ and $8.2 \times 10^{-33}  \: {\rm cm^{6}} 
\: {\rm s^{-1}}$, respectively. These values are roughly an order of 
magnitude larger than those given by the \citet{abn02} rate coefficient, 
and about 30\% lower than the values given by the \citet{pss83} rate
coefficient. 

More recently, \citet{fh07} have argued in favor of a rate coefficient
\begin{equation}
k_{\rm TB1, FH} = 1.44 \times 10^{-26} T^{-1.54} \: {\rm cm^{6}} \: {\rm s^{-1}},
\end{equation}
which they derived from the rate coefficient of the inverse process ($\mHt$ collisional
dissociation by atomic hydrogen, reaction CD9) by using the principle of detailed 
balance. This rate coefficient is approximately six times larger than the 
\citet{pss83} rate coefficient at $T = 1000 \: {\rm K}$, or approximately ninety times 
larger than the \citet{abn02} rate coefficient. Unfortunately, the accuracy of a rate 
coefficient derived
using detailed balance depends upon the accuracy with which the rate coefficient 
of the inverse process is known. In this case, that accuracy is poor, as the
$\mHt$ collisional dissociation rate coefficient is not well constrained by experiment at
low temperatures (i.e., $T < 2000 \: {\rm K}$) owing to its small size at these
temperatures. \citet{fh07} used the \citet{jgc67} fit to the collisional dissociation
rate coefficient, but if we instead use the calculated rate coefficient from \citet{msm96}, then a much
smaller three-body $\mHt$ formation rate coefficient is obtained, which can be fit to
within $\sim 20\%$ by \citep{glo08}
\begin{equation}
k_{\rm TB1, GL} = 7.7 \times 10^{-31} T^{-0.464} \: {\rm cm^{6}} \: {\rm s^{-1}}.
\end{equation}
There is thus an uncertainty of almost two orders of magnitude in the 
rate coefficient for reaction TB1. In our simulations, we adopt the \citet{abn02} rate 
coefficient as our default value, but in \S\ref{res:chem} we examine the effect 
of using a different rate coefficient.

The rate coefficient for three-body $\mHt$ formation in collisions where
$\mHt$ is the third body (reaction TB2) is known with greater precision, 
but nevertheless substantial uncertainty remains. \citet{pss83} quote a rate 
coefficient 
\begin{equation}
k_{\rm TB2, PSS} = 6.9 \times 10^{-30} T^{-1.0} \: {\rm cm^{6}} \: {\rm s^{-1}}
\end{equation}
for this reaction, again taken from \citet{jgc67}, while \citet{cw83} recommend 
instead 
\begin{equation}
k_{\rm TB2, CW} = 2.8 \times 10^{-31} T^{-0.6} \: {\rm cm^{6}} \: {\rm s^{-1}}. 
\end{equation}
\citet{fh07} assume that the ratio of $k_{\rm TB2}$ to $k_{\rm TB1}$ is the
same as that measured by \citet{jgc67}, i.e., one-eighth. Therefore, their
rate coefficient for reaction TB2 is:
\begin{equation}
k_{\rm TB2, FH} = 1.8 \times 10^{-27} T^{-1.54} \: {\rm cm^{6}} \: {\rm s^{-1}},
\end{equation}
The same assumption applied to the rate coefficient from \citet{glo08} gives
\begin{equation}
k_{\rm TB2, GL} = 9.625 \times 10^{-32} T^{-0.464} \: {\rm cm^{6}} \: {\rm s^{-1}}.
\end{equation}
Finally,  calculations by \citet{schw88} using orbital resonance theory find a 
rate coefficient that is about a factor of two lower than the \citet{cw83} values, 
but \citet{schw90} shows that one of the assumptions underlying his own
orbital resonance calculations is invalid, and provides revised values, obtained 
from a master  equation approach, that agree well with the \citet{cw83} 
recommendation.

These rate coefficients agree to within a factor of a few at $T = 3000 \: {\rm K}$, 
consistent with the scatter in experimental determinations of the
rate coefficient at this temperature \citep{cw83}, but differ more by more than an
order of magnitude at low temperatures. In our simulations, we use 
the \citet{pss83} rate coefficient as our default value, but we examine in 
\S\ref{res:chem} the effect of altering $k_{\rm TB2}$.

We also included three-body formation of $\mHt$ via collisions
with $\He$ (reaction TB3), using a rate coefficient from \citet{wk75}. This 
reaction has not been included in previous treatments of the evolution of dense 
primordial gas and so we wished to assess its effects. We found that  
reaction TB3 could be responsible for anywhere between 0.1\% and 
10\% of the total three-body $\mHt$ formation rate, depending on the 
temperature, the $\mHt$ abundance and the choice of rate coefficients for 
reactions TB1 and TB2. Moreover, this estimate does not take 
into account the uncertainty in the rate of reaction TB3,  which we have 
been unable to quantify, but which should probably be assumed to 
be comparable to the uncertainty in the other three-body rates. Thus, 
although it probably never dominates, reaction TB3 should be included 
if accurate modeling of $\mHt$ formation in dense gas is desired. 

\subsubsection{Deuterated three-body reactions (TB4--TB9,  TB11--TB13,
TB17--TB31, \& TB34--TB35, Table~\ref{tab:TB})}
In view of the large uncertainties present in the rate coefficients of
many of the three-body reaction rates (particularly for reactions
TB1 and TB2, as discussed above), we consider that the most 
prudent course of action when estimating rates for the deuterated 
forms of these reactions is simply to adopt the same values as
for the non-deuterated reactions. Any uncertainty introduced by
this assumption is likely dwarfed by the uncertainties arising from
our poor knowledge of the non-deuterated reaction rates. We note
that \citet{fh07} follow a similar course of action in their study 
of three-body $\mHt$ and $\hd$ formation in primordial gas.

\subsubsection{Destruction of $D_{2}$ by collision with H (reaction IX20, 
Table~\ref{tab:IX})}
Our fit to the data collated by \citet{mie03} for this reaction is accurate
to within a few percent over the temperature range of the tabulated data,
$200 \le T \le 2200 \: {\rm K}$. Outside of this range, our fit may be
significantly inaccurate (although at low temperatures, the reaction
rate is small enough that any inaccuracy is unlikely to be important).

\subsubsection{Photodissociation of $H_{2}$ and HD
(reactions BP7 \& BP8, Table~\ref{phototab})}
\label{h2hdph}
Table~\ref{phototab} lists the rates of these reactions in optically thin gas,
given our assumed incident UV spectrum (see \S\ref{photochem_sec}). 
In optically thick gas,
however, self-shielding of $\mHt$ by $\mHt$ and HD by HD
can significantly reduce both of these rates, by factors 
$f_{\rm sh, H_{2}}$ and $f_{\rm sh, HD}$ respectively. In static,
isothermal gas, these self-shielding factors can be calculated 
approximately using the prescription of \citet{db96} together
with an appropriate set of molecular data, provided that one
knows the $\mHt$ and HD column densities. In gas which is
in motion, with internal velocities comparable to or larger than
the thermal velocity of the gas, the \citet{db96} treatment breaks 
down, and one must use approaches that are either less
accurate or more computationally expensive, as discussed in
\citet{gj07}. However, in the one-zone calculations presented
here, we know neither the $\mHt$ and HD column densities,
nor the velocity structure of the gas, and so including even
a highly approximate treatment of the effects of self-shielding is 
problematic. In our runs with a non-zero UV background, we
therefore consider two limiting cases: one in which self-shielding
is highly efficient and $f_{\rm sh, H_{2}} = f_{\rm sh, HD} = 0$ 
throughout the run, and one in which it is ineffective, and
we remain in the optically thin limit throughout (i.e., 
$f_{\rm sh, H_{2}} = f_{\rm sh, HD} = 1$). These two
limiting cases bracket the true behaviour of the gas.

\subsubsection{Formation and destruction of $\lihhp$ (reactions RA20, DR19, DR20, DR21 \& CD26)}
\label{lihhp_chem}
To date, the $\lihhp$ ion has attracted little attention in the astrochemical
literature. \citet{kd78} considered the reaction chain
\begin{eqnarray}
\lip + \mHt & \rightarrow & \lihhp + \gamma,  \\
\lihhp + \me & \rightarrow & \lih + \mH 
\end{eqnarray}
(reactions RA20 and DR20 in our chemical model)
as a possible source of $\lih$ in the interstellar medium, but showed
that even if the rate coefficient for the radiative association were 
assumed to be very large ($k_{\rm RA20} \sim 10^{-16} \: {\rm cm^{3}} \:
{\rm s^{-1}}$), 
the resulting $\lih$ abundance would be far too small to be 
observable. More recently, \citet{sld96} considered the $\lihhp$
ion in their comprehensive study of the lithium chemistry of the
primordial intergalactic medium, but again reached the conclusion
that its abundance would be very small, and so chose not to include
it in their chemical model. However, to the best of our knowledge,
there has been no previous investigation of the role that this
ion may play in regulating the fractional ionization of very dense
primordial gas. 

Previous work focussing on modelling the fractional ionization at
high densities \citep{ms04,ms07}, in the context of the study of ambipolar
diffusion in dense population III prestellar cores, has shown that
once the free electron fraction falls below $x \sim 10^{-10}$, 
ionized lithium takes over from ionized hydrogen as the primary
positive ion in the gas. It is therefore
important to ensure that all of the major loss routes for $\lip$ are
represented in the chemical model. In addition to photorecombination 
(reaction PR4), $\lip$ can be removed from the gas by a number of
reactions with atomic or molecular hydrogen:
\begin{eqnarray}
\lip + \mH & \rightarrow & \li + \Hp, \\
\lip + \mH & \rightarrow & \lihp + \gamma, \\
\lip + \mHt & \rightarrow & \li + \mHtp, \\
\lip+ \mHt & \rightarrow & \lihp + \mH, \\
\lip + \mHt & \rightarrow & \lih + \Hp, \\
\lip + \mHt & \rightarrow & \lihhp + \gamma.
\end{eqnarray}
Most of these processes are highly endothermic, and so are not
competitive with photorecombination even when $x$ is small. 
However, the two radiative association reactions are exothermic
and deserve closer scrutiny. Radiative association with atomic
hydrogen (reaction RA10) has been included in a number of previous
models of primordial gas chemistry \citep[e.g.][]{sld96,gp98}, and 
accurate quantal calculations of the rate coefficient for  this reaction 
are available \citep{dks96,ggg96}. However, at the densities of interest 
in the present case, efficient three-body formation of $\mHt$ ensures that
the hydrogen is primarily molecular rather than atomic, and so 
renders radiative association with $\mHt$ (reaction RA20) the more
important reaction. Unfortunately, we have been unable to locate
any calculation of the rate coefficient of this reaction. \citet{kd78}
quote an upper limit of $k_{\rm RA20} = 10^{-16} \: {\rm cm^{3}} \: {\rm s^{-1}}$,
while \citet{sld96} quote an upper limit of $k_{\rm RA20} = 10^{-17} \: {\rm cm^{3}} 
\: {\rm s^{-1}}$, but the true rate coefficient could be orders of magnitude 
smaller. In our reference model, we make the conservative assumption
that the rate coefficient is of the same order of magnitude as that
for reaction RA10, and hence adopt a value $k_{\rm RA20} = 10^{-22} 
\: {\rm cm^{3}} \: {\rm s^{-1}}$. We investigate the effects of adopting 
a larger value in section \S\ref{res:chem:lih2p}. 

The inclusion of reaction RA20 in our chemical network necessitates the
inclusion of additional reactions: the dominant destruction mechanisms 
for $\lihhp$. Unfortunately, there has been very little theoretical or experimental
study of any reactions involving $\lihhp$ and so it is not even clear which processes
dominate. The best studied destruction process is dissociative recombination
(reactions DR19, DR20 and DR21):
\begin{eqnarray}
\lihhp + \me & \rightarrow & \li + \mHt \\
 & \rightarrow & \lih + \mH \\
  & \rightarrow & \li + \mH + \mH.
\end{eqnarray}
\citet{thomas06} have studied this process experimentally using the CRYRING 
heavy ion storage ring, and have reported preliminary results regarding the 
branching ratio of the reaction, but have not yet reported any value for the
total rate. C.\ Greene and collaborators are currently involved in a theoretical
calculation of the total rate coefficient, but again have yet to publish any results.
However, their preliminary findings suggest a total rate coefficient that is about
2.5--3 times larger than that for the dissociative recombination of $\htp$.
(C. Greene, private communication). We therefore adopt a total rate coefficient 
$2 \times 10^{-7} \left( \frac{T}{300} \right)^{-1/2} \: {\rm cm^{3}} \: {\rm s^{-1}}$ 
for the dissociative recombination 
of $\lihhp$, and use the values quoted by \citet{thomas06} for the branching ratios.

Other reactions that could be important destruction mechanisms for $\lihhp$
include
\begin{eqnarray}
\lihhp + \mH & \rightarrow & \lihp + \mHt \\
\lihhp + \mH & \rightarrow & \lip + \mHt + \mH \\
\lihhp + \mHt & \rightarrow & \lip + \mHt + \mHt \\
\lihhp + \He & \rightarrow & {\rm LiHe^{+}} + \mHt.
\end{eqnarray}
None of these reactions appears to have previously been studied in the
astrochemical literature, and so their rates are unknown. For simplicity, 
therefore, we include only a single representative example from this set
of reactions, namely the collisional dissociation of $\lihhp$ by $\mHt$
(reaction CD26). As we are primarily interested in the role of $\lihhp$ within the highly 
molecular dense core, it is likely that this reaction will dominate, unless
its rate coefficient is unusually small. In our reference model, we adopt
a rate coefficient of $k_{\rm CD26} = 1.0 \times 10^{-9} \expf{-}{3250}{T}
\: {\rm cm^{3}} \: {\rm s^{-1}}$
for this reaction; however, in \S\ref{res:chem:lih2p} we examine the
effect of adopting a smaller value.

\subsection{Photochemistry}
\label{photochem_sec}
To compute rates for the photochemical reactions listed in 
Table~\ref{phototab}, we assume that the gas is illuminated
by an external background radiation field with the spectral shape
of a $10^{5} \: {\rm K}$ black-body at energies $h\nu < 13.6 \: 
{\rm eV}$, and which is zero at higher energies. This choice of
spectrum is motivated by the fact that the brightest population III
stars are expected to have high effective temperatures, 
$T_{\rm eff} \simeq 10^{5} \: {\rm K}$ \citep{coj00}, while the 
cutoff at the Lyman limit is intended to account for the effects of
absorption by neutral hydrogen in the intergalactic medium.
We quantify the strength of this background radiation field in 
terms of the flux at the Lyman limit, $J(\nu_{\alpha}) = 10^{-21} 
J_{21} \: {\rm erg} \: {\rm s^{-1}} \: {\rm cm^{-2}} \: {\rm Hz^{-1}} 
\: {\rm sr^{-1}}$. The rates listed in Table~\ref{phototab} are 
computed assuming that $J_{21} = 1.0$, but scale linearly with
$J_{21}$ and so can easily be rescaled for other values of the
background radiation field strength.

\subsection{Cosmic rays}
\label{cosmic_rays}
If cosmic rays are present, then they will directly ionize some
species and indirectly photoionize and photodissociate others.
Direct ionization is simple to treat, and the appropriate rates
are listed in Table~\ref{cr_ion_tab}, normalized by the cosmic ray 
ionization rate of atomic hydrogen, $\zeta_{\rm H}$, which we treat 
as a free parameter. However, the indirect effects of the cosmic
rays are harder to model accurately. 

The basic physics is straightforward, and was first discussed by
\citet{pt83}. They noted that the secondary electrons produced by
cosmic ray ionizations are energetic enough to excite the electronic 
states of $\mHt$, and that the subsequent radiative decay of these
excited states would produce ultraviolet photons. In the Galactic
context, the mean free path of these photons is small, and so the
cosmic-ray induced photochemistry can be modelled as a purely 
local process \citep[see e.g.,][]{gldh89}. 

In the population III star formation context in which we are interested,
however, there are two main factors that complicate matters. First, the 
$\mHt$ fraction in the gas is small at densities $n \ll 10^{10} \: {\rm cm^{-3}}$,
and so most of the secondary electrons produced by the cosmic rays
lose energy by exciting and ionizing atomic hydrogen, rather than molecular hydrogen.
Second, an accurate treatment of the propagation of the photons 
produced by excited $\mH$ and $\mHt$ is far more involved than in 
the Galactic case. The continuum opacity of metal-free gas is very small at 
most densities of interest \citep{lcs91}, owing to the absence of dust absorption, 
and so the majority of the photons produced by cosmic-ray induced excitation of 
$\mHt$ have large mean free paths. On the other hand, Lyman-$\alpha$ 
photons produced by the excitation of atomic hydrogen have small mean 
free paths, but scatter many, many times before escaping the gas 
\citep[see e.g.,][]{dhs06}. In neither case is it a good approximation to
assume that all of the photons are absorbed locally in the gas, and so the
simple local treatment developed for Galactic dark clouds no longer applies.

If we consider only the effects of emission from $\mHt$, then to compute 
$R_{\rm X}$, the photoionization (or photodissociation) rate per unit 
volume of species X, one must use an equation of the form
\begin{equation}
R_{\rm X}(\mathbf{x}) =  \frac{1}{4\pi} \int_{0}^{\infty} \sigma_{\rm X}(\nu)
\int_{V} \frac{\epsilon(\nu, \mathbf{x^\prime}) e^{-\tau(\nu, \mathbf{x^\prime}, \mathbf{x})}}
{|\mathbf{x^\prime} - \mathbf{x}|^{2}} \: {\rm d}\mathbf{x^\prime} \: {\rm d}\nu.
\label{rx_full}
\end{equation}
where the volume $V$ that we integrate over corresponds to the entirety of the
protogalactic core,
and where $\tau(\nu, \mathbf{x^\prime}, \mathbf{x})$ is the optical depth of the
gas between point $\mathbf{x}$ and point $\mathbf{x^\prime}$
at a frequency $\nu$. The photon emissivity $\epsilon(\nu, \mathbf{x^\prime})$ is
given in this case by 
\begin{equation}
\epsilon(\nu, \mathbf{x^\prime}) = P_{\mHt}(\nu) \zeta_{\mHt} n_{\mHt},
\end{equation}
where  $P_{\mHt}(\nu) \, {\rm d}\nu$ is the probability that the 
cosmic ray ionization of $\mHt$ leads to the production of a 
photon with a frequency in the range $\nu \rightarrow \nu + {\rm d}\nu$,
and  $\zeta_{\mHt}$ is the cosmic ray ionization rate of $\mHt$.

For gas at the center of a spherically symmetric protogalactic core, we can
simplify Equation~\ref{rx_full} to 
\begin{equation}
R_{\rm X} =  \int_{0}^{\infty} \sigma_{\rm X}(\nu)
\int_{0}^{R} \epsilon(\nu, r) e^{-\tau(\nu, r)}
\: {\rm d}r \: {\rm d}\nu.
\label{rx_sym}
\end{equation}
where $\tau(\nu, r)$ is the optical depth between the center of the halo
and gas at a radius $r$, $\epsilon(\nu, r)$ is the emissivity at $r$,
and $R$ is the core radius.  However, even after making
this simplification, calculation of $R_{\rm X}$ still requires more information
than we have available in our one-zone calculation, namely the radial 
profiles of density, temperature and chemical abundances, which 
determine both $\tau(\nu, r)$ and $\epsilon(\nu, r)$. In their absence,
we are forced to approximate. 

If we assume that the protogalactic core has a density structure with a `core 
plus halo' form, i.e.,
\begin{equation}
n(r) = \left\{ \begin{array}{lr}
n_{\rm c} & r < r_{\rm c} \\
n_{\rm c} (r_{c} / r)^{\alpha}, & r > r_{\rm c}
\end{array} \right.
\end{equation}
then provided that $\alpha > 1$, the integral in Equation~\ref{rx_full}
will be dominated by the contribution from the core of the density profile.
If we further assume that the core is chemically homogeneous, and 
that $\tau(\nu, r) \sim 0$, then we can approximate Equation~\ref{rx_full} as
\begin{equation}
R_{\rm X} \simeq  \int_{0}^{\infty} \sigma_{\rm X}(\nu)
r_{\rm c} P_{\mHt}(\nu) \zeta_{\mHt} n_{\rm c, H_{2}}  \: {\rm d}\nu,
\label{rx_approx}
\end{equation}
where $n_{\rm c, H_{2}}$ is the number density of $\mHt$ within the core. 
Note that even if the point we are considering is not directly at the center
of the core, Equation~\ref{rx_approx} remains a reasonable approximation
to $R_{\rm X}$, provided that we are considering a point within $r_{\rm c}$,
and that $r_{\rm c} \ll R$. Provided that our approximations hold, 
Equation~\ref{rx_approx} allows us to reduce what is formally a non-local 
problem into one that can be treated as if it were local.

To properly include the effects of hydrogen excitation, one would have
to solve for the radiative transfer of the Lyman alpha photons within
the collapsing protostellar core. However, as the outcome would be
highly sensitive to the assumed density and velocity profiles of the gas,
which are not available from our one-zone calculation, the wisdom of
performing such a detailed calculation for each of our simulations that
include cosmic rays is questionable; we run the risk of getting an answer
that is completely determined by our assumptions, and that therefore
is not robust. Instead, we have chosen a more conservative course, 
and have attempted to put limits on the effects of the Lyman alpha
photons by considering two limiting cases: one in which they do not
propagate significantly into the core of the protogalaxy, and do not
contribute to $R_{\rm X}$ (which is thus given in this case by 
Equation~\ref{rx_approx} above), and 
another in which the optical depth of the gas to the Lyman
alpha photons is negligible, and $R_{\rm X}$ is given by a generalization
of Equation~\ref{rx_approx}:
\begin{equation}
R_{\rm X} \simeq  \int_{0}^{\infty} \sigma_{\rm X}(\nu)
r_{\rm c} \left( P_{\mHt}(\nu) \zeta_{\mHt} n_{\rm c, H_{2}} +
P_{\mH}(\nu) \zeta_{\mH} n_{\rm c, H} \right) \: {\rm d}\nu,
\label{rx_approx_2}
\end{equation}
where  $P_{\mH}(\nu) \, {\rm d}\nu$ is the probability that the
cosmic ray ionization of $\mH$ leads to the production of a 
photon with a frequency in the range $\nu \rightarrow \nu + {\rm d}\nu$,
$\zeta_{\mH}$ is the cosmic ray ionization rate of $\mH$, and
$n_{\rm c, H}$ is the number density of atomic hydrogen in the core

To evaluate $R_{\rm X}$, it remains necessary to specify $r_{\rm c}$. In our 
calculations, we assume, following \citet{om00} and \citet{om05}, that 
$r_{\rm c}$ is given by the current  Jeans length. 

In Table~\ref{cr_pd_tab}, we list estimated values for 
$\sigma_{\rm X, eff, \mHt} = \int_{0}^{\infty} \sigma_{\rm X}(\nu) P_{\mHt}(\nu)
\: {\rm d}\nu$ and $\sigma_{\rm X, eff, \mH} = \int_{0}^{\infty}  
\sigma_{\rm X}(\nu) P_{\mH}(\nu) \: {\rm d}\nu $
for both photoionization and photodissociation for a number of
different chemical species. To compute these values, we assumed
that all of the photons produced by excited hydrogen are emitted
in the Lyman-$\alpha$ line, implying that $P_{\mH} = \delta (\nu_{\alpha} 
- \nu)$, where $\nu_{\alpha}$ is the frequency of Lyman-$\alpha$ and
$\delta$ is the Dirac delta function. For $P_{\mHt}$, we used estimated
values based on the emission spectra given in \citet{sdl87}; note that these
are likely accurate only to within a factor of a few. Given these values, the rates 
for the cosmic-ray induced photoionization and photodissociation of 
these species in our model cores can be calculated using 
Equation~\ref{rx_approx} or \ref{rx_approx_2}, as appropriate.

\subsection{Neglected processes}
\label{neglect}
Although the chemical network presented in this paper is, to the
best of our knowledge, the most comprehensive network used to
date to simulate primordial gas chemistry, there remain a large
number of possible reactions that we have not included. 
Below, we discuss which types of processes have been omitted,
and why.

\begin{enumerate}
\item We do not include the formation of $\mHt$ or $\DD$ by the radiative
association of ground state atomic hydrogen or deuterium, respectively, on
the grounds that the rate coefficients for these processes are negligible.

\item We have not included reactions that involve electronically excited 
atomic hydrogen (as considered in \citealt{lb91} or \citealt{rdb93}, for 
instance). We justify this omission by noting that the population of the
$n=2$ electronic level of atomic hydrogen will be very small at the densities 
and temperatures considered in this work, on account of the large Einstein
coefficient associated with the Lyman-$\alpha$ transition and the large
energy separation of the $n=1$ and $n=2$ levels. For similar reasons, we 
have not included any reactions that require electronically excited deuterium, 
helium or lithium.

\item We have restricted the range of chemical species considered
to those with three or fewer atoms. In principle, the formation of larger 
species is possible -- for instance, ${\rm H_{5}^{+}}$ can be formed 
from $\htp$ by the radiative association reaction \citep{pau95}
\begin{equation}
\htp + \mHt \rightarrow {\rm H_{5}^{+}} + \gamma.
\end{equation}
However, the chemical abundances of the three-atom species in our
model are very small, and we expect the abundance of even
larger species to be much smaller still. It therefore seems unlikely
that they will play any significant role in the cooling or chemistry of
the gas.

\item We have omitted any photochemical reactions that require photons
with energies greater than $13.6 \: {\rm eV}$, under the assumption that
any such photons emitted by external sources of radiation will be absorbed
in the intergalactic medium, or in the interstellar medium of the protogalaxy,
before reaching the particular collapsing core under study. Moreover, since
we consider only the initial collapse of the core, internal sources of radiation,
such as a central protostar, fall outside of the scope of this work

\item We do not include processes that have negligible reaction rates at
all temperatures treated in this work. This includes the production of 
doubly-ionized helium, ${\rm He^{2+}}$, or doubly or triply ionized lithium, 
${\rm Li^{2+}}$ and ${\rm Li^{3+}}$ by collisional ionization \citep[see e.g.,][]{lsd02}, 
which are therefore omitted from the chemical model.

\item We have ignored the effects of stimulated radiative association and
stimulated radiative attachment, i.e., reactions of the form
\begin{eqnarray}
{\rm X} + {\rm Y} + \gamma_{\rm b} & \rightarrow & {\rm XY} + \gamma + \gamma_{\rm b}, \\
{\rm X} + \me + \gamma_{\rm b} & \rightarrow & {\rm X^{-}} + \gamma + \gamma_{\rm b},
\end{eqnarray}
where $\gamma_{\rm b}$ represents a background photon. The influence of stimulated 
radiative association or attachment on the production of various molecules in primordial 
gas ($\lih$, HD, $\Hm$, ${\rm Li^{-}}$ and $\hehp$) has been investigated 
\citep{sd97a,sd97b,sd98,zsd98}, generally for the case of a black-body radiation field.
However, significant effects are found only for radiation temperatures $T_{\rm rad} >
500 \: {\rm K}$, much larger than the CMB temperature at the redshifts of interest in
this study. The background radiation fields considered in \S\ref{res:back} have the
same shape below $h\nu = 13.6 \: {\rm eV}$ as a $10^{5} \: {\rm K}$ black-body, but 
have intensities that are orders of magnitude weaker than a true black-body radiation
field with this temperature, and so are also unimportant in this context. Therefore,
it is clear that the influence of these simulated processes will be negligible.

\item We have omitted collisional processes such as
\begin{equation}
\mHt + \mHt \rightarrow \mH + \mH + \mH + \mH,
\end{equation}
or 
\begin{equation}
\mH + \mH \rightarrow \Hp + \Hp + \me + \me,
\end{equation}
that have energy thresholds corresponding to temperatures significantly higher 
than those considered in this work, as we do not expect these processes to play
an important role in low temperature gas.

\item We do not include dissociative charge transfer reactions involving $\mHtp$
or its isotopologues, e.g.
\begin{eqnarray}
\mH + \mHtp & \rightarrow & \mH + \mH + \Hp, \\
\mH + \hdp  & \rightarrow & \mH + \mD + \Hp,
\end{eqnarray}
because at the temperature of interest in this study, the cross-sections for these
processes are far smaller than those for the equivalent non-dissociative charge 
transfer reactions \citep{kr02,ks03}
\begin{eqnarray}
\mH + \mHtp & \rightarrow & \mHt + \Hp, \\
\mH + \hdp  & \rightarrow & \hd + \Hp.
\end{eqnarray}

\item We have not included charge transfer from $\Hep$ to ${\rm Li}$
\begin{equation}
\Hep + \li \rightarrow \He + \lip,
\end{equation}
or its inverse
\begin{equation}
\He + \lip \rightarrow \Hep + \li.
\end{equation}
The first of these reactions is unimportant in comparison to charge transfer from
$\Hp$ owing to the low $\Hep$ abundances we find in our simulations. The second
reaction is negligible at $T < 10000 \: {\rm K}$ due to its large endothermicity.

\item We have omitted all collisional dissociation reactions caused
by minor molecules or ions, e.g.\ HD, Li, LiH, etc. Collisional dissociation
reactions involving HD, such as
\begin{equation}
\mHt + \hd \rightarrow \mH + \mH + \hd,
\end{equation}
will be unimportant compared to the analogous reactions involving $\mHt$,
while reactions of the form
\begin{equation}
{\rm XY} + {\rm Li} \rightarrow {\rm X} + {\rm Y} + {\rm Li},
\end{equation}
will be unimportant due to the very small abundance of lithium relative to
hydrogen.

\item We have ignored a large number of possible three-body
processes: specifically, every process that involves any species
other than $\mH$, $\mHt$ or $\He$ as the third body. At the densities
at which three-body reactions become significant, the abundances of 
these three species are orders of magnitude higher than the 
abundances of any other species, and so it is easy to justify the
omission of these minor contributions.

\item We do not include transfer reactions involving two ions of the same charge, e.g.\
\begin{equation}
\mHtp + \mHtp \rightarrow \htp + \Hp,
\end{equation}
as the mutual Coulomb repulsion of the ions renders these reactions ineffective at
the temperatures considered in this work.

\item We do not include the double ionization of $\mHt$ or He by cosmic rays, i.e.\
\begin{eqnarray}
\mHt + {\rm C.R.} & \rightarrow & \Hp + \Hp + \me + \me \\
\He + {\rm C.R.} & \rightarrow & {\rm He^{++}} + \me + \me,
\end{eqnarray}
as the fraction of cosmic ray ionization events leading to these outcomes is expected
to be very small \citep{gl73}. 

\item We assume that in mutual neutralization reactions involving $\htp$ or one of
its isotopologues, complete breakup of the ion is unlikely to occur; this is in line with
e.g.\ the detailed chemical network of \citet{lmm00}, which includes the processes
\begin{equation}
\htp + \Hm \rightarrow \mHt + \mHt
\end{equation}
and
\begin{equation}
\htp + \Hm \rightarrow \mHt + \mH + \mH,
\end{equation}
but not
\begin{equation}
\htp + \Hm \rightarrow \mH + \mH + \mH + \mH.
\end{equation}

\item A number of possible reactions involving ${\rm LiH_{2}^{+}}$ have been omitted, as
have any reactions involving ${\rm LiHD^{+}}$ or ${\rm LiD_{2}^{+}}$. As we have already 
discussed in \S\ref{lihhp_chem}, we know very little regarding the values of the rate coefficients 
for many of the most important formation and destruction processes for ${\rm LiH_{2}^{+}}$, 
rendering its abundance highly uncertain. In view of this large uncertainty, there is little to be
gained by adding in additional, equally uncertain but less important processes involving 
${\rm LiH_{2}^{+}}$, or by considering the chemistry of the deuterated forms of the molecular 
ion.

\end{enumerate}

\section{Thermal processes}
\label{cool}
Our model of the thermal behaviour of the gas includes the effects of heating
and cooling from a large number of different radiative and chemical processes.
A full list of the processes included is given in Table~\ref{cool_model}, 
while a more detailed discussion is given below.

\begin{table}
\caption{Processes included in our thermal model \label{cool_model}}
\begin{tabular}{llcc}
\hline
Species & Process & Collision partner(s) & Refs.\ \\
\hline
{\bf Cooling:} & & & \\
$\mHt$ & Rovibrational lines & $\mH$, $\mHt$, $\He$, $\Hp$, $\me$ & 1, 2 \\
$\mHt$ & Collision-induced emission & $\mHt$ & 2 \\
$\hd$ & rovibrational lines  & $\mH$ & 3 \\
$\lih$ & rovibrational lines  & $\mH$, $\mHt$ & 4 \\
$\htp$ & rovibrational lines & $\mH$, $\mHt$, $\He$, $\me$ & 5 \\
$\mH$ & resonance lines & $\me$ & 6 \\
CMB photons & Compton scattering & $\me$ & 6 \\
$\mHtp$, $\hdp$, $\ddp$ & Rovibrational lines & $\mH$, $\me$ & 7 \\
Minor species & Rovibrational lines & $\mH$, $\mHt$, $\He$, $\me$ & 8 \\
$\Hp$ & Recombination & $\me$ & 9 \\
$\Hep$ & Recombination & $\me$ & 10 \\
$\mH$ & Collisional ionization & $\me$ & 11 \\
$\mHt$ & Collisional dissociation & $\mH$, $\mHt$, $\He$, $\me$  & 12 \\
$\Hp$ & Charge transfer (reaction CT19) & $\mHt$ & 13 \\
\hline
{\bf Heating:} & & & \\
$\mHt$ & Formation & --- & 12 \\
$\mHt$ & Photodissociation & --- & 14 \\
$\mHt$ & Ultraviolet pumping & --- & 15 \\  
Cosmic rays & Ionization/excitation & --- & 16 \\ 
\hline
\end{tabular}
\medskip
\\
{\bf Note:}
See the appropriate subsections in \S\ref{cool} for details of how we have decided which
collision partners to include. \\
{\bf References:}
1 -- \citet{wf07}; 
2 -- \citet{ra04}; 
3 -- \citet{lna05};
4 -- \citet{gp98};
5 -- \citet{nmt96}, \citet{oe04};
6 -- \citet{black81}, \citet{cen92};
7 -- See \S\ref{h2p_cooling};
8 -- See \S\ref{minor_cool};
9 -- \citet{fer92};
10 -- \citet{hs98};
11 -- \citet{j87};
12 -- See \S\ref{chem} and  \S\ref{cool:react};
13 -- \citet{skhs04};
14 -- \citet{bd77};
15 -- \citet{bht90};
16 -- \citet{gl78} \\
\end{table}

\subsection{$\mathbf{H_{2}}$ cooling}
In our treatment of $\mHt$ rovibrational cooling at low densities, we include
the effects of collisions between $\mHt$ and $\mH$, $\mHt$, $\He$, $\Hp$ and $\me$, 
using fitting formulae from \citet{ga08}. At high densities, we use the standard
LTE cooling function \citep[see e.g.,][]{hm79}. We 
assumed the usual value of 3:1 for the $\mHt$ ortho-para ratio, which
\citet{ga08} have demonstrated is a good approximation for
all temperatures in the range in which $\mHt$ cooling is important. 
Although the revised treatment of $\mHt$ cooling presented by 
\citet{ga08} can make a significant difference to the thermal evolution
of the gas in some circumstances -- notably, in gas with a substantial
fractional ionization -- we do not expect it to have a significant impact
on our current results, as at the densities of greatest interest in this paper,
$\mHt$ is well within the LTE cooling regime.

An important source of inaccuracy at high densities is the treatment
of the opacity of the $\mHt$ emission lines. In our models, we
follow \citet{ra04} and model optically thick $\mHt$ cooling with the 
expression
\begin{equation}
 \Lambda_{\rm \mHt, thick} = \Lambda_{\rm \mHt, thin} \times {\rm min} 
 \left[1, (n/n_{0})^{ -\beta} \right],
\end{equation}
where  $ \Lambda_{\rm \mHt, thick}$ and $ \Lambda_{\rm \mHt, thin}$ are 
the optically thick and optically thin cooling rates, respectively, $n$ is the
number density of hydrogen nuclei, and where $n_{0} = 8 \times 10^{9} 
\: {\rm cm^{-3}}$ and $\beta = 0.45$. This approximation works well for 
modelling gas at the center of a collapsing core, but is less accurate 
when used to treat $\mHt$ cooling in the surrounding envelope 
(N.\ Yoshida,  private communication). \citet{yoha06} present a more accurate 
approach based on the computation of escape probabilities for each 
individual $\mHt$ line using the Sobolev approximation. However,
this treatment requires dynamical information, in the form of the local
velocity gradient, that is not available in any meaningful form in our 
simple models, and so for our current study we must be content with
the \citet{ra04} approach.

At very high densities ($n > 10^{14} \: {\rm cm^{-3}}$),  cooling from $\mHt$
becomes dominated by collision-induced emission (CIE). When an $\mHt$
molecule collides with a hydrogen or helium atom, or another $\mHt$
molecule, the particles involved briefly act as a `supermolecule' with a
non-zero electric dipole, which has a high probability of emitting a photon.
Because the collision time is very short, the resulting collision-induced
emission lines are very broad, and typically merge into a continuum. In 
our model, we model the CIE cooling rate with a power-law approximation
taken from \citet{ra04}, valid for gas in which $x_{\mHt} > 0.5$:
\begin{equation}
\Lambda_{\rm CIE} = 4.578 \times 10^{-49} T^{4} n_{\mHt} n 
\: {\rm erg} \: {\rm s^{-1}} \: {\rm cm^{-3}}.
\end{equation}
Although cooling from collision-induced emission contributes only 12\%
of the total cooling at the density reached at the end of our simulation,
$n_{\rm f} = 3 \times 10^{13} \: {\rm cm^{-3}}$, we have verified in test
runs that at higher densities it very rapidly 
becomes the dominant form of cooling, justifying our decision to end
our simulations at this point.

\subsection{$\mathbf{H_{3}^{+}}$ cooling}
\label{h3p_cool_rate}
In the LTE limit, it is straightforward to calculate the $\htp$ cooling rate using
the data presented by \citet{nmt96}. In our simulations, we use tabulated 
values computed directly from the \citeauthor{nmt96} line list. However, 
for the convenience of readers, we also provide an analytical fit of the form
\begin{equation}
\log \left[ \frac{\Lambda_{\rm \htp, LTE}}{n_{\htp}} \right] = \sum_{i=0}^{8} a_{i} (\log T)^{i},
\label{h3p_fit_eq}
 \end{equation}
where $\Lambda_{\rm \htp, LTE}$ is the $\htp$ cooling rate per unit volume
in the LTE limit, and where the values of the $a_{i}$ coefficients are listed in Table~\ref{h3p_coeffs}. 
This fit is accurate to within 25\% for temperatures in the range 
$20 < T < 400 \: {\rm K}$ and to within a few percent for
temperatures in the range $400 < T < 10000 \: {\rm K}$. At high temperatures, 
the \citet{nmt96} line list is known to be incomplete, and so for 
$T > 3000 \: {\rm K}$, we may systematically underestimate the cooling 
due to $\htp$. However, as we shall see in \S\ref{results}, we never find 
gas at these temperatures in the regime where $\htp$ is potentially 
important, and so this incompleteness will not affect our results.

\begin{table}
\caption{Numerical coefficients used in our analytical fit to the 
$\htp$ cooling rate \label{h3p_coeffs}}
\begin{tabular}{lrrr}
\hline
& \multicolumn{1}{c}{LTE} & \multicolumn{1}{c}{LTE} & \multicolumn{1}{c}{$n \rightarrow 0$} \\
& $20 < T < 400 \: {\rm K}$ & $ 400 < T < 10000 \: {\rm K}$ & $20 < T < 10000 \: {\rm K}$ \\
\hline
$a_{0}$ & $-1.6583133 \times 10^{4}$ & $9.5033824 \times 10^{3}$ & $-7.9192725$ \\
$a_{1}$ & $5.0808831 \times 10^{4}$ &$-1.7832745 \times 10^{4}$ & $-43.505799$ \\
$a_{2}$ & $-5.9475456 \times 10^{4}$ & $1.2847118 \times 10^{4}$ & $41.100652$ \\ 
$a_{3}$ & $2.8459331 \times 10^{4}$ & $-3.9079919 \times 10^{3}$ & $-17.327161$ \\
$a_{4}$ & $1.9988968 \times 10^{3}$& $-2.8286326  \times 10^{1}$ & $3.3895649$ \\
$a_{5}$ & $-8.6370305 \times 10^{3}$ & $3.7394515 \times 10^{2}$ & $-0.24931287$ \\
$a_{6}$ & $4.0429912  \times 10^{3}$ & $-1.1130317  \times 10^{2}$ & 0.0 \\
$a_{7}$ & $-8.2863818  \times 10^{2}$ & $1.4187579  \times 10^{1}$ & 0.0 \\
$a_{8}$ & $6.5975582 \times 10^{1}$ & $-6.9969136 \times 10^{-1}$ & 0.0 \\
\hline
\end{tabular}
\end{table}

At densities where $\htp$ is not in LTE, the calculation of the $\htp$ cooling 
rate presents more of a problem. A commonly used approximation for
dealing with the cooling from molecular species \citep[see e.g.,][]{hm79} is 
to compute the cooling at a density $n$ using the expression 
\begin{equation}
\Lambda = \frac{\Lambda_{\rm LTE}}{1 + (n_{\rm cr} / n)},
\end{equation}
where $\Lambda_{\rm LTE}$ is the LTE cooling rate per unit volume, and where
the critical density $n_{\rm cr}$ is given by  $n_{\rm cr} / n = \Lambda_{\rm LTE} / 
\Lambda_{n \rightarrow 0}$, where $\Lambda_{n \rightarrow 0}$ is the 
cooling rate per unit volume in the $n \rightarrow 0$ limit. For 
$n \ll n_{\rm cr}$ and $n \gg n_{\rm cr}$, this expression is highly accurate, 
while for $n \sim n_{\rm cr}$ it does a reasonable job of capturing the basic
behaviour, at a far smaller computational cost than a full level population
calculation would require. We adopt this approximation in our treatment 
of $\htp$ cooling, reducing the problem of calculating $\Lambda_{\htp}$ to 
one of calculating $\Lambda_{\htp, n \rightarrow 0}$. Here, however, we 
hit a problem. To compute  $\Lambda_{\htp, n \rightarrow 0}$, we must evaluate
\begin{equation}
\label{htp_low_basic}
\Lambda_{\htp, n \rightarrow 0} = n_{\htp} \sum_{j} C_{0j} E_{0j},
\end{equation}
where $C_{0j}$ is the rate of collisional excitation from the $\htp$ ground 
state\footnote{The $(J,K) = (1,1)$ rotational level of the vibrational ground 
state; occupation of the $(J,K) = (0,0)$ level is forbidden by the Pauli
exclusion principle}, here denoted as level 0, to an excited
level $j$, and $E_{0j}$ is the energy difference between level 0 and level
$j$. The collisional excitation rate for transitions from $0 \rightarrow j$ is 
simply
\begin{equation}
 C_{0j} = q_{0j, \mH} n_{\mH} + q_{0j, \mHt} n_{\mHt} + q_{0j, \He} n_{\He} +
 q_{0j, \me} n_{\me},   \label{htp_low_2}
\end{equation}
where $q_{0j, \mH}$, $q_{0j, \mHt}$, $q_{0j, \He}$ and $q_{0j, \me}$ are the
collisional excitation rate coefficients for collisions with $\mH$, $\mHt$, $\He$ and 
$\me$, respectively; $n_{\mH}$, $n_{\mHt}$, $n_{\He}$ and $n_{\me}$ are the 
corresponding particle number densities; and where we have ignored the 
effect of collisions with protons (which are unimportant in the case of a positively
charged ion such as $\htp$) or with minor ionic or molecular species such as
$\Hm$ or $\hd$. The difficulty in computing the collisional terms, and hence the 
low density limit of the $\htp$ cooling rate, arises because most of the required 
collisional excitation rate coefficients are unknown. \citet{ft03} give rate coefficients 
for the collisional excitation of a number of low-lying rotational states by collisions with
electrons, but in the high density, low ionization conditions of interest in this study, 
collisions with electrons are unimportant, and analogous datasets for collisions 
with $\mH$, $\He$ or $\mHt$ are not available.

To deal with this problem, we have used an approach introduced by \citet{oe04}.
They computed rate coefficients for rotational transitions in $\htp$ caused by collisions 
with $\mHt$ by making use of the principle of detailed balance and by assuming that
the collisional transitions are completely random (i.e., that they obey no selection
rules). These assumptions led them to suggest rate coefficients of the form
\begin{equation}
 q_{ij} = K_{ij} \sqrt{\frac{g_{j}}{g_{i}}} \exp \left(-\frac{E_{j} - E_{i}}{2kT} \right)
\end{equation}
for transitions between an initial level $i$ and final level $j$, where $g_{i}$
and $g_{j}$, are the statistical weights of levels $i$ and $j$, respectively,
$E_{i}$ and $E_{j}$ are the corresponding level energies, and $K_{ij}$ is a 
normalizing factor given by
\begin{equation}
K_{ij} = C \left\{ 1 + \sum_{m} \left(\frac{g_{m}}{\sqrt{g_{j}g_{i}}}\right)^{1/2}
 \exp \left[-\frac{E_{m} - (1/2)(E_{j} + E_{i})}{2kT} \right] \right\}^{-1},
\end{equation}
where $C$ is the total collision rate, which is independent of $i$ and $j$,
and where the summation does not include levels $i$ or $j$. Although \citet{oe04} 
consider only pure rotational transitions, the same scheme can be used to treat 
ro-vibrational transitions. 

In our treatment, we assume that the Oka \& Epp scheme can be used to treat collisions 
with atomic hydrogen and helium as well as $\mHt$, and hence are able to
determine the temperature dependence of the set of collisional excitation rate coefficients
for each collider ($ q_{0j, \mH}$, $ q_{0j, \mHt}$ and  $q_{0j, \He}$); we ignore
collisions with electrons, on the grounds of the very small electron abundance 
that exists at the densities where $\htp$ cooling is potentially  important.
Using these collisional excitation rate coefficients, we can then construct $C_{0j}$ 
via Equation~\ref{htp_low_2}, from which
 $\Lambda_{\htp, n \rightarrow 0}$ follows via Equation~\ref{htp_low_basic}. 
The overall normalization of the cooling rate remains uncertain, as it depends on 
the $\htp$ number density, and on the total collision rates with each of $\mH$,
$\mHt$ and $\He$, which we can write as $C_{\rm H}$, $C_{\mHt}$ and $C_{\He}$.
If we define the total collision rate $C$ to be the sum of these three unknowns
\begin{equation}
C = C_{\mH} + C_{\mHt} + C_{\He},
\end{equation}
and write the low density $\htp$ cooling rate as
\begin{equation}
\Lambda_{\htp, n \rightarrow 0} = L_{\htp, n \rightarrow 0} n_{\htp},
\end{equation}
then it is easy to show that the combination $L_{\htp, n \rightarrow 0} / C$ is 
completely determined. We have computed an analytical fit to this quantity, using a fit
of the form of Equation~\ref{h3p_fit_eq}. The fitting coefficients are listed in 
Table~\ref{h3p_coeffs}. This fit is accurate to within 1\% over the temperature 
range $20 < T < 10000 \: {\rm K}$. To convert from $L_{\htp, n \rightarrow 0} / C$
to $L_{\htp, n \rightarrow 0}$, we must fix the size of our remaining free parameter, 
the total collision rate $C$. In most of our simulations, we assume that $C$ is given by 
\begin{equation}
C = 2.2 \times 10^{-9} n_{\mH} + 1.9 \times 10^{-9} n_{\mHt} +  8.1 \times 10^{-10} n_{\He} \: {\rm s^{-1}}
\label{cref}
\end{equation}
which is the sum of the Langevin rates for collisions between $\htp$
and $\mH$, $\mHt$ and $\He$, respectively. These Langevin rates were computed using 
polarizabilities for He and $\mHt$ taken from \citet{hb93}; the exact value was used for H. 
The true value of $C$ is unlikely to be very much larger than this, but could be considerably 
smaller, and so in \S\ref{res:C} we examine the sensitivity of our results to our choice of 
value for $C$. Finally, we note that as both $\Lambda_{\rm \htp, LTE}$ and 
$\Lambda_{\htp, n \rightarrow 0}$ are directly proportional to $n_{\htp}$, the $\htp$
critical density is independent of $n_{\htp}$. Thus, once $C$ is specified, $n_{\rm cr}$
can be trivially computed using our numerical fits.

We assume that the $\htp$ emission remains optically thin throughout our 
simulations. For a subsonic collapse in which the effect of large-scale velocity 
gradients are unimportant compared to the local Doppler broadening of the 
emission lines, the optical depth at line center corresponding to a given emission 
line can be written as  \citep{mm84}
\begin{equation}
\tau_{ji} = \frac{g_{j}}{g_{i}} \frac{c^{2}}{8 \pi \nu_{ij}^{2}}
A_{ji}  \frac{N_{i}}{\pi^{1/2} \Delta \nu_{\rm D}},
\end{equation}
where $g_{j}$ and $g_{i}$ are the statistical weights of levels $i$ and
$j$, $\nu_{ij}$ is the frequency of the transition from level $j$ to level $i$, 
$A_{ij}$ is the corresponding spontaneous radiative transition rate,
$N_{i}$ is the column density of absorbers in level $i$, and 
$\Delta \nu_{\rm D} = (\nu_{ij} / c) (2kT / m)^{1/2}$ is the Doppler width of 
the line, where $m$ is the mass of the $\htp$ ion. For simplicity, we
have neglected the effects of stimulated emission. 
Illustrative values for $\nu_{ij}$ and $A_{ij}$ for a strong vibrational
transition are $\nu_{ij} \simeq 8.46 \times 10^{13} \: {\rm Hz}$ and
$A_{ij} \simeq 94 \: {\rm s^{-1}}$ \citep{nmt96}, and so for this transition
\begin{equation}
\tau_{ji} \sim 10^{-15} N_{i},
\end{equation}
where we have assumed a gas temperature of 1000~K. Therefore, in this 
example, the line becomes optically thick only once $N_{i} \simgreat 10^{15} 
\: {\rm cm^{-2}}$. As many of the $\htp$ emission lines are considerably weaker
than this example, and as the column density of $\htp$ ions in any particular
level $i$ can clearly be no larger than the total $\htp$ column density, $N_{\htp}$,
it is safe to conclude from this analysis that optical depth effects are unlikely
to significantly affect the $\htp$ cooling rate until 
$N_{\htp} > 10^{15} \: {\rm cm^{-2}}$.

We now investigate whether we expect our models to reach this column density in
$\htp$. Our one-zone dynamical model does not contain any information about the 
overall structure of the collapsing core and so does not 
predict $N_{\htp}$ directly. However, based on the results of more detailed 
numerical simulations \citep{abn02,yoha06}, we assume that the protostellar core 
has a density profile that is well approximated by a power law $n(r) \propto r^{-2.2}$, 
and that it is collapsing subsonically. With this assumed density profile,
the column density of hydrogen nuclei along a radial ray from a point $r$ to the edge 
of the core is given by
\begin{eqnarray}
N_{\rm H, tot}(r) & = & \int_{r}^{r_{\rm core}} n(r^{\prime}) {\rm d}r^{\prime}, \\
 & = & \frac{5}{6} r n(r) \left[ 1 - \left(\frac{r}{r_{\rm core}} \right)^{1.2} \right],
 \label{NHtot}
\end{eqnarray}
where $n(r)$ is the number density of hydrogen nuclei at $r$ and $r_{\rm core}$
is the radius of the core. As we shall see in \S\ref{results}, the $\htp$ abundance
in the collapsing gas typically varies only slightly with density below some
threshold density $n_{\rm thr}$ and then declines sharply for $n > n_{\rm thr}$. 
The value of $n_{\rm thr}$ depends on factors such as the cosmic ray ionization 
rate and the speed of the collapse, but even in the most extreme models
(e.g., run CR5; see \S\ref{res:cosmic}) $n_{\rm thr} = 10^{11} \: {\rm cm^{-3}}$, 
while in general it is much smaller. Therefore, almost all of the contribution to
$N_{\htp}$ comes from gas at densities $n < n_{\rm thr}$, and hence at radii
$r > r_{\rm thr}$, where $r_{\rm thr}$ is the radius such that $n(r_{\rm thr}) = n_{\rm thr}$.
If we assume thar $r_{\rm thr} \ll r_{\rm core}$, or in other words that the density
distribution of the collapsing core extends to densities $n \ll n_{\rm thr}$, then
Equation~\ref{NHtot} tells us that the column density of hydrogen nuclei between
$r_{\rm thr}$ and the edge of the core is approximately
\begin{equation}
N_{\rm tot}(r_{\rm thr}) \simeq \frac{5}{6} r_{\rm thr} n_{\rm thr}.
\end{equation}
Denoting the fractional abundance of $\htp$ at $r_{\rm thr}$ as $x_{\htp}(r_{\rm thr})$,
and assuming that it remains constant for $r > r_{\rm thr}$, we can therefore write the
$\htp$ column density between $r_{\rm thr}$ and the edge of the core as
\begin{equation}
N_{\htp}(r_{\rm thr}) \simeq  \frac{5}{6}  x_{\htp}(r_{\rm thr})  r_{\rm thr} n_{\rm thr}.
\end{equation}

For a typical protogalactic core, simulations have shown that a density $n_{\rm thr} = 
10^{11} \: {\rm cm^{-3}}$ corresponds to a radius $r_{\rm thr} \simeq 10^{15} \: {\rm cm}$ 
\citep[see e.g.,][]{abn02,yoha06}. Furthermore, the results presented in \S\ref{res:cosmic} 
demonstrate that in run CR5,  
$x_{\htp}(r_{\rm thr}) \simeq 4.4 \times 10^{-11}$. We therefore obtain 
$N_{\htp}(r_{\rm thr}) \simeq 4 \times 10^{15} \: {\rm cm^{-2}}$. At $r > r_{\rm thr}$,
the $\htp$ column density is smaller, but at $r < r_{\rm thr}$, it does not grow
significantly larger. In this particular case, the core may be marginally optically
thick, albeit only in the strongest lines. However, in most of our models, $n_{\rm thr}$
is significantly smaller, and $x_{\htp}(r_{\rm thr})$ is orders of magnitude smaller.
In these runs, it is clear that the core remains optically thin.

\subsection{$\mathbf{HD}$ cooling}
To model $\hd$ cooling, we use the cooling function of \citet{lna05}. Although formally
valid only in the temperature range $100 < T < 2 \times 10^{4} \: {\rm K}$, we have
compared its behaviour at lower temperatures with an explicit calculation of the 
cooling rate made using radiative de-excitation rates from \citet{arv82} and collisional rates
extrapolated from those computed by \citet{wgf07}. We find that the \citet{lna05} rate 
remains reasonably accurate down to temperatures as low as 50~K, with errors no 
greater than 20\%, and that even at $T = 30 \: {\rm K}$ it remains accurate to within 
a factor of two. At temperatures $T \gg 100 \: {\rm K}$, the \citet{lna05} cooling rate
slightly underestimates the effects of HD cooling compared to the newer calculations
of \citet{wgf07}, presumably owing to the more accurate vibrational excitation rates
used in the latter study, but the differences are never greater than about 50\%,
and in any case occur in the
temperature regime in which $\mHt$ cooling dominates. The breakdown of the 
\citet{lna05} fit at very high temperatures ($T > 20000 \: {\rm K}$) is unimportant, 
as the gas in our models never reaches this temperature.

To correctly  model the effects of $\hd$ cooling at low temperatures, it is 
necessary to take the effects of the cosmic microwave background into account. 
We do this approximately, by using a modified HD cooling rate, 
$\Lambda_{\rm HD}^{\prime}$, defined as
\begin{equation}
\Lambda_{\rm HD}^{\prime} = \Lambda_{\rm HD}(T) - \Lambda_{\rm HD}(T_{\rm CMB})
\end{equation}
where $\Lambda_{\rm HD}(T)$ and $ \Lambda_{\rm HD}(T_{\rm CMB})$ are the 
unmodified HD cooling rates at the gas temperature $T$ and the CMB temperature
$T_{\rm CMB}$, respectively.

The quoted range of densities for which the \citet{lna05} cooling function is valid
is $1 < n < 10^{8} \: {\rm cm^{-3}}$. To extend the range of the cooling function
to densities $n < 1 \: {\rm cm^{-3}}$, we assume that at these densities the HD
cooling rate scales proportionately to the number density of HD times the number
density of colliders (i.e.\ as $n^{2}$), and hence that
\begin{equation}
\Lambda_{\rm HD}(n = n^{\prime}) = (n^{\prime})^{2} \Lambda_{\rm HD}(n=1)
\end{equation}
for $n^{\prime} \le 1 \: {\rm cm^{-3}}$, where $\Lambda_{\rm HD}(n)$ is the HD cooling
rate per unit volume (with units ${\rm erg} \: {\rm cm^{-3}} \: {\rm s^{-1}}$) at gas number 
density $n$. To extend the cooling function to high densities, $n > 10^{8} \: {\rm cm^{-3}}$, 
we assume that the HD molecule is in LTE and hence that the HD cooling rate per unit volume 
is independent of the number density of colliders and scales linearly with $n$; or in other words, 
that $\Lambda_{\rm HD} / n_{\rm HD}$ is independent
of $n$. In view of the fact that $1 \ll n_{\rm cr, HD} \ll 10^{8} \: {\rm cm^{-3}}$, where 
$n_{\rm cr, HD}$ is the HD critical density, both of these assumptions appear
well justified.

The Lipovka~et~al.~(2005) cooling function only includes the effects of collisions between
$\hd$ and $\mH$. However, \citet{flpr00} have shown that the influence of the $\mHt$/$\mH$
ratio on the $\hd$ cooling rate is very small, and so the Lipovka~et~al.~(2005) cooling
function should remain reasonably accurate even after the molecular fraction becomes
large. Moreover, collisions between $\hd$ and other species (electrons, $\Hp$, etc.)
can be neglected compared to collisions with $\mH$ on account of the much larger 
abundance of the latter at the gas densities of interest.

Finally, we assume that the $\hd$ rovibrational lines remain optically thin throughout 
all of our runs. In practice, this is probably not the case: the strongest $\hd$ lines will 
become saturated once the $\hd$ column density exceeds $N_{\hd} = 10^{22} 
\: {\rm cm^{-2}}$, and an analysis similar to that performed for $\htp$ in the 
previous section suggests that this will occur in our model cores once 
$n \simgreat 3 \times 10^{13} \: {\rm cm^{-3}}$. However, $\hd$ is only a minor 
coolant at these high densities, and so we can safely neglect optical depth effects 
on $\hd$ cooling without significantly affecting the thermal evolution of the gas in our 
models.

\subsection{$\mathbf{LiH}$ cooling}
To treat cooling from $\lih$, we use the cooling function given in \citet{gp98}. This is 
the low density limit of the $\lih$ cooling rate and so is strictly valid only for gas
densities significantly below the $\lih$ critical density, $n_{\rm cr, \lih}$. However, 
the transition probabilities for the rotational and vibrational transitions of $\lih$ are
very large, on account of the molecule's large dipole moment. This means that the
$\lih$ critical density is large, $n_{\rm cr, \lih} \simeq 10^{12} \: {\rm cm^{-3}}$ 
\citep{ls84}, and so the \citet{gp98} cooling function is a reasonable 
choice over most of the range of densities covered by our simulations. At very high 
densities, we would expect $\lih$ to begin to reach LTE, and our approximation to 
break down; at these densities, our continued use of the \citet{gp98} cooling function
means that we will overestimate the effectiveness of $\lih$ cooling. Despite this, 
we find $\lih$ cooling to be ineffective at all densities (see \S\ref{res:basic} below), 
suggesting that if we were to use a more accurate treatment of $\lih$ cooling at high 
densities it would not  significantly alter our conclusions.

In principle, we should adjust the $\lih$ cooling rate to account for the effects of the
CMB, just as we do for the $\hd$ cooling rate. However, as $\lih$ cooling proves to
be unimportant at all densities, this correction is also unimportant and its omission does 
not significantly affect the thermal evolution of the gas.

We assume that the $\liD$ cooling rate is the same as the $\lih$ cooling rate. 
While this is a crude approximation, in practice the $\liD$ abundance is so small 
that its contribution is always negligible and thus the choice of $\liD$ cooling
rate is unimportant.

\subsection{$\mathbf{H_{2}^{+}}$, $\mathbf{HD^{+}}$ and $\mathbf{D_{2}^{+}}$ cooling}
\label{h2p_cooling}
To treat cooling from these molecular ions, we use the same approach
as in \citet{ga08}. At low densities, most $\mHtp$ cooling occurs due to collisions
with free electrons and neutral hydrogen atoms; collisions with $\He$ and $\mHt$
excite $\mHtp$ at comparable rates to collisions with $\mH$ \citep{rd82}, but are 
unimportant due to the low abundances of these species relative to atomic
hydrogen. To model the cooling due to
collisions with electrons, we use the vibrational excitation rate coefficients of \citet{st93},
while for collisions with neutral hydrogen, we use a fit to the \citet{ss78} rate coefficient
provided to us by D.\ Galli (private communication); note that this is a factor of ten smaller
than the rate coefficient given in \citet{gp98}, owing to a normalization error in the latter
paper.

At high densities, the vibrational levels of $\mHtp$ will be in LTE. In this
regime, the cooling rate is given approximately by
\begin{equation}
\Lambda_{\rm \mHtp,  LTE} = 2.0 \times 10^{-19} T^{0.1} \expf{-}{3125}{T} n_{\mHtp}.
\end{equation}
This fit is from \citet{ga08} and includes contributions from all vibrational states 
$v \le 8$ (higher vibrational states are not expected to contribute significantly
at the temperatures of interest in this work). 
It was computed using level energies from \citet{kh06} and radiative 
transition rates from \citet{pdp83}. The effects of rotational excitation were not included, 
but are unlikely to change this expression by a large amount, owing to the
very small Einstein coefficients associated with pure rotational transitions in
$\mHtp$.

At intermediate densities, we assume that the $\mHtp$ vibrational cooling
rate is given approximately by the function
\begin{equation}
\Lambda_{\mHtp} = \frac{\Lambda_{\rm \mHtp, LTE}}{1 + 
n_{\rm cr} / n},
\label{h2p_full}
\end{equation}
where $n_{\rm cr} / n = \Lambda_{\rm H_{2}^{+}, LTE} / \Lambda_{\rm H_{2}^{+}, n \rightarrow 0}$
is the $\mHtp$ critical density, and where $\Lambda_{\rm H_{2}^{+}, n \rightarrow 0}$ is the 
$\mHtp$ cooling rate in the low density limit. This is given by
\begin{equation}
 \Lambda_{\rm H_{2}^{+}, n \rightarrow 0} =  \left[ L_{\rm \mHtp, e} n_{\me} 
 + L_{\rm \mHtp, H} n_{\mH} \right] n_{\mHtp}
\end{equation}
where $L_{\rm \mHtp, e}$ and $L_{\rm \mHtp, H}$ are the cooling rates per $\mHtp$ ion
per unit collider density for collisions with electrons and atomic hydrogen, respectively,
taken from \citet{st93} and \citet{ss78} as noted above.

To model cooling from vibrational transitions in $\hdp$, we assume, in the absence
of better information, that the low density cooling rate is the same as that used for
$\mHtp$. However, since $\hdp$ has much larger radiative transition rates than
$\mHtp$, the LTE cooling rate for $\hdp$ is much larger than that for $\mHtp$.
Accordingly, we use the following functional fit for the $\hdp$ LTE cooling rate:
\begin{equation}
\Lambda_{\rm \hdp, LTE} = 1.09 \times 10^{-11} T^{0.03} \expf{-}{2750}{T} n_{\hdp}
\end{equation}
at temperatures $T \le 1000 \: {\rm K}$ and
\begin{equation}
\Lambda_{\rm \hdp, LTE} = 5.07 \times 10^{-12} T^{0.14} \expf{-}{2750}{T} n_{\hdp}
\end{equation}
at $T > 1000 \: {\rm K}$. These fits are from \citet{ga08} and were calculated using
$\hdp$ level energies from \citet{kh06} and radiative transition rates from \citet{phb79}. 
For densities between the low density and LTE limits, we again use a function of the 
form of Equation~\ref{h2p_full} to compute the cooling rate.

Finally, to model $\ddp$ cooling, we simply assume that the same rates apply 
as for $\mHtp$ cooling. In practice, the very small size of the typical $\ddp$ 
abundance renders this process completely unimportant.

\subsection{Other radiative coolants}
\label{minor_cool}
In addition to the coolants discussed above, we also include a treatment of
cooling from a number of other minor molecular ions that are present in the gas.
Specifically, we include the effects of cooling from $\hhdp$, $\hddp$, $\dtp$,
$\hehp$, $\hedp$, $\hehep$ , $\lihp$, $\lidp$ and $\lihhp$. Since appropriate collisional
data are not available for most of these species, we treat their contribution to the
cooling rate in an extremely simple fashion. We assume that the contribution to 
the cooling rate made by a species $i$ with number density $n_{i}$ can be written as
\begin{equation}
\Lambda_{i} = kT \left( \sum_{c} C_{ic} n_{c} \right) n_{i},
\end{equation}
where $n_{c}$ is the number density of a collider $c$, $C_{ic}$ is the rate coefficient
for inelastic collisions between $i$ and $c$, and where we sum over all possible
colliders. For collisions with $\mH$, $\mHt$ or $\He$, we assume that $C_{ic}$ is
given by the Langevin rate, while for collisions with electrons we conservatively
assume that $C_{ic} = 10^{-6} \: {\rm cm^{3}} \: {\rm s^{-1}}$, which is comparable 
to the total rate coefficients found for other molecules, such as $\htp$ \citep{ft03}. 
Collisions with all other species can be and are neglected.

In constructing this approximation we have assumed that each collision with $i$ 
transfers an amount of energy $kT$, all of which is subsequently radiated. In 
practice, this procedure is likely to significantly overestimate the cooling provided
by $i$, for several reasons. For one thing, it is not clear that the mean amount of
energy transferred in a collision will always be $\sim kT$, since collisions that
transfer $\Delta E \ll kT$ are possible while collisions that transfer $\Delta E \gg kT$ 
are highly unlikely. More importantly, this procedure neglects effects such as the
collisional de-excitation of excited levels that will significantly limit cooling at high 
densities.  However, these simplifications are unlikely to significantly affect our 
results, since even when we use these overestimates for the cooling rates, we
find that the contribution of these minor species to the total cooling rate is 
negligible (see \S\ref{results}).

As well as these minor coolants, we also include two forms of cooling that are
of great importance in hot, ionized gas. The first is cooling from electron impact 
excitation of atomic hydrogen (Lyman-$\alpha$ cooling). We treat this using a 
rate from \citet{cen92}, which is itself based on a rate in \citet{black81}. However, 
we note that for temperatures $T < 8000 \: {\rm K}$,  Lyman-$\alpha$ cooling is 
completely negligible, and so it does not significantly affect the outcome of the 
simulations presented in this paper.

The second process is cooling due to the Compton scattering of CMB photons by 
free electrons. This is also treated using a rate from 
\citet{cen92}, but again plays very little role in the thermal evolution of the gas, 
since it is important primarily at low densities  ($n \simless 1 \: {\rm cm^{-3}}$), 
even when the gas is initially highly ionized.

Finally, we note that we do not include the effects of cooling from $\DD$.
As $\DD$ is a homonuclear molecule, it suffers from the same drawbacks
that $\mHt$ does with regard to low temperature cooling. However, as the
results in \S\ref{res:basic} demonstrate, it generally has a chemical abundance 
that is many orders of magnitude smaller than $\mHt$. Thus, in contrast to
HD, it appears highly unlikely that $\DD$ cooling is ever significant.

\subsection{Radiative heating}
We include two forms of radiative heating that can be significant if a strong 
ultraviolet background is present. The first is the photodissociation of $\mHt$.
We calculate the $\mHt$ photodissociation rate as discussed in \S\ref{chem}, 
and then, following \citet{bd77}, we assume that each photodissociation deposits 
$0.4 \: {\rm eV}$ of heat into the gas. Note that although the photodissociation of
other ionic and molecular species (e.g., $\mHtp$, $\hd$) will also heat the 
gas, they are unimportant when compared to $\mHt$ photodissociation 
owing to the low abundances of the other species relative to $\mHt$.

The second form of radiative heating included in our thermal model arises
due to the population of excited vibrational states of $\mHt$ produced by radiative
pumping by the UV field. At high densities, this leads to heating of the gas,
as most of the excited molecules undergo collisional de-excitation. We adopt 
a radiative pumping rate that is 8.5 times larger than the photodissociation 
rate \citep{db96}, and assume that each excitation transfers an average of 
$2 \, (1 + n_{\rm cr}/n)^{-1} \: {\rm eV}$ to the gas \citep{bht90}, where 
$n_{\rm cr}$ is the $\mHt$ critical density, calculated as discussed in 
\S \ref{chem} above. 

\subsection{Cosmic ray heating}
Following \citet{gl78}, we assume that each primary ionization deposits 
$20 \: {\rm eV}$ of energy into the gas, giving us a heating rate
\begin{equation}
 \Gamma_{cr} = 3.2 \times 10^{-28} \left( \frac{\zeta}{10^{-17} \: 
 {\rm s}^{-1}} \right) n \: {\rm ergs} \: {\rm s}^{-1} \: {\rm cm}^{-3},
\end{equation}
where $\zeta = \sum_{i} \zeta_{i}$ and we sum over all species listed
in Table~\ref{cr_ion_tab}.

\subsection{Chemical heating and cooling}
\label{cool:react}
Any exothermic chemical reaction will potentially heat the gas, while any 
endothermic reaction will cool it. In practice, however, the effect of most
reactions on the gas temperature is small, and only in a few cases do we
need to take chemical heating or cooling into account.  

In highly ionized gas, cooling due to the recombination of hydrogen and
helium can be a significant effect, particularly at temperatures which are
too low for Lyman-$\alpha$ cooling to be effective. However, recombination 
cooling becomes ineffective once the fractional ionization of the gas falls
below $x \sim 0.01$, and it therefore plays no role at the high gas densities 
of interest in this study.

Other forms of chemical cooling included in our model occur due to the
collisional dissociation of $\mHt$ (reactions CD9, CD10, CD11 and CD12), the
destruction of $\mHt$ by charge transfer with $\Hp$ (reaction CT2), and 
the collisional ionization of hydrogen and helium (reactions CI1 \&  CI2). 
Cooling from these processes may be of some importance at very early
times in simulations starting at high temperatures ($T \simgreat 10^{4} \: {\rm K}$),
but in general the gas temperature is too low for these 
sources of cooling to be significant.

As far as chemical heating is concerned, the most significant process is
$\mHt$ formation heating. When $\mHt$ is formed by reaction AD1 or reaction CT1, it 
preferentially forms in an excited vibrational state, with an energy comparable to 
the exothermicity of the reaction (3.73~eV for reaction AD1, 1.83~eV for reaction CT1). 
In low density gas, this energy is simply radiated away, but for 
$n > 10^{4} \: {\rm cm^{-3}}$, most is instead converted into thermal energy by 
collisional de-excitation of the newly formed $\mHt$. $\mHt$ formation via 
$\Hm$ or $\mHtp$ therefore acts as a minor heat source in gas with 
$n > 10^{4} \: {\rm cm^{-3}}$. 

Three-body formation of $\mHt$ also heats the gas, since the third body in the
collision generally carries away additional energy equal to the binding energy
of the new $\mHt$ molecule, 4.48~eV. In our reference simulation, this is the 
dominant heat source for densities $5 \times 10^{10} < n  < 2 \times 10^{12} \: 
{\rm cm^{-3}}$.

\section{Results}
\label{results}
Our simple one-zone dynamical model of gravitationally collapsing primordial 
gas contains a number of free parameters. To fully explore the role of $\htp$
cooling and its sensitivity to these free parameters, it is necessary to perform
a large number of calculations. However, discussion of the results of all of 
these calculations to the same level of detail would not only be extremely
tedious, but would also run the risk of obscuring our main results. Therefore,
we proceed by first discussing in detail in \S\ref{res:basic}
the results  of a single calculation -- our reference model, hereafter denoted as 
computational run REF -- before highlighting in the subsequent sections
the differences in outcome (if any) that result from alterations in our free
parameters. Full details of all of the runs discussed here can be found in
Table~\ref{runlist}.

\begin{table*}
\begin{minipage}{126mm}
\caption{List of simulations run \label{runlist}}
\begin{tabular}{ccccccccc}
\hline
Run & $n_{\rm i}$ (${\rm cm^{-3}}$) & $T_{\rm i}$ (K) & $x_{\Hp}$ & 
$J_{21}$ & $\zeta_{\rm H}$ (s$^{-1}$) & $C / C_{\rm ref}$  & $\eta$ & Notes \\
\hline
REF & 1.0 & 1000 & $2.2 \times 10^{-4}$ & 0.0 & 0.0 & 1.0 & 1.0 & \\
C1 & 1.0 & 1000 &$2.2 \times 10^{-4}$ &  0.0 & 0.0 & 0.1 & 1.0 & \\
C2 & 1.0 & 1000 &$2.2 \times 10^{-4}$ &  0.0 & 0.0 & 10.0 & 1.0 & \\
CR1 & 1.0 & 1000 &$2.2 \times 10^{-4}$ &  0.0 & $10^{-20}$ & 1.0 & 1.0 & \\
CR2 & 1.0 & 1000 &$2.2 \times 10^{-4}$ &  0.0& $10^{-19}$ & 1.0 & 1.0 & \\
CR3 & 1.0 & 1000 &$2.2 \times 10^{-4}$ &  0.0 & $10^{-18}$ & 1.0 & 1.0 & \\
CR4 & 1.0 & 1000 &$2.2 \times 10^{-4}$ &  0.0 & $10^{-17}$ & 1.0 & 1.0 & \\
CR5 & 1.0 & 1000 &$2.2 \times 10^{-4}$ &  0.0 & $10^{-16}$ & 1.0 & 1.0 & \\
CR6 & 1.0 & 1000 &$2.2 \times 10^{-4}$ &  0.0 & $10^{-16}$ & 1.0 & 1.0 & No $\htp$ cooling \\
CR7 & 1.0 & 1000 &$2.2 \times 10^{-4}$ &  0.0 & $10^{-20}$ & 1.0 & 1.0 & No PT mechanism \\
CR8 & 1.0 & 1000 &$2.2 \times 10^{-4}$ &  0.0 & $10^{-18}$ & 1.0 & 1.0 & No PT mechanism \\
CR9 & 1.0 & 1000 &$2.2 \times 10^{-4}$ &  0.0 & $10^{-16}$ & 1.0 & 1.0 & No PT mechanism \\
CR10 & 1.0 & 1000 &$2.2 \times 10^{-4}$ &  0.0 & $10^{-20}$ & 1.0 & 1.0 & `Maximal' PT mechanism \\
CR11 & 1.0 & 1000 &$2.2 \times 10^{-4}$ &  0.0 & $10^{-18}$ & 1.0 & 1.0 & `Maximal' PT mechanism \\
CR12 & 1.0 & 1000 &$2.2 \times 10^{-4}$ &  0.0 & $10^{-16}$ & 1.0 & 1.0 & `Maximal' PT mechanism \\
UV1 & 1.0 & 1000 &$2.2 \times 10^{-4}$ &  $10^{-4}$ & 0.0 & 1.0 & 1.0 & Optically thin \\
UV2 & 1.0 & 1000 &$2.2 \times 10^{-4}$ &  $10^{-2}$ & 0.0 & 1.0 & 1.0 & Optically thin \\
UV3 & 1.0 & 1000 &$2.2 \times 10^{-4}$ &  1.0 & 0.0 & 1.0 & 1.0 & Optically thin \\
UV4 & 1.0 & 1000 &$2.2 \times 10^{-4}$ &  $10^{-4}$ & 0.0 & 1.0 & 1.0 & $f_{\rm sh, H_{2}} = f_{\rm sh, HD} = 0$  \\
UV5 & 1.0 & 1000 &$2.2 \times 10^{-4}$ &  $10^{-2}$ & 0.0 & 1.0 & 1.0 & $f_{\rm sh, H_{2}} = f_{\rm sh, HD} = 0$ \\
UV6 & 1.0 & 1000 &$2.2 \times 10^{-4}$ &  1.0 & 0.0 & 1.0 & 1.0 & 
$f_{\rm sh, H_{2}} = f_{\rm sh, HD} = 0$ \\
N1 & 0.03 & 1000 &$2.2 \times 10^{-4}$ &  0.0 & 0.0 & 1.0 & 1.0 & \\
N2 & 30 & 1000 &$2.2 \times 10^{-4}$ &  0.0 & 0.0 & 1.0 & 1.0 & \\
T1 & 1.0 & 100 &$2.2 \times 10^{-4}$ &  0.0 & 0.0 & 1.0 & 1.0 & \\
T2 & 1.0 & 10000 &$2.2 \times 10^{-4}$ &  0.0 & 0.0 & 1.0 & 1.0 & \\
X1 & 1.0 & 1000 &$10^{-6}$ &  0.0 & 0.0 & 1.0 & 1.0 & \\
X2 & 1.0 & 1000 &$10^{-2}$ &  0.0 & 0.0 & 1.0 & 1.0 &  \\
X3 & 1.0 & 1000 &1.0 &  0.0 & 0.0 & 1.0 & 1.0 &  \\
EL1 & 1.0 & 1000 &$2.2 \times 10^{-4}$ &  0.0 & 0.0 & 1.0 & 1.0 & No D \\
EL2 & 1.0 & 1000 &$2.2 \times 10^{-4}$ &  0.0 & 0.0 & 1.0 & 1.0 & No Li \\
EL3 & 1.0 & 1000 &$2.2 \times 10^{-4}$ &  0.0 & 0.0 & 1.0 & 1.0 & No D or Li \\
RA & 1.0 & 1000 &$2.2 \times 10^{-4}$ &  0.0 & 0.0 & 1.0 & 1.0 & $k_{\rm RA18}$ from ref. 1 \\
AR1 & 1.0 & 1000 & $2.2 \times 10^{-4}$ & 0.0 & 0.0 & 1.0 & 1.0 & See \S\ref{res:chem:ad1} \\
AR2 & 1.0 & 1000 & $2.2 \times 10^{-4}$ & 0.0 & 0.0 & 1.0 & 1.0 & See \S\ref{res:chem:ad1} \\
3B1 & 1.0 & 1000 &$2.2 \times 10^{-4}$ &  0.0 & 0.0 & 1.0 & 1.0 & $k_{\rm TB1}$ from ref. 2 \\
3B2 & 1.0 & 1000 &$2.2 \times 10^{-4}$ &  0.0 & 0.0 & 1.0 & 1.0 & $k_{\rm TB1}$ from ref. 3 \\
3B3 & 1.0 & 1000 &$2.2 \times 10^{-4}$ &  0.0 & 0.0 & 1.0 & 1.0 & $k_{\rm TB2}$ from ref. 4 \\
3B4 & 1.0 & 1000 &$2.2 \times 10^{-4}$ &  0.0 & 0.0 & 1.0 & 1.0 & $k_{\rm TB2}$ from ref. 3 \\
LP1 & 1.0 & 1000 &$2.2 \times 10^{-4}$ &  0.0 & 0.0 & 1.0 & 1.0 & See \S\ref{res:chem:lih2p} \\
LP2 & 1.0 & 1000 &$2.2 \times 10^{-4}$ &  0.0 & 0.0 & 1.0 & 1.0 & See \S\ref{res:chem:lih2p} \\
LP3 & 1.0 & 1000 &$2.2 \times 10^{-4}$ &  0.0 & 0.0 & 1.0 & 1.0 & See \S\ref{res:chem:lih2p} \\
DYN1 & 1.0 & 1000 &$2.2 \times 10^{-4}$ &  0.0 & 0.0 & 1.0 & 0.6 & \\
DYN2 & 1.0 & 1000 &$2.2 \times 10^{-4}$ &  0.0 & 0.0 & 1.0 & 0.3 & \\
DYN3 & 1.0 & 1000 &$2.2 \times 10^{-4}$ &  0.0 & 0.0 & 1.0 & 0.1 & \\
\hline
\end{tabular}
\medskip
\\
{\bf References}: 1 -- \citet{sld98}; 2 -- \citet{pss83}; 3 -- \citet{fh07}; 4 -- \citet{cw83} \\
\end{minipage}
\end{table*}

\subsection{The role of $\mathbf{H_{3}^{+}}$ cooling}
\label{res:basic}
\begin{figure}
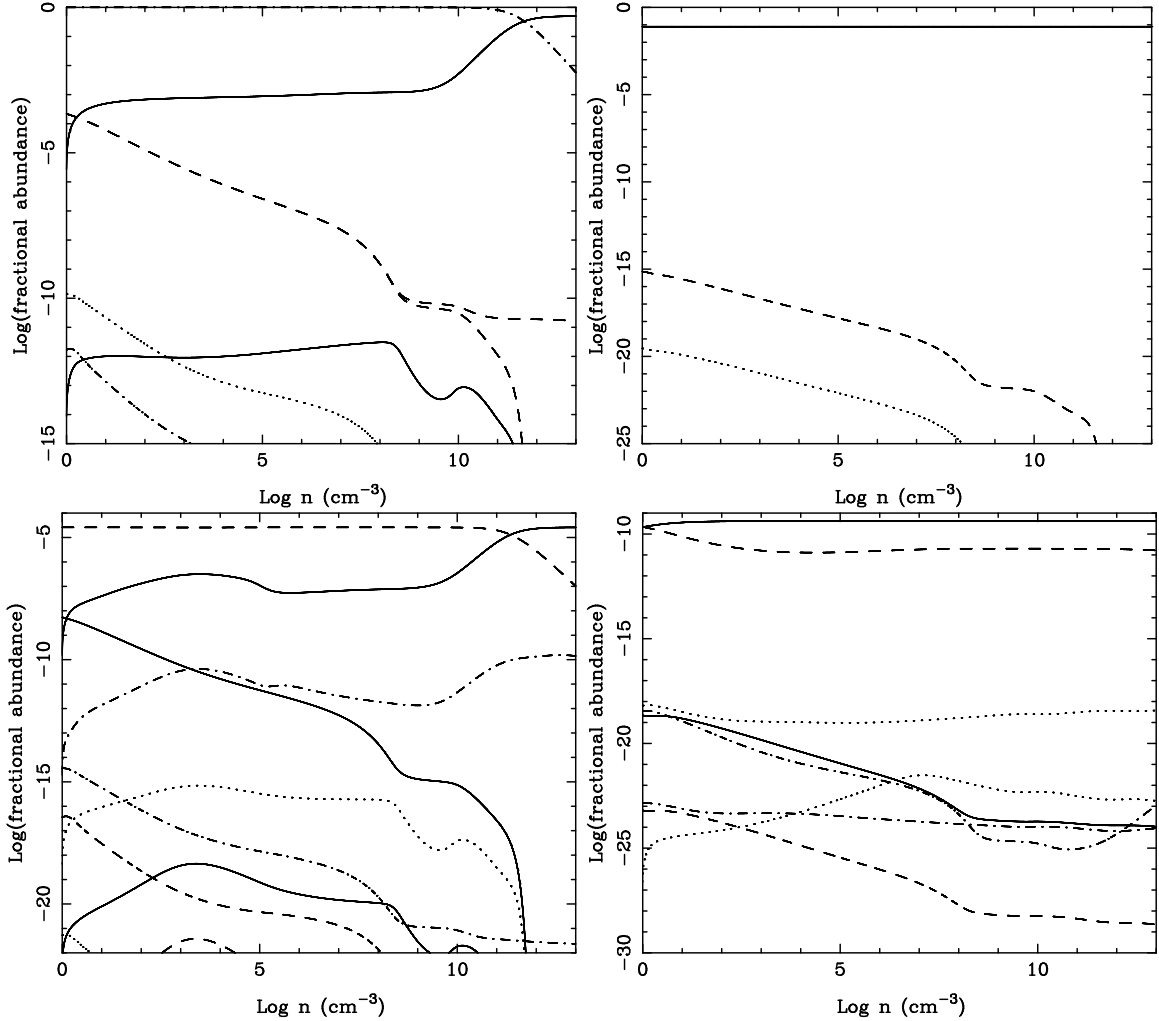

\centering
\epsfig{figure=f1a.eps,width=16pc,angle=270,clip=}  
\epsfig{figure=f1b.eps,width=16pc,angle=270,clip=}  
\epsfig{figure=f1c.eps,width=16pc,angle=270,clip=} 
\epsfig{figure=f1d.eps,width=16pc,angle=270,clip=} 
\caption{(a) Chemical evolution of the gas in our 
reference calculation. Fractional abundances are
plotted for $\mHt$ (upper solid line), $\htp$ (lower
solid line), $\me$ (upper dashed line), $\Hp$ (lower
dashed line), $\mH$ (upper dash-dotted line), 
$\mHtp$ (lower dash-dotted line) and $\Hm$ (dotted line).
(b) As (a), but showing the fractional abundances of 
$\He$ (solid line), $\hehp$ (dashed line) and $\hedp$ 
(dotted line). The abundances of $\Hep$ and $\hehep$ 
remained negligibly small throughout the simulation
and are not plotted here.
(c) As (a), but for the fractional abundances of $\hd$
(upper solid line), $\mD$ (upper dashed line), $\DD$
(upper dash-dotted line), $\Dp$ (central solid line),
$\hhdp$ (upper dotted line), $\Dm$ (lower dash-dotted
line), $\hddp$ (lower solid line), $\hdp$ (central dashed
line), $\ddp$ (lower dotted line) and $\dtp$ (lower dashed
line).
(d) As (a), but for the fractional abundances of $\li$
(upper solid line), $\lip$ (upper dashed line), $\lih$
(upper dotted line), $\Limi$ (upper dash-dotted line), 
$\lihhp$ (lower dotted line), $\lihp$ (lower solid line), 
$\liD$ (lower dash-dotted line) and $\lidp$ (lower dashed line).
\label{abund_ref}
}
\end{figure}

We begin our study by investigating the outcome of our
reference calculation, run REF, whose parameters are indicated in
Table~\ref{runlist}. In Figure~\ref{abund_ref} we show how the fractional
abundances of 28 of our 30 chemical species vary with density
during the course of the collapse.  For clarity, we have divided 
these species into four sets on the basis of the elements that 
they contain, and illustrate the evolution of each set separately
in Figures~\ref{abund_ref}a--\ref{abund_ref}d. The two species
that are not plotted -- $\Hep$ and $\hehep$ -- have abundances that 
remain negligibly small throughout the calculation.
  
Figure~\ref{abund_ref} demonstrates that the evolution of the $\htp$
abundance passes through four distinct phases. In the
first phase, at $n \simless 10^{8} \: {\rm cm^{-3}}$, the ratio of $\htp$
to $\mHt$ remains approximately constant: $x_{\htp} / x_{\mHt} \sim
10^{-9}$. This is a consequence of the balance between two main
processes: the formation of $\htp$ by the radiative association of
$\mHt$ and $\Hp$ (reaction RA18) and its destruction by dissociative
recombination (reactions DR4 and DR5). If we assume that these reactions 
dominate the formation and destruction of $\htp$, then the 
corresponding equilibrium abundance of $\htp$ is given by
\begin{equation}
x_{\htp}  = \frac{k_{\rm RA18} n_{\Hp} x_{\mHt}}{(k_{\rm DR4} + k_{\rm DR5}) n_{\rm e}},
\end{equation}
which reduces to
\begin{equation}
x_{\htp} = \frac{k_{\rm RA18}}{(k_{\rm DR4} + k_{\rm DR5})} x_{\mHt} \simeq 10^{-9} x_{\mHt}
\end{equation}
if $x_{\rm e} = x_{\Hp}$, which is a very good approximation at these densities,
as Figure~\ref{abund_ref}a demonstrates.

\begin{figure}
\centering
\epsfig{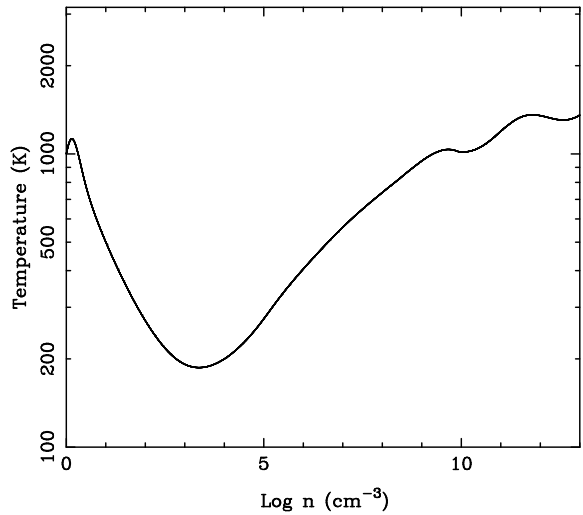}   
\caption{Temperature evolution as a function of gas density
in our reference calculation, run REF. \label{temp-ref}}
\end{figure}

The second phase in the evolution of the $\htp$ abundance begins at a density
of around $10^{8} \: {\rm cm^{-3}}$, when there is a sudden decrease in the $\htp$ 
abundance. This decline is caused by the fact that at these densities, dissociative 
recombination is no longer the only significant destruction mechanism. Increasingly,
$\htp$ is also destroyed by reaction TR17:
\begin{equation}
\htp + \mH \rightarrow \mHtp + \mHt.
\end{equation}
Although this reaction is endothermic, the temperature of the gas at these densities
($T \simgreat 800 \: {\rm K}$; see Figure~\ref{temp-ref}) is high enough to make this 
mechanism significant in comparison to reactions RA18, DR4 and DR5, owing to the 
very low fractional ionization of the gas.

The third phase occurs at $n \sim 10^{10} \: {\rm cm^{-3}}$ as the decline in the
$\htp$ abundance is briefly halted by an increase in the $\htp$ formation
rate. This is caused by the increase in the $\mHt$ abundance at these densities,
which itself is driven by the onset of efficient three-body formation of $\mHt$.

Finally, at a density $n \simgreat 10^{11} \: {\rm cm^{-3}}$, the $\htp$ 
abundance decreases once more, owing to the rapid loss of the few
remaining free $\Hp$ ions from the gas, and the consequent 
disruption of the major $\htp$ formation mechanisms. The reaction 
responsible for this loss of $\Hp$ ions is the same as the reaction
driving the formation of $\htp$, namely RA18. If this were the only process
operating, then it would convert all of the $\Hp$ in the gas into $\htp$ in
a time $t_{\rm conv}$, given approximately by
\begin{equation}
t_{\rm conv} \sim \frac{1}{k_{\rm RA18} n_{\mHt}}.
\end{equation}
Comparing this timescale with the free-fall timescale, $t_{\rm ff} \simeq 1.4 \times 10^{15} n^{-1/2}
\: {\rm s}$, and taking $k_{\rm RA18} = 10^{-16} \: {\rm cm^{3}} \: {\rm s^{-1}}$,
we find that $t_{\rm conv} < t_{\rm ff}$ if $n > 50 / x_{\mHt}^{2}$. For $x_{\mHt} \sim 10^{-3}$,
this gives a critical density $n_{\rm conv} \sim 5 \times 10^{7} \: {\rm cm^{-3}}$. Therefore, in the
absence of any other effects, conversion of $\Hp$ to $\htp$ should occur rapidly once the number
density exceeds $n_{\rm conv}$. In practice, however, a second effect intervenes.
The steady increase in the gas temperature at these densities soon results in reaction TR17
becoming a major destruction mechanism for $\htp$, as noted above. Destruction of $\htp$ by reaction 
TR17 produces $\mHtp$ ions, most of which are then destroyed by reaction CT3
\begin{equation}
\mHtp + \mH \rightarrow \mHt + \Hp,
\end{equation}
producing $\Hp$ ions. Therefore, most of the $\Hp$ ions that are removed from the gas by
reaction RA18 are replaced by this chain of reactions. A net loss of $\Hp$ from the gas
occurs only if the $\htp$ ion produced by reaction RA18 is destroyed by dissociative 
recombination, rather than by reaction TR17. The proportion of the $\htp$ destroyed by
dissociative recombination is given by
\begin{equation}
 f_{\rm DR} \sim \frac{(k_{\rm DR4} + k_{\rm DR5}) 
 n_{\rm e}}{(k_{\rm DR4} + k_{\rm DR5}) n_{\rm e} + k_{\rm TR17} 
 n_{\mH}},
\end{equation}
where we have assumed that dissociative recombination (reactions DR4 and DR5) and 
reaction TR17 are the only significant processes destroying $\htp$. Consequently,
the net rate at which $\Hp$ ions are removed from the gas is a factor $f_{\rm DR}$
slower than was assumed in our calculation of $t_{\rm conv}$ above, and hence the
actual timescale on which the majority of the $\Hp$ ions are removed is given by
\begin{equation}
t_{\rm loss} = \frac{1}{f_{\rm DR}} t_{\rm conv}.
\end{equation}
Now, $t_{\rm loss}$ depends on the electron density through $f_{\rm DR}$, and so as
long as $\Hp$ is the dominant source of free electrons, decreasing its 
abundance increases $t_{\rm loss}$, thereby preventing rapid removal of the $\Hp$
ions from the gas. However, once the $\Hp$ abundance falls below $\sim 10^{-11}$, 
it is singly ionized lithium,  $\lip$, that becomes the dominant positive ion. At
this point, further decreases in $x_{\Hp}$ have very little effect on $t_{\rm loss}$.
Taking $x_{\me} = 10^{-11}$ and $T = 1000 \: {\rm K}$, and assuming that 
$(k_{\rm DR4} + k_{\rm DR5}) n_{\rm e} \ll k_{\rm TR17} n_{\mH}$, we find 
that $f_{\rm DR} \simeq 10^{-4} x_{\mH}^{-1}$, and hence $t_{\rm loss} \simeq 10^{20} x_{\mH} 
/ n_{\mHt} \: {\rm s}$. Thus, at the point at which $\lip$ first becomes the dominant
positive ion, which occurs around $n \sim 10^{9} \: {\rm cm^{-3}}$ in our reference
simulation, $t_{\rm loss} \gg t_{\rm ff}$. However, the wholesale conversion of $\mH$
to $\mHt$ by three-body reactions that begins to set in at around this density rapidly
decreases $t_{\rm loss}$, and at $n \sim 10^{11} \: {\rm cm^{-3}}$ it becomes shorter
than the free-fall timescale of the gas. At this point, most of the remaining $\Hp$ ions
are lost from the gas, following which $\htp$ formation largely ceases. Since the destruction 
of $\htp$ by reactions DR4, DR5 and TR17 is unaffected by the fall-off in the $\Hp$ abundance,
the end result is a very rapid fall-off in the $\htp$ abundance.

\begin{figure}
\centering
\epsfig{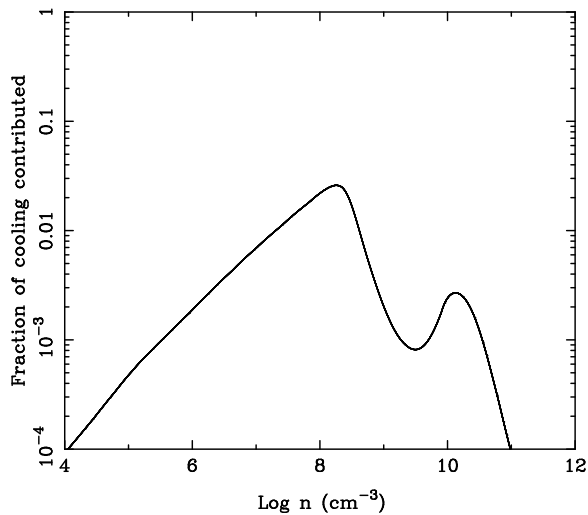}  
\caption{Ratio of the $\htp$ cooling rate to the total cooling rate,
plotted as a function of density, for run REF. \label{cool_contrib_h3p}}
\end{figure}

To determine whether the small amount of $\htp$ that forms in the
gas is enough to significantly affect its thermal evolution, we have 
compared the $\htp$ cooling rate per unit volume to the total
cooling rate per unit volume. The results are plotted in 
Figure~\ref{cool_contrib_h3p}. We see that  even at its moment
of peak effectiveness, which occurs at $n \sim 10^{8} \: {\rm cm^{-3}}$,
$\htp$ contributes no more than about 3\% of the total cooling rate.
At lower densities, the $\htp$ ions, which are not yet in LTE, 
contribute less of the cooling  because they undergo fewer collisions.
At higher densities, on the other hand, the effectiveness of $\htp$
is reduced by the significant decrease in its chemical abundance,
even though each individual $\htp$ ion contributes more cooling
than at $n = 10^{8} \: {\rm cm^{-3}}$.

\begin{figure}
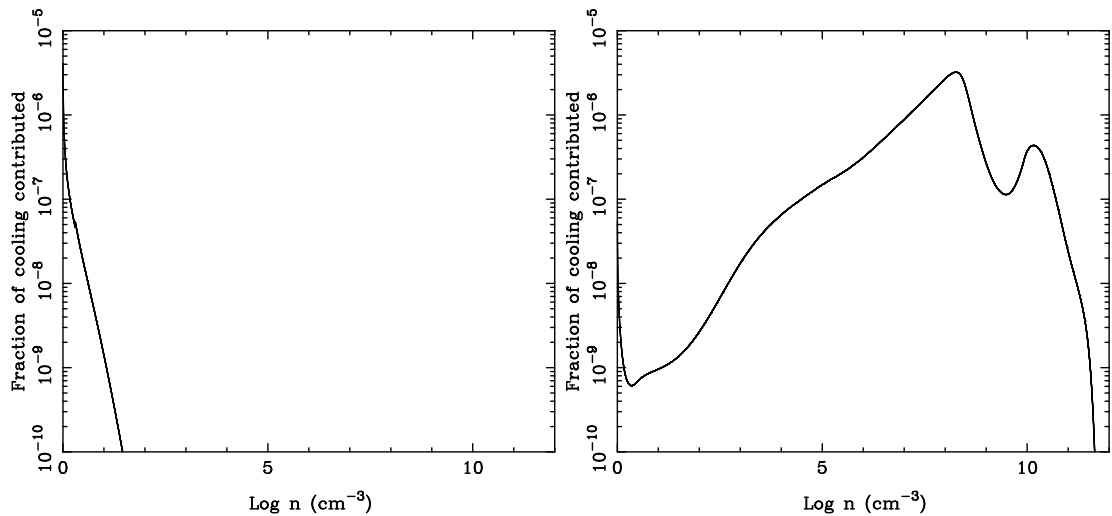

\centering
\epsfig{figure=f4a.eps,width=16pc,angle=270,clip=} 
\epsfig{figure=f4b.eps,width=16pc,angle=270,clip=} 
\caption{(a) As Figure~\ref{cool_contrib_h3p}, but for cooling from
$\mHtp$ and its deuterated counterparts.
Note the difference in horizontal and vertical scales from 
Figure~\ref{cool_contrib_h3p}.
(b) As (a), but for the sum of the contributions of the minor coolants
discussed in \S\ref{minor_cool}.
\label{cool_contribs_other}}
\end{figure}

If $\htp$ is ineffective, then what about other potential coolants,
such as $\lih$, $\mHtp$ and its deuterated counterparts, or other
ions such as $\hehp$ or $\hhdp$? As far as $\lih$ is concerned,
\citet{mon05} have already shown that far too little forms for it
ever to be a significant coolant, a result which we confirm (c.f., 
our Fig.~\ref{abund_ref}d with their Fig.\ 1b). The contributions
made by the other possible coolants are assessed in 
Figures~\ref{cool_contribs_other}a and \ref{cool_contribs_other}b,
where for convenience we plot only the sum of the contributions
from two sets of species. For the ions in one of these sets ($\mHtp$, $\hdp$ 
and $\ddp$; Fig.~\ref{cool_contribs_other}a), we have cooling functions that 
should be at least reasonably accurate; for those in the other set ($\hhdp$, $\hddp$, $\dtp$,
$\hehp$, $\hedp$, $\hehep$, $\lihp$, $\lidp$ and $\lihhp$; Figure~\ref{cool_contribs_other}b), 
we use the highly approximate treatment described in \S\ref{minor_cool}.
It is clear from the figures that none of these species contribute
significantly to the total cooling rate, which is unsurprising given
their extremely small abundances throughout the range of densities
examined here (see also Figure~\ref{abund_ref}).

\subsection{Sensitivity to the choice of $\mathbf{H_{3}^{+}}$ collision rate}
\label{res:C}
As we saw in the previous sub-section, one of the factors preventing $\htp$
from becoming a dominant coolant in our reference calculation is the fact
that its abundance begins to decrease, owing to the increasing importance
of destruction by collisions with hydrogen atoms, before its cooling rate has 
reached its LTE limit. If the $\htp$ ion were to reach LTE earlier than we have
assumed -- in other words, if the low density limit of its cooling rate were to
be larger -- then $\htp$ cooling would have more effect. We have therefore
explored the effect of altering $C$, the total $\htp$ collisional excitation rate
coefficient that is the single free parameter in our treatment of $\htp$
cooling. The value of $C$ in our reference model, hereafter $C_{\rm ref}$,
is given by Equation~\ref{cref}. Increasing it or decreasing it compared to
this value has the effect of, respectively, increasing or decreasing the low-density $\htp$
cooling rate; or, equivalently, decreasing or increasing the critical density
at which $\htp$ reaches LTE.

\begin{figure}
\centering
\epsfig{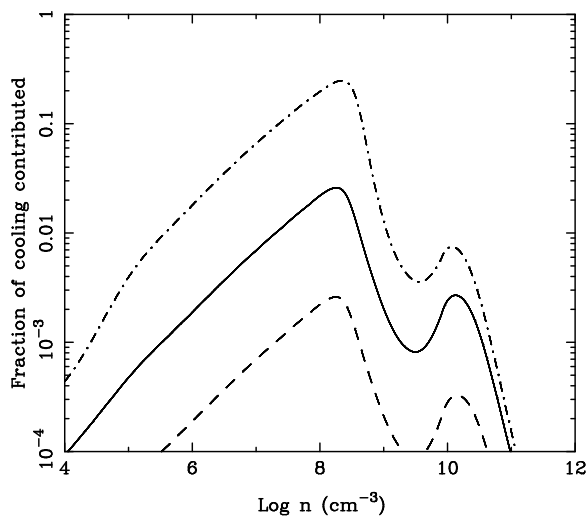}  
\caption{Ratio of the $\htp$ cooling rate to the total cooling rate,
plotted as a function of density, for runs in which the total $\htp$ 
collision rate $C$ was varied. Results are plotted for runs REF (solid line), 
C1 (dashed line) and C2 (dash-dotted line), corresponding to
$C = C_{\rm ref}$, $0.1 \, C_{\rm ref}$ and $10 \, C_{\rm ref}$,
respectively.
\label{contrib-C}}
\end{figure}

\begin{figure}
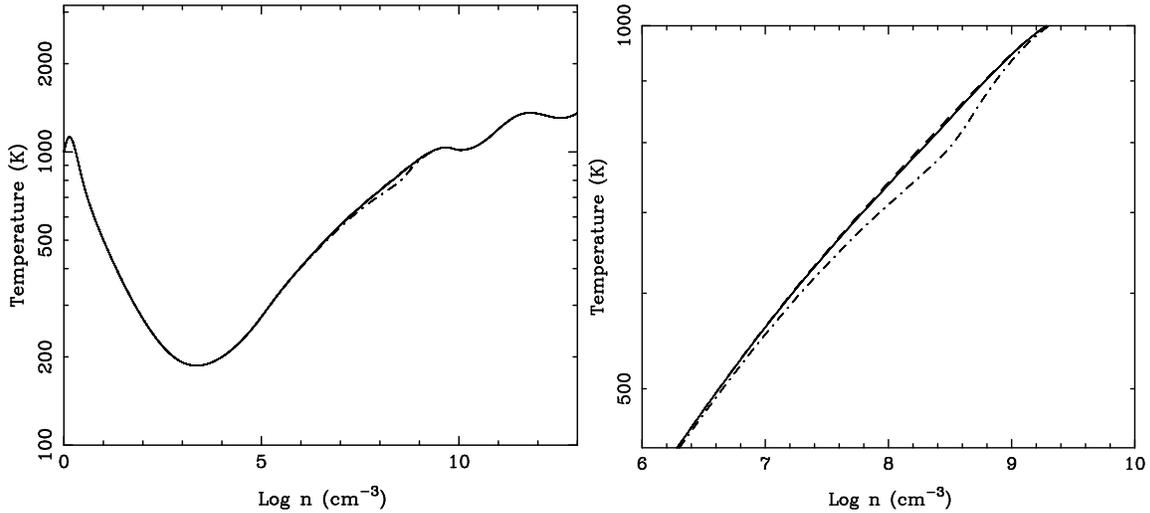

\centering
\epsfig{figure=f6a.eps,width=16pc,angle=270,clip=}   
\epsfig{figure=f6b.eps,width=16pc,angle=270,clip=}  
\caption{(a) Temperature evolution as a function of gas density 
for runs in which the total $\htp$ collision rate $C$ was varied.
Results are plotted for runs REF (solid line), C1 (dashed line) and 
C2 (dash-dotted line). Note that the solid and dashed lines are not
distinguishable in this plot.
(b) As (a), but focussing on a smaller range of densities and
temperatures, to better show the difference between the runs.
The results of runs REF and C1 remain barely distinguishable.
\label{temp-C}}
\end{figure}

In Figure~\ref{contrib-C}, we show the effect that increasing or decreasing
$C$ by a factor of ten has on the contribution made by $\htp$ to the total
cooling rate. As one would expect, the result is rather dramatic. In particular,
it is clear that if $C = 10 \: C_{\rm ref}$, then $\htp$ cooling {\em does} contribute
significantly to the total cooling rate around densities $n \sim 10^{8} \: 
{\rm cm^{-3}}$. However, such a large value for $C$ seems unrealistic, 
given that in our expression for $C_{\rm ref}$, we are already assuming that
collisions occur at the Langevin rate. Collisional excitation by electrons
could in ideal circumstances give one a large value for $C$, but
it is clear from Figure~\ref{abund_ref} that the electron abundance is
orders of magnitude too low in the present case for collisions with
electrons to be important. 
Moreover, even if $C$ were as large as $10 \: C_{\rm ref}$, the extra
cooling provided by the $\htp$ ions would have only a small effect on
the temperature evolution, as Figure~\ref{temp-C} demonstrates.

On the other hand, if $C$ is smaller than we have assumed, then $\htp$
cooling has even less effect. Therefore, despite the uncertainties in our
treatment of $\htp$ cooling at low densities, our main result -- that $\htp$
cooling is, in general, unimportant -- seems robust.

\subsection{Influence of cosmic rays}
\label{res:cosmic}
In \S\ref{res:basic} we saw that a major reason for the dramatic decrease in the $\htp$ 
abundance at gas densities $n > 10^{8} \: {\rm cm^{-3}}$ is the loss of the remaining $\Hp$ 
ions from the gas. In the absence of any $\Hp$ ions, $\htp$ can no longer
be produced directly by reaction RA18 (formerly, the dominant production mechanism), 
while its production from reactions involving $\mHtp$ or $\hehp$ is also disrupted, 
as the main formation routes for these species also depend upon the availability of $\Hp$.
Clearly, therefore, one way of significantly enhancing the abundance of $\htp$ at
high densities would be to provide a source of $\Hp$ ions at high densities. Alternatively,
the $\htp$ abundance could also be enhanced if there were a suitable source of $\mHtp$ 
ions in dense gas. 

One mechanism capable of producing both $\Hp$ and $\mHtp$ ions in dense gas
is the partial ionization of the gas by an external flux of cosmic rays. In view of the
many uncertainties and unknowns regarding the composition, energy spectrum
and energy density of the cosmic rays produced by the earliest supernovae, 
summarized in \citet{sb07} and \citet{jce07}, we use a highly simplified treatment
of their effects. We assume that all of the uncertainties can be folded into a single
free parameter, $\zeta_{\mH}$, the cosmic ray ionization rate of atomic hydrogen,
and that the cosmic ray ionization rates of other atoms and molecules have the
same scaling with  $\zeta_{\mH}$ as they are commonly assumed to have 
in the local ISM. Secondary effects
resulting from the Prasad-Tarafdar mechanism are treated as outlined in 
\S\ref{cosmic_rays}; note that we assume in this first set of models that Lyman-$\alpha$
photons make a negligible contribution to the secondary photochemical rates.

\begin{figure}
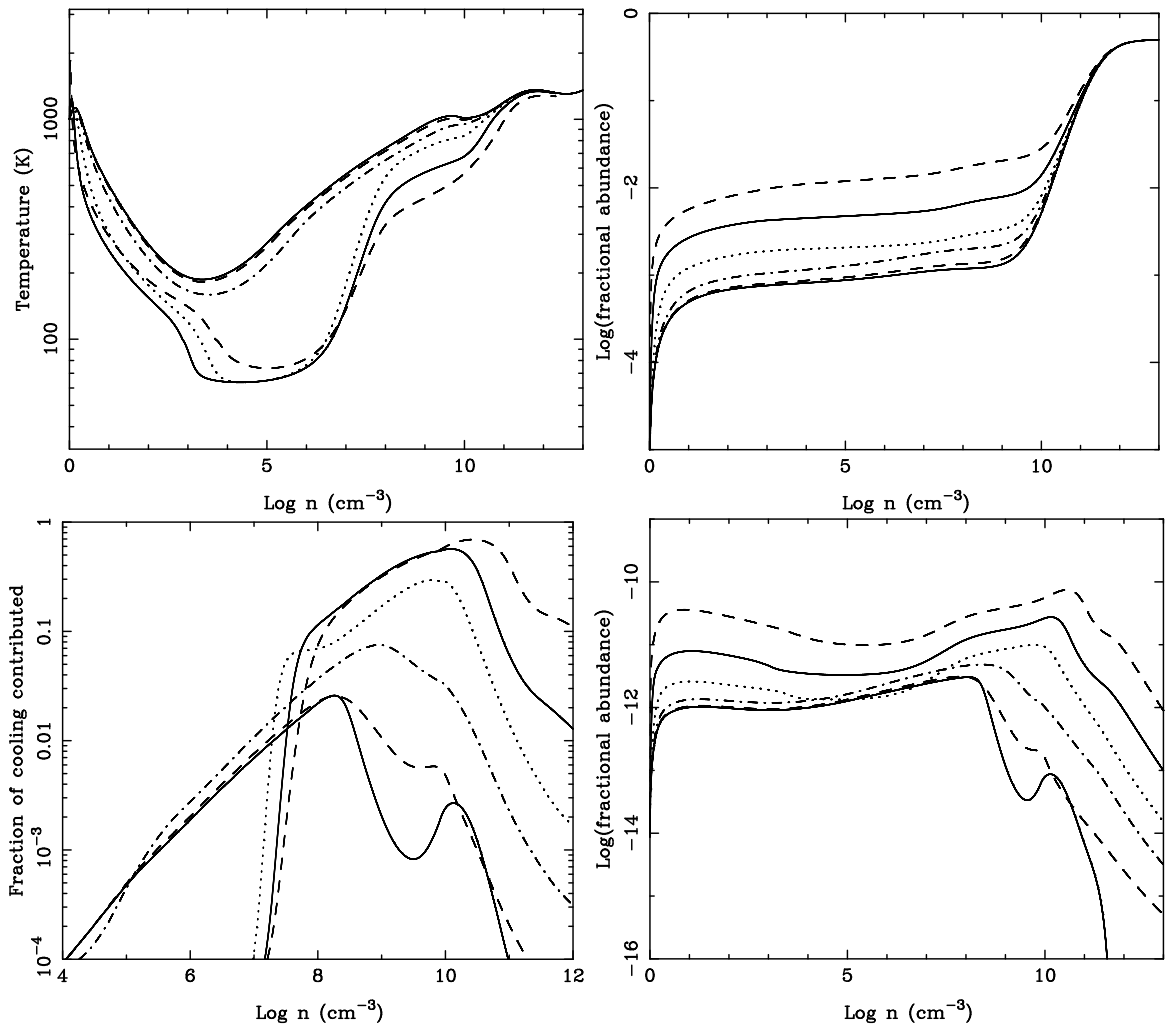

\centering
\epsfig{figure=f7a.eps,width=16pc,angle=270,clip=}  
\epsfig{figure=f7b.eps,width=16pc,angle=270,clip=}  
\epsfig{figure=f7c.eps,width=16pc,angle=270,clip=}  
\epsfig{figure=f7d.eps,width=16pc,angle=270,clip=}  
\caption{(a) Temperature evolution as a function of gas density 
for runs in which the cosmic ray ionization rate $\zeta_{\mH}$ was varied.
Results are plotted for runs REF (upper solid line), CR1 (upper dashed line),
CR2 (dash-dotted line), CR3 (dotted line), CR4 (lower solid line) and 
CR5 (lower dashed line), corresponding to $\zeta_{\mH} = 0.0$,  $10^{-20}$, 
$10^{-19}$,  $10^{-18}$,  $10^{-17}$ and $10^{-16} \: {\rm s^{-1}}$, respectively.
(b) As (a), but showing the evolution of the $\mHt$ abundance with density
in the same set of runs. Note that in this plot the lower solid and dashed lines
correspond to runs REF and CR1, respectively, while the upper solid and dashed lines
correspond to runs CR4 and CR5, respectively.
(c) As (b), but showing the ratio of the $\htp$ cooling rate to the total cooling rate.
The lower solid and dashed lines on the right hand side of the plot correspond
to runs REF and CR1, respectively, the dash-dotted and dotted lines to runs 
CR2 and CR3, respectively, and the upper solid and dashed lines on the right 
hand side of the plot to runs CR4 and CR5, respectively.
(d) As (c), but showing the evolution of the $\htp$ abundance with density
in the same set of runs.
\label{sense-zeta}}
\end{figure}

In Figure~\ref{sense-zeta}a, we show how the temperature evolution of the gas
changes as we increase $\zeta_{\mH}$. We show in the figure results from
five runs that included cosmic rays:  CR1, CR2, CR3, CR4 and CR5,  with
$\zeta_{\mH} = 10^{-20}$, $10^{-19}$,  $10^{-18}$,  $10^{-17}$ and 
$10^{-16} \: {\rm s^{-1}}$, respectively. We also plot the temperature evolution
in our reference run REF, for the purposes of comparison. We see that as 
we increase the cosmic ray ionization rate, the gas gets colder. In particular,
for $\zeta_{\rm H} \ge 10^{-18} \: {\rm s^{-1}}$, the gas is able to cool to 
$T < 100 \: {\rm K}$, indicative of the fact that in these runs, HD becomes the
dominant low-temperature coolant. This is a simple consequence of the
ionization produced by the cosmic rays: the additional free electrons allow
more $\mHt$ to be produced than in our reference run (see Figure~\ref{sense-zeta}b), and 
so the gas can cool to lower temperatures. \citet{sb07} find a similar effect
in their recent study of the effects of cosmic rays on primordial star formation,
and also show that the effect of the cosmic rays on the temperature evolution 
becomes significant once $\zeta_{\rm H} \ge 10^{-19} \: {\rm s^{-1}}$; we find
a slightly larger critical value here, possibly due to the differences in our treatment
of $\mHt$ cooling.

In Figure~\ref{sense-zeta}c, we show how the increase in $\zeta_{\mH}$ affects
the contribution that $\htp$ cooling makes to the total cooling rate. We see that
as the ionization rate increases, $\htp$ cooling becomes steadily less effective
at low densities, particularly for $\zeta_{\rm H} \ge 10^{-18} \: {\rm s^{-1}}$, owing
to the growing importance of HD cooling at these densities. Above $n \sim
10^{8} \: {\rm cm^{-3}}$, however, the contribution from $\htp$ cooling increases
with increasing $\zeta_{\rm H}$. As Figure~\ref{sense-zeta}d illustrates, this is a 
consequence of a significant increase in the high density $\htp$ abundance in 
these runs compared to run REF, which is an expected consequence of the
greater availability of $\Hp$ and $\mHtp$ at high densities in the runs with
non-zero $\zeta_{\rm H}$.

Figure~\ref{sense-zeta}c also demonstrates that if $\zeta_{\mH} \ge 10^{-18} 
\: {\rm s^{-1}}$, then $\htp$ is responsible for $> 10\%$ of the total cooling
over several orders of magnitude in gas density. Moreover, if $\zeta_{\mH} \ge 
10^{-17} \: {\rm s^{-1}}$, then there is a brief period in which it provides
$> 50\%$ of the total cooling. Thus, for cosmic ray ionization rates of this
order of magnitude, $\htp$ cooling is clearly significant and $\htp$ may even 
be the dominant source of cooling at densities $n \sim 10^{10}$--$10^{11}
\: {\rm cm^{-3}}$.

How plausible is it that the cosmic ray ionization rate in primordial gas will
be as large as $10^{-17} \: {\rm s^{-1}}$? This value is comparable with the
standard estimates for the cosmic ray ionization rate in dense gas in the
local ISM \citep[see e.g.,][]{bpwm99,vv00}, and lower than recent estimates
of the rate in diffuse gas \citep[see e.g.,][]{mac03}, and so this value is not {\em prima facie}
unreasonable. However, our requirement that the cosmic rays penetrate 
to very high gas densities means that they must be highly energetic. If we
use the same simple model for our collapsing protostellar core as in 
\S\ref{h3p_cool_rate}, then at $n \sim 10^{10} \: {\rm cm^{-3}}$, the core radius
is $r \sim 10^{15} \: {\rm cm}$, and the column
density of the core is $N \sim 10^{25} \: {\rm cm^{-2}}$. To penetrate to
this depth, the cosmic rays must have energies of at least $100 \: {\rm MeV}$
\citep{sb07}, which means that given reasonable assumptions regarding
the shape of the cosmic ray energy spectrum, the main contribution to the
cosmic ray ionization rate will come from cosmic rays with roughly this
energy.  Following \citet{sb07}, we can estimate the required energy density
in 100~MeV cosmic rays as
\begin{equation}
U_{\rm CR, 100 MeV}  \simeq 4 \times 10^{-13} 
\left( \frac{\zeta_{\mH}}{10^{-17} \: {\rm s^{-1}}} \right) \: {\rm erg}
\: {\rm cm^{-3}}.
\end{equation}

If we further assume that most high-redshift cosmic rays are 
produced in the supernova remnants left by pair-instability 
supernovae \citep{hw02}, then \citet{sb07} show that the
total cosmic ray energy density $U_{\rm CR}$ is related to
the cosmological star-forming rate per unit comoving volume,
$\Psi_{*}$, by
\begin{eqnarray}
U_{\rm CR}(z) & = & 2 \times 10^{-15} \: {\rm erg} \: {\rm cm^{-3}} \: 
\left(\frac{p_{\rm CR}}{0.1} \right)
\left(\frac{E_{\rm SN}}{10^{52} \: {\rm erg}}\right) \left( \frac{1+z}{21}
\right)^{3/2} \nonumber \\
 & & \times \left(\frac{f_{\rm PISN}}{2 \times 10^{-3} \: {\rm M_{\odot}^{-1}}}
\right) \left( \frac{\Psi_{*}}{2 \times 10^{-2} \: {\rm M_{\odot}} \: {\rm yr^{-1}}
\: {\rm Mpc^{-3}}} \right),
\end{eqnarray}
where $p_{\rm CR}$ is the fraction of the supernova explosion energy,
$E_{\rm SN}$, that is used to accelerate cosmic rays, and $f_{\rm PISN}$
is the number of pair-instability supernovae per solar mass of stars formed.
Given reasonable values for $p_{\rm CR}$, $E_{\rm SN}$ and $f_{\rm PISN}$,
this relationship implies that to produce a cosmic ray energy density of
order $10^{-13}  \: {\rm erg} \: {\rm cm^{-3}}$, we require a star formation
rate per unit volume of order $1 \: {\rm M_{\odot}} \: {\rm yr^{-1}} \: {\rm Mpc^{-3}}$,
two to three orders of magnitude larger than current estimates of the
population III star formation rate \citep{yahs03,bl06}. Moreover,
this estimate assumes that essentially all of the cosmic ray energy density is
in 100~MeV cosmic rays; if we allow for a significant fraction of cosmic
rays with smaller energies, as are required to produce the ionization in low
density gas in our simplified model, then the required cosmic star formation
rate increases still further. 

From this argument, we can conclude that any extragalactic cosmic ray
background will be too small to produce the effect that we desire. How
about local sources of cosmic rays? \citet{sb07} show that much higher
energy densities can be produced close to individual supernova
remnants, but to get an energy density of  $10^{-13}  \: {\rm erg} 
\: {\rm cm^{-3}}$ one would have to be within $\sim 10 \: {\rm pc}$ of the 
remnant, near enough that the gas would have been strongly processed 
by the ultraviolet radiation of the supernova progenitor \citep{gb01,wan04,susa07}.
Consequently, this scenario for producing a high cosmic ray
ionization rate also does not appear promising.

\begin{figure}
\centering
\epsfig{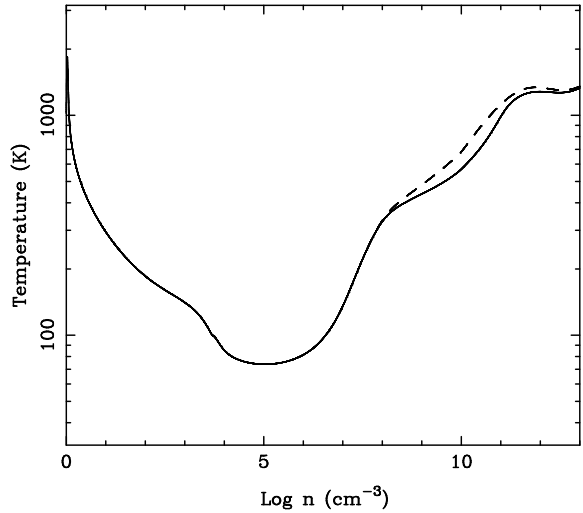}  
\caption{Temperature evolution as a function of gas density in runs CR5
(solid line) and CR6 (dashed line). Both runs share the same set of input
parameters, including a cosmic ray ionization rate $\zeta_{\mH} = 10^{-16}
\: {\rm s^{-1}}$, but in run CR6, $\htp$ cooling has been artificially disabled.
\label{temp-cr-h3p}}
\end{figure}

Furthermore, even if we assume that it is possible to maintain a
large $\zeta_{\rm H}$ at high densities, and that $\htp$ cooling does 
briefly become dominant, it is possible to show that its effects on the 
temperature evolution of the gas remain small. In Figure~\ref{temp-cr-h3p}, we 
compare the temperature evolution in runs CR5 and CR6. In both
runs, we have set $\zeta_{\rm H} = 10^{-16} \: {\rm s^{-1}}$, but in 
run CR6 we have artificially disabled $\htp$ cooling. We see that
at densities $10^{8} \simless n \simless 10^{12} \: {\rm cm^{-3}}$
the temperature in run CR5 is smaller than the temperature in
run CR6, as expected. However, the difference is relatively small,
and the temperature evolution is qualitatively similar in both cases.

Finally, we have examined the importance of cosmic-ray induced
photoionization and photodissocation (the Prasad-Tarafdar
mechanism, discussed in section~\ref{cosmic_rays}) 
by peforming several additional simulations. In runs 
CR7, CR8 and CR9, we took $\zeta_{\mH} = 10^{-20}$, $10^{-18}$ 
and  $10^{-16} \: {\rm s^{-1}}$, respectively, but neglected the effects 
of the Prasad-Tarafdar mechanism completely. In runs CR10,
CR11 and CR12, we adopted the same cosmic ray ionization rates,
but maximized the effects of the Prasad-Tarafdar mechanism by 
assuming that the Lyman-$\alpha$ photons produced by secondary
excitations of atomic hydrogen could propagate freely into the core of 
the protogalaxy, and could contribute to the total secondary photoionization 
and photodissociation rates there (cf.\ our standard treatment, where
we assume that the Lyman-$\alpha$ photons are unable to penetrate
into the core).

\begin{figure}
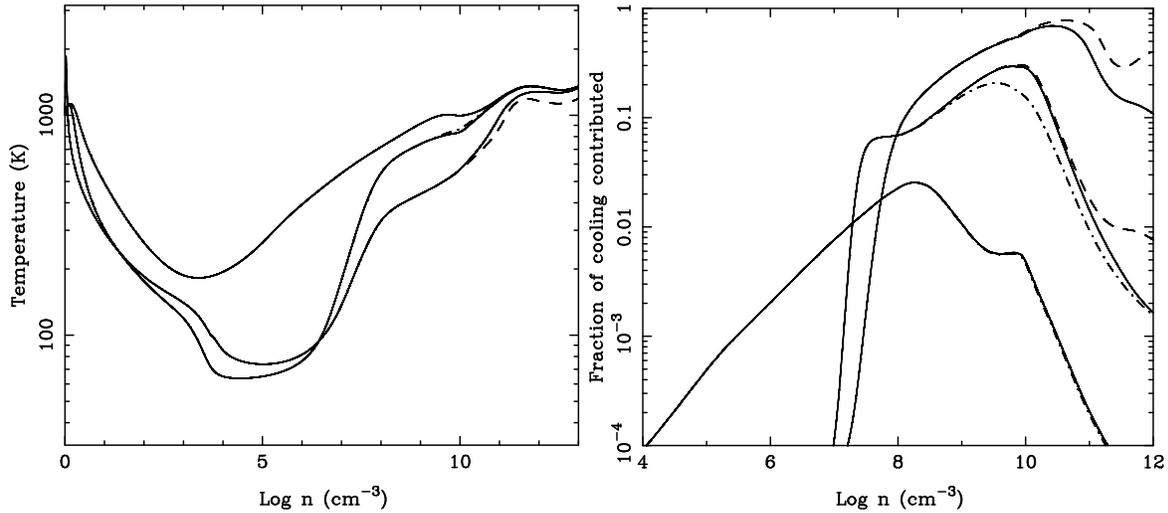

\centering
\epsfig{figure=f9a.eps,width=16pc,angle=270,clip=}  
\epsfig{figure=f9b.eps,width=16pc,angle=270,clip=}  
\caption{(a) Sensitivity of the temperature evolution to our treatment of the
Prasad-Tarafdar mechanism. Solid lines correspond to runs using our
default treatment, dashed lines to runs that neglect its effect entirely, and
dash-dotted lines to runs that maximize its effects by including the effects
of Lyman-$\alpha$ photons as if the gas were optically thin to them. The
lower, middle and upper sets of curves correspond to runs with $\zeta_{\mH} 
= 10^{-20}$, $10^{-18}$ and $10^{-16} \: {\rm s^{-1}}$, respectively. Note
that many of the lines in this figure are indistinguishable.
(b) As (a), but showing the ratio of the $\htp$ cooling rate to the total 
cooling rate in the same set of runs.  The lower, middle and upper 
solid lines on the right-hand side of this plot correspond to runs CR1,
CR3 and CR5, respectively, the lower, middle and upper dashed lines 
on the same side correspond to runs CR7, CR8 and CR9, and  while
the lower, middle and upper dash-dotted lines on that side correspond
to runs CR10, CR11 and CR12. (Note that the results of runs CR5 and 
CR12 are indistinguishable in the plot, as are the results of runs CR1, 
CR7 and CR10). \label{cr-nopt}}
\end{figure}

In Figure~\ref{cr-nopt}a we compare the temperature
evolution of the gas in these six runs with the evolution in runs
CR1, CR3, and CR5, which have the same values of $\zeta_{\mH}$,
but which include the effects of cosmic-ray induced photoprocesses.
In Figure~\ref{cr-nopt}b, we show a similar comparison of the ratio
of $\htp$ cooling to total cooling in these runs.

Below $n = 10^{8} \: {\rm cm^{-3}}$, the Prasad-Tarafdar mechanism
has no discernable effect on the evolution of the gas. At higher densities,
however, its effect is to suppress $\htp$ cooling. More specifically,
cosmic-ray induced photodissociation of $\mHtp$ (reaction CP3)
reduces the amount of $\htp$ formed via reaction TR3, leading to
a reduction in the $\htp$ abundance at these densities, and hence
an overall reduction in the effectiveness of $\htp$ cooling. Nevertheless,
the effect of the enhanced cooling in runs including its effects is
slight, as can be seen from Figure~\ref{cr-nopt}a.

Figure~\ref{cr-nopt} also allows us to assess the impact of the inaccuracy
in our treatment of the Lyman-$\alpha$ photons photons produced by 
secondary excitations of atomic hydrogen. Our results indicate that when
$\zeta_{\rm H}$ is small, the Prasad-Tarafdar mechanism has little effect
on the amount of $\htp$ cooling that occurs, and so any inaccuracies in 
our treatment of it are unimportant. In high-$\zeta_{\rm H}$ runs, the
Prasad-Tarafdar mechanism is more important, but we find that we obtain
very similar results with and without the inclusion of the Lyman-$\alpha$ 
photons, and so any inaccuracy in their treatment is again unimportant.

\subsection{Influence of a radiation background}
\label{res:back}
In most of our calculations, we have assumed that any external sources of radiation
have a negligible effect on the evolution of our collapsing protogalactic gas. At the
epoch corresponding to the formation of the very first stars, this assumption is
well-justified: the cosmic microwave background does not significantly affect the 
gas chemistry at redshifts $z < 100$ and no other sources of radiation yet exist.
Once population III star formation has begun, however, the situation changes.
Radiation from massive population III stars or from their remnants can affect
primordial gas through a variety of mechanisms, as discussed in detail in the
recent reviews by \citet{cf05} and \citet{cia08}. In this section, we explore how
an external radiation field can influence the thermal evolution of primordial 
gas and affect the role of $\htp$ cooling.

\subsubsection{Ultraviolet radiation}
One of the most important forms of radiative feedback in the high-redshift Universe
is the build-up of a soft ultraviolet background at photon energies $h\nu < 13.6 \: 
{\rm eV}$. Photons from this background that are absorbed in the Lyman and
Werner band transitions of $\mHt$ can cause photodissociation, and since 
these photons can  propagate to large cosmological distances through the
intergalactic medium, the strength of the background and the size of the 
associated photodissociation rate can both become considerable. This soft
UV background is therefore expected to have a significant effect on the
evolution of primordial gas within protogalaxies (\citealt{hrl97,har00}; but see
also \citealt{wa07} and \citealt{on08} for evidence that the effects of the Lyman-Werner 
background may be less important than previously supposed).

To investigate the impact of such an ultraviolet background on our results, we 
have run several models with non-zero backgrounds: runs UV1, UV2, UV3,
UV4, UV5 and UV6. As previously noted in \S\ref{photochem_sec}, we assume that the 
spectral shape of the background is that of a diluted $10^{5} \: {\rm K}$ 
black-body, with a sharp cutoff at $13.6 \: {\rm eV}$. The only free parameter 
is then the normalization of this spectrum. We choose to normalize it by 
specifying its strength at the Lyman limit. In runs UV1, UV2 and UV3, the
field strength is $J_{21} = 10^{-4}$, $10^{-2}$, and 1.0, respectively,
where $J_{21}$ is the flux at the Lyman limit in units of $10^{-21} \:
{\rm erg} \: {\rm s^{-1}} \: {\rm cm^{-2}} \: {\rm Hz^{-1}} \: {\rm sr^{-1}}$.
In these three runs, we assume that self-shielding by
$\mHt$ and HD is not effective, 
and that the gas remains optically thin to the external radiation field 
throughout the simulation. In runs UV4, UV5 and UV6, the field strength
is the same as in runs UV1, UV2 and UV3, respectively, but we assume
that $\mHt$ and HD self-shielding is so effective that the $\mHt$ and HD
photodissociation rates are negligible. The true behaviour of the gas
lies between these two limiting cases. 

\begin{figure}
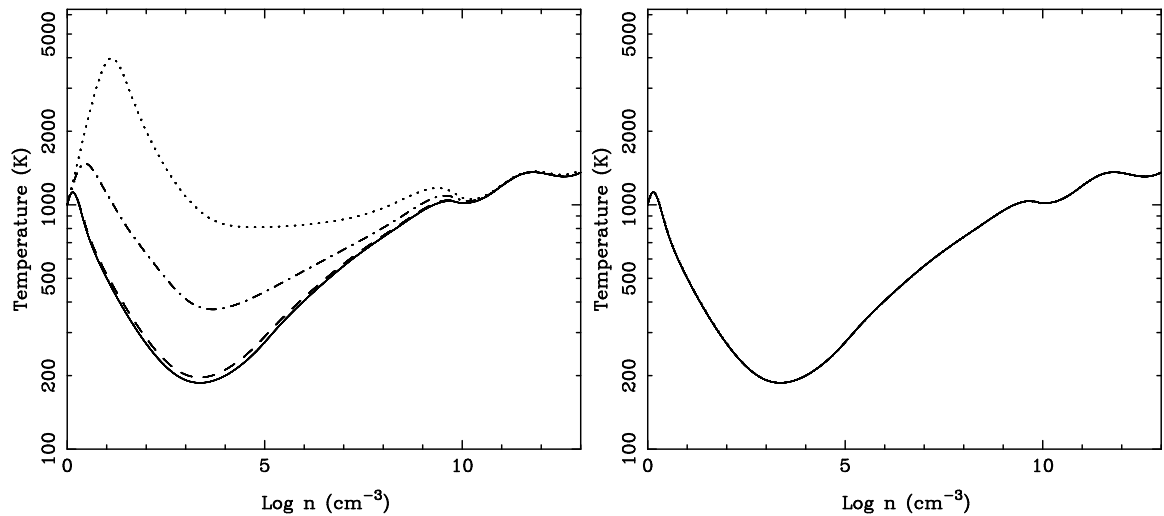

\centering
\epsfig{figure=f10a.eps,width=16pc,angle=270,clip=}   
\epsfig{figure=f10b.eps,width=16pc,angle=270,clip=}   
\caption{(a) Temperature evolution as a function of gas density
in runs REF (solid line), UV1 (dashed line), UV2 (dash-dotted line)
and UV3 (dotted line). The strength of the ultraviolet background
in these runs was $J_{21} = 0.0$, $10^{-4}$, $10^{-2}$ and 1.0,
respectively, and the gas was assumed to remain optically thin
throughout its evolution.
(b) As (a), but for runs REF, UV4, UV5 and UV6. Runs UV4, UV5
and UV6 had the same UV background field strengths as
runs UV1, UV2 and UV3, respectively, but in this case we 
assumed that the gas was optically thick in the Lyman-Werner
lines of $\mHt$ and HD. The results of the four runs are 
indistinguishable in the plot.
\label{temp-UV}}
\end{figure}

In Figure~\ref{temp-UV}a, we show how the gas temperature evolves in optically
thin runs UV1, UV2 and UV3, as well as in run REF for comparison. 
The corresponding behaviour in runs UV4, UV5 and UV6 is shown in
Figure~\ref{temp-UV}b. In the optically thin runs, the effect of the ultraviolet
background is to increase the temperature of the gas at early times; quite
dramatically so in the case of run UV3, where the minimum temperature
reached by the gas is $T \sim 900 \: {\rm K}$, compared to only $T \sim 200
\: {\rm K}$ in our reference calculation. This temperature increase is an
obvious consequence of the photodissociation of $\mHt$ in low density
gas, as can be seen clearly by comparing these results with those from
the runs in which $\mHt$ photodissociation was assumed to be negligible.

\begin{figure}
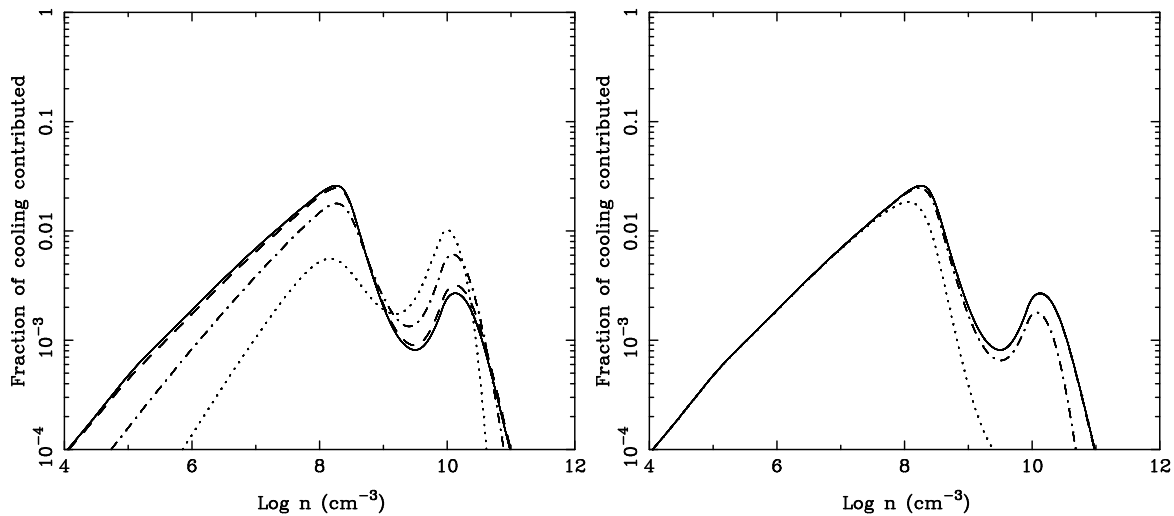

\centering
\epsfig{figure=f11a.eps,width=16pc,angle=270,clip=}   
\epsfig{figure=f11b.eps,width=16pc,angle=270,clip=}   
\caption{(a) Ratio of the $\htp$ cooling rate to the total
cooling rate in runs REF (solid line), UV1 (dashed line),
UV2 (dash-dotted line) and UV3 (dotted line).
(b) As (a), but for runs REF (solid line), UV4 (dashed line),
UV5 (dash-dotted line) and UV6 (dotted line).
\label{contrib-UV}}
\end{figure}

As far as $\htp$ cooling is concerned, Figure~\ref{contrib-UV} demonstrates
that it remains ineffective in both sets of runs. In the optically thin
runs, $\mHt$ dissociation at early times reduces the effectiveness of 
$\mHt$ cooling, but also significantly reduces the $\htp$ abundance.
The net effect is to reduce the amount of cooling coming from $\htp$
to below the level that it has in our reference run. At densities 
$n \simgreat 10^{9} \: {\rm cm^{-3}}$, on the other hand, the effect of 
the UV background is to enhance cooling from $\htp$. This occurs
because the $\Hp$ abundance does not decline so quickly in the 
runs with the UV background, owing to the higher gas temperature,
and so more free protons are available for making $\htp$ at very high 
densities. This effect boosts the contribution of $\htp$ to the cooling rate 
in run UV3 by about a factor of four compared to our reference 
run. Despite this, however, the contribution from $\htp$ remains unimportant.

In runs UV4, UV5 and UV6, the behaviour of the contribution from $\htp$
is somewhat different. The $\mHt$ abundance in these runs evolves in
almost the same manner as in run REF, as does the temperature. Therefore
at early times, there is no difference in the $\htp$ contribution. At
$n > 10^{7} \: {\rm cm^{-3}}$, however, a difference does become apparent
between run REF and runs UV5 and UV6, with the $\htp$ contribution
falling off faster the more the strength of the UV background is increased.
This behaviour is again a result of a change in the behaviour of the $\Hp$
abundance at high densities. In this case, the $\Hp$ abundance falls off
faster at high density when the UV field strength is increased. This occurs
because the ultraviolet background maintains a higher $\lip$ fraction in
the gas than in our reference calculation. Because the $\lip$ abundance
is larger, it contributes more free electrons to the gas, and so the fraction
of $\htp$ ions that are destroyed by dissociative recombination in the high
density regime becomes larger. Consequently, fewer of the $\Hp$ ions
destroyed by reaction RA18 are recycled by reaction TR17, and so the
$\Hp$ and $\htp$ abundances fall off more rapidly than in our reference
model. A similar effect is not seen in the optically thin runs because it is more 
than offset by the effects of the higher gas temperature, which increases the 
rate of reaction TR17, thereby decreasing the fraction of $\htp$ ions that are 
destroyed by dissociative recombination.

\subsubsection{X-rays}
\label{res:X}
X-rays are another form of radiation that can affect the evolution of primordial
gas. By providing an additional source of ionization in dense gas, they can
promote $\mHt$ formation \citep{hrl97,har00,gb03,mba03}, in much the same 
manner as the cosmic rays considered in section~\ref{res:cosmic}.  However, 
just as in the case of cosmic rays, for X-rays to materially affect the importance of $\htp$ 
cooling, they must be able to penetrate deeply into the collapsing protostellar 
core, to densities  $n \sim 10^{10} \: {\rm cm^{-3}}$ or more, corresponding to 
column densities $N \sim 10^{25} \: {\rm cm^{-2}}$ or more. As a core with this
column density is opaque even to hard X-ray photons, the photon flux required
to produce a significant photoionization rate is considerable. 

Given reasonable assumptions regarding the shape of any high-redshift hard 
X-ray background, the dominant contribution to the photoionization rate at 
$n = 10^{10} \: {\rm cm^{-3}}$ comes from X-ray photons with energies 
$E \sim 3$~${\rm keV}$ \citep{ysd98}. The hydrogen ionization cross-section
at this energy is approximately $10^{-24} \: {\rm cm^{2}}$, and so for a column 
density $N = 10^{25} \: {\rm cm^{-2}}$, the gas has an optical depth $\tau \sim 10$ 
for a 3~keV photon. Each of the photons that is absorbed is responsible for roughly 
100 ionizations, once the effects of secondary ionization are taken into account \citep{dyl99}. 
Therefore, a flux $F_{\rm X}$ of 3~keV photons incident on the cloud exterior produces 
an ionization rate $R_{\rm X}$ given approximately by
\begin{eqnarray}
R_{\rm X} & \sim & 100 \left( 10^{-24} F_{\rm X} \Delta \nu \right) e^{-10} \: {\rm s^{-1}}, \nonumber \\
 & \sim & 1.1 \times 10^{-9} F_{\rm X} \: {\rm s^{-1}},
\end{eqnarray}
where the second line follows if we assume that $\Delta \nu \sim 1 \: {\rm keV} / h$.
Inverting this expression, we obtain
\begin{equation}
F_{\rm X}  \sim  9 \times 10^{-9}  \left(\frac{R_{\rm X}}{10^{-17} \: {\rm s^{-1}}} \right)
{\rm photons} \: {\rm s^{-1}} \: {\rm cm^{-2}} \: {\rm Hz}^{-1}.
\end{equation}
This corresponds to an X-ray background field strength at $E = 3 \: {\rm keV}$ of
\begin{equation}
I_{\rm X} = 4.2 \times 10^{-17} \left(\frac{R_{\rm X}}{10^{-17} \: {\rm s^{-1}}} \right)
\: {\rm erg} \: {\rm s^{-1}} \: {\rm cm^{-2}} \: {\rm Hz^{-1}},
\end{equation}
which is orders of magnitude larger than any plausible range of values for the
high-redshift X-ray background \citep[see e.g.,][]{gb03}. We can therefore
rule out an extragalactic X-ray background as the source of the required
hard X-ray photons.

Furthermore, although higher X-ray fluxes can be maintained close to strong
X-ray sources such as miniquasars \citep{har00,km05}, even in this case it
is difficult to produce a significant ionization rate. For example, the luminosity
at $3 \: {\rm keV}$ of the model miniquasars considered by \citet{km05}, 
assuming their hardest spectral model, is $L_{\rm X} \simeq 10^{22} \: {\rm erg} \:
{\rm s^{-1}} \: {\rm Hz}^{-1}$. To see a flux $I_{\rm X} = 4.2 \times 10^{-17}
\: {\rm erg} \: {\rm s^{-1}} \: {\rm cm^{-2}} \: {\rm Hz^{-1}}$ from this miniquasar,
one must therefore be within a distance
\begin{equation}
r \simeq \left( \frac{L_{\rm X}}{I_{\rm X}} \right)^{1/2} = 5.0 \: {\rm pc}
\end{equation}
of it. Gas this close to the miniquasar would have been strongly processed
by the ultraviolet radiation of its progenitor, and is not a promising place to
expect to find further star formation.

We therefore consider it likely that the hard X-ray flux seen by most 
collapsing protostellar cores will be far too small to significantly affect
the production of $\htp$ at high densities. Moreover, even if somehow
a sufficiently large flux was produced, we would expect its effects to
be very similar to those of the cosmic rays considered in \S\ref{res:cosmic}.
Accordingly, we do not consider it necessary or time-efficient to examine
the effects of a hard X-ray flux in any greater detail.

\subsection{Sensitivity to initial conditions}
\label{res:IC}
\subsubsection{Altering the initial density and temperature}
\begin{figure}
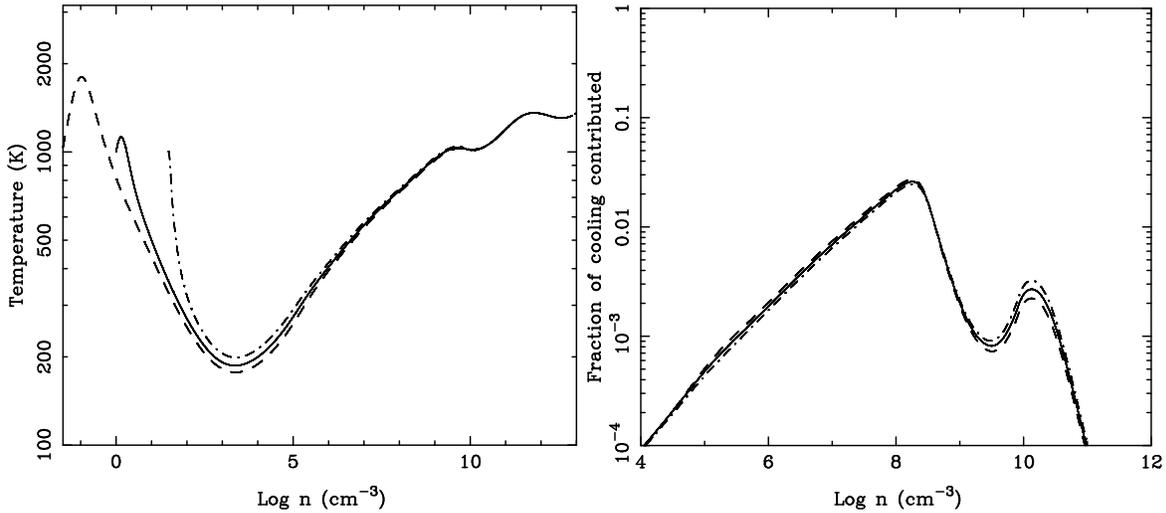

\centering
\epsfig{figure=f12a.eps,width=16pc,angle=270,clip=}  
\epsfig{figure=f12b.eps,width=16pc,angle=270,clip=}  
\caption{(a) Temperature evolution of the gas for various different initial densities.
Results are plotted for runs REF, N1 and N2, which have initial densities
$n_{\rm i} = 1 \: {\rm cm^{-3}}$ (solid line), $n_{\rm i} = 0.03 \: {\rm cm^{-3}}$ (dashed line) 
and $n_{\rm i} = 30 \: {\rm cm^{-3}}$ (dash-dotted line), respectively. 
(b) As (a), but showing the ratio of the $\htp$ cooling rate to the total cooling rate
for the three different models.  \label{n-sense}}
\end{figure}

In order to verify that our main results are not sensitive to the initial temperature
or density assumed in our models, we have performed several calculations
with different initial densities or temperatures. 
In runs N1 and N2, we set $n_{\rm i} = 0.03$ and $30 \: {\rm cm^{-3}}$, respectively, 
while keeping all of the other input parameters fixed. The effect that this has on the 
thermal state of the gas is illustrated in Figure~\ref{n-sense}a, where we compare
the temperature evolution in runs N1 and N2 with the evolution in our reference
calculation, run REF. It is clear from the Figure that the temperature evolution of
the three runs is strongly convergent, in line with previous findings \citep[see
e.g.,][]{pss83,om00}. Consequently, it comes as no surprise to find that
the contribution that $\htp$ makes to the cooling in the three runs is not greatly
affected by the choice of $n_{\rm i}$, as shown in Figure~\ref{n-sense}b.

\begin{figure}
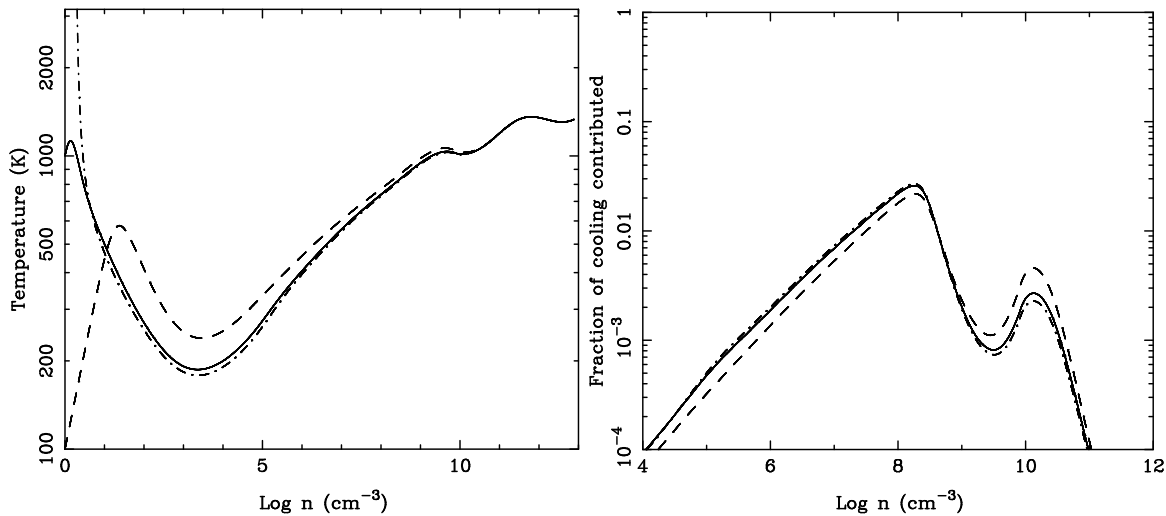

\centering
\epsfig{figure=f13a.eps,width=16pc,angle=270,clip=}  
\epsfig{figure=f13b.eps,width=16pc,angle=270,clip=}  
\caption{(a) Temperature evolution of the gas for various different initial temperatures.
The results plotted are from runs REF, T1 and T2, which had initial temperatures
of  $T_{\rm i} = 1000 \: {\rm K}$ (solid line), $T_{\rm i} = 100 \: {\rm K}$ (dashed line) 
and $T_{\rm i} = 10000 \: {\rm K}$ (dash-dotted line), respectively.
(b) As (a), but showing the ratio of the $\htp$ cooling rate to the total cooling rate
for the three different models.  \label{T-sense}}
\end{figure}

In runs T1 and T2, we set $T_{\rm i} = 100$ and $10000 \: {\rm K}$, respectively, 
and performed a similar comparison, which is illustrated in Figure~\ref{T-sense}.
Again we find that the results of runs T1, T2 and our reference run REF converge
well, although the differences in this case are slightly larger than those that occur
when $n_{\rm i}$ is varied.

\subsubsection{Altering the initial fractional ionization}
\begin{figure}
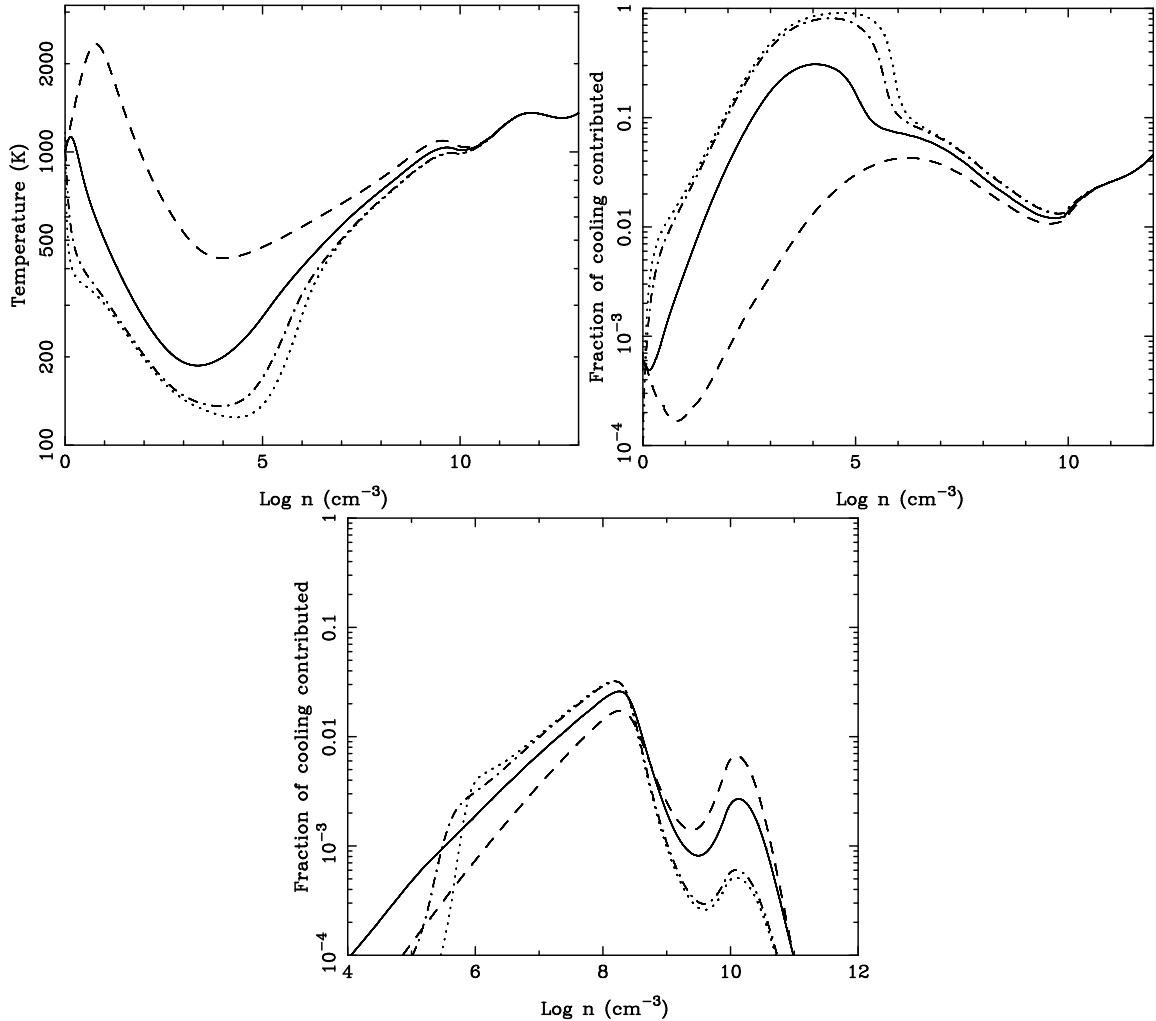

\centering
\epsfig{figure=f14a.eps, width=16pc,angle=270,clip=} 
\epsfig{figure=f14b.eps,width=16pc,angle=270,clip=} 
\epsfig{figure=f14c.eps,width=16pc,angle=270,clip=} 
\caption{(a) Temperature evolution of the gas in runs REF (solid line),
X1 (dashed line), X2 (dash-dotted line), and X3 (dotted line), which had
initial $\Hp$ abundances of $x_{\Hp} = 2.2 \times 10^{-4}$, 
$10^{-6}$, $10^{-2}$ and $1.0$, respectively.
(b) As (a), but showing the ratio of the HD cooling rate to the total cooling rate
in these four runs.
(c) As (b), but for the ratio of the $\htp$ cooling rate to the total cooling rate.
\label{x-sense}}
\end{figure}

We have also explored the effect of altering the initial fractional ionization of the
gas. In runs X1, X2 and X3, we set the initial $\Hp$ abundance to $10^{-6}$,
$10^{-2}$ or 1.0, respectively. We also rescaled the initial $\Dp$ abundances by
a similar amount. However, the initial $\Hep$ abundance was not altered.
As Figure~\ref{x-sense}a demonstrates, altering the initial fractional ionization 
in this way has a dramatic effect on the temperature evolution of the gas.
In run X1, the low abundances of free electrons and of $\Hp$ delay the formation
of $\mHt$ and limit the amount that can form. The gas therefore undergoes a
period of adiabatic heating that lasts for much longer than in our reference 
calculation. Furthermore, once enough $\mHt$ has formed to cool the gas, the
gas temperature remains significantly higher than in run REF. However, the
two runs eventually converge at $n \sim 10^{10} \: {\rm cm^{-3}}$, as at this
density three-body processes dominate the formation of $\mHt$, and so the
$\mHt$ abundance, and hence the thermal evolution of the gas, are no longer
sensitive to the fractional ionization.

In runs X2 and X3, on the other hand, the enhanced initial ionization has the 
effect of promoting the formation of $\mHt$, and cooling the gas more than in
our reference run. Moreover, the cooling provided by this extra $\mHt$ is
sufficient to lower the gas temperature to a level at which chemical fractionation
between HD and $\mHt$ becomes highly effective. The resulting boost in the HD 
abundance allows it to dominate the gas cooling and to cool the gas to a lower
temperature than could be reached by $\mHt$ cooling alone. Note, however,
that in contrast to the results of previous studies 
\citep[see e.g.,][]{nu02,no05,jb06,yokh07}, the gas does not reach the CMB
temperature. This is a consequence of the relatively rapid rate of collapse
assumed here (c.f. \S\ref{res:dyn}) and of the revised treatment of $\mHt$
cooling used in this work, which tends to render $\mHt$ cooling less 
effective, as explored in more detail in \citet{ga08}. In any case, the period of HD 
dominance lasts for only a short
time, as illustrated in Figure~\ref{x-sense}b. As the HD level populations near their 
LTE values,  HD cooling becomes
less effective and the gas starts to warm. As it warms, chemical fractionation
becomes less effective and the HD:$\mHt$ ratio declines. Once the gas
has warmed to $T \sim 200 \: {\rm K}$, which occurs at a density between
$10^{5}$ and $10^{6} \: {\rm cm^{-3}}$, the amount of HD remaining in the gas is
no longer sufficient to maintain HD as the dominant coolant; $\mHt$ becomes
dominant once more, with HD thereafter relegated to a minor role.

In Figure~\ref{x-sense}c, we show how the ratio of the $\htp$ cooling rate
to the total cooling rate varies in these runs. We see that if the initial
fractional ionization is lowered, the contribution of $\htp$ to the cooling is
lowered at $n < 10^{9} \: {\rm cm^{-3}}$ and increased at $n > 10^{9} \:
{\rm cm^{-3}}$ relative to our reference calculation.  The lowered importance
of $\htp$ cooling at low densities  is a result of the higher gas temperature:
the greater temperature sensitivity of the $\mHt$ cooling rate compared
to the $\htp$ cooling rate makes the former more effective in comparison
to the latter as the temperature is raised. At high densities, the gas 
temperatures converge, and the difference between the runs has a
different cause: the lower $\mHt$ abundance (discussed at the beginning of
this subsection) increases the time required
to convert all of the $\Hp$ to $\htp$. For this reason, the $\Hp$ abundance
in $n \sim 10^{9}$--$10^{10} \: {\rm cm^{-3}}$ gas in the low ionization run
is higher than in the reference run, with the result that the $\htp$ abundance
and the $\htp$ contribution to the cooling are also marginally higher. 

If the initial fractional ionization is increased, an interesting effect occurs.
The contribution of $\htp$ to the total cooling rate is considerably
suppressed at densities $n < 10^{6} \: {\rm cm^{-3}}$ and 
$n > 10^{8.5} \: {\rm cm^{-3}}$ compared to the contribution in our 
reference run, but is slightly {\em enhanced} at densities
$10^{6} < n < 10^{8.5} \: {\rm cm^{-3}}$. 
The reduction in the effectiveness of $\htp$ at low densities in these runs
is a result of the previously noted strong enhancement of the HD cooling
rate at $n < 10^{6} \: {\rm cm^{-3}}$, as can clearly be seen by comparing
Figures~\ref{x-sense}b and \ref{x-sense}c. On the other hand, the suppression
of $\htp$ cooling at $n > 10^{8.5} \: {\rm cm^{-3}}$ results from the rapid loss
of $\Hp$ in the gas driven by reaction RA18, 
which occurs more rapidly than in the reference run owing to
the greater $\mHt$ abundance in these runs. Between these two density 
regimes, there is a small range of densities in which the $\htp$:$\mHt$ ratio 
remains relatively large, and where the gas temperature is $\sim 300$--500~K.
At these temperatures, the ratio of the $\htp$ to $\mHt$ cooling rates is
larger than at $T = 1000 \: {\rm K}$, owing to the greater temperature 
dependence of the $\mHt$ cooling rate, but the temperature is not low 
enough for the HD abundance to be significantly enhanced by chemical
fractionation. These conditions are therefore close to ideal for $\htp$ cooling, 
and the fact that even in this case the $\htp$ contributes no more than a few percent 
of the total cooling helps to strengthen our conclusion that it is generally
of little or no importance.

\subsubsection{Changing the elemental composition}
\label{res:elem}
\begin{figure}
\centering
\epsfig{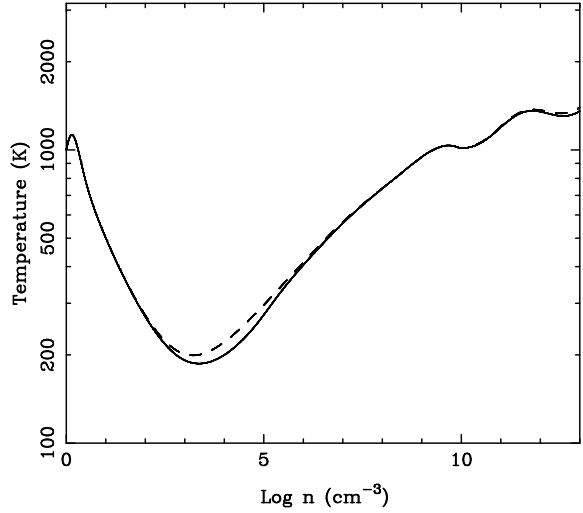}  
\caption{Temperature evolution as a function of density in runs REF
(solid line), which had the standard cosmological deuterium abundance, 
and EL1 (dashed line), in which the deuterium abundance was set 
to zero. \label{noD-temp}}
\end{figure}

Finally, we have examined the effect of changing the elemental composition
of the gas by removing all of the deuterium (run EL1), lithium (run EL2) or both
(run EL3). Although not physically realistic, these runs do provide a 
convenient way to examine the roles that the deuterium and lithium play in
the overall thermal evolution of the gas. In Figure~\ref{noD-temp}, we show how
the gas temperature evolves in run EL1 in comparison to our reference
calculation, run REF. We see that the omission of deuterium has a noticeable
effect on the temperature evolution at densities $10^{3} < n < 10^{6} \: {\rm cm^{-3}}$,
and a very slight effect on the temperature at densities $n > 10^{12} \: {\rm cm^{-3}}$.
Examination of the contribution of HD cooling to the total cooling rate in run REF
(plotted as the solid line in Figure~\ref{x-sense}b) demonstrates that at these densities, 
HD contributes
significantly to the total cooling rate; indeed, at its peak at $n \sim 10^{4} \: {\rm cm^{-3}}$,
it contributes almost a third of the total cooling. At first sight, this result is rather 
surprising, as it is often assumed that HD cooling is unimportant in primordial gas
unless the gas has a large initial fractional ionization, as in runs X2 or X3 discussed 
above. However, it appears that the conventional wisdom is wrong on this point;
our results here are consistent with those of previous studies that have included 
HD \citep[see e.g.,][]{bcl02,mon05}, and show that although HD is never the {\em 
dominant} coolant, it does contribute enough to the total cooling rate at densities
$n \sim 10^{4}$--$10^{5} \: {\rm cm^{-3}}$ to warrant inclusion in future models
of population III star formation.

\begin{figure}
\centering
\epsfig{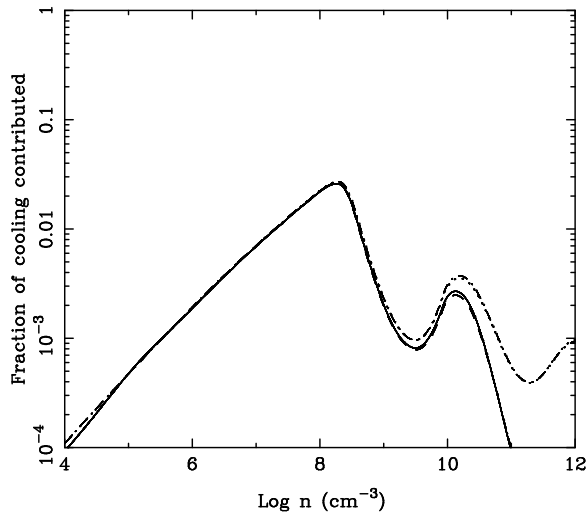}  
\caption{Contribution of $\htp$ cooling to the total cooling rate in runs
with no deuterium (EL1; dashed), no lithium (EL2; dash-dotted) and no
deuterium or lithium (EL3; dotted), along with the results of our reference
run REF (solid) for comparison. Note that the dashed and solid lines are
barely distinguishable from each other; similarly, the dotted and 
dash-dotted lines are not distinguishable in this plot. \label{contrib-elem}}
\end{figure}

In run EL2, the temperature evolution is essentially the same as in run REF, 
while in run EL3, the evolution is the same as in run EL1, indicating that lithium
does not play a significant role in the cooling of the gas, in agreement with
the results presented in \S\ref{res:basic}; note that runs EL2 and EL3 are not
plotted in Figure~\ref{noD-temp}, as they would not be distinguishable from
the existing lines.
In Figure~\ref{contrib-elem}, we investigate the size of the contribution that 
$\htp$ makes to the total cooling rate in runs EL1, EL2 and EL3; we also plot 
the result from run REF for purposes of comparison. The most obvious point to note
is that in the runs without lithium, namely EL2 and EL3, the $\htp$
contribution no longer falls off sharply at high densities, although it remains
too small to be significant. This is easy to understand, given our previous
discussion of the chemistry of the gas at high densities and low
fractional ionizations (see \S\ref{res:basic}).  As previously noted, in the
absence of lithium, the net rate of removal of $\Hp$ ions from the gas via
reaction RA18 decreases as the fractional ionization decreases, since an 
increasing fraction of the $\htp$ created by reaction RA18 is converted 
back to $\mHt$ and $\Hp$ by reactions TR17 and CT3, instead of being
destroyed by reactions DR4 and DR5. 
Therefore, the $\Hp$ removal timescale, $t_{\rm loss}$, remains longer 
than the free-fall timescale throughout the simulation, and the $\Hp$ abundance falls
off gradually at high densities; the rapid fall-off that occurs in our reference
run once $t_{\rm loss} < t_{\rm ff}$ does not take place. Consequently, the 
corresponding rapid fall-off in the $\htp$ abundance also does not occur, as the
$\htp$ formation rate never becomes negligible.

Although the presence or absence of lithium does not affect the
conclusions of our current study, as in either case cooling from $\htp$ 
is negligible, it is clear that if one is interested in determining the fractional
ionization of the gas accurately at very high densities, it is vital to include
$\li$ and $\lip$ in the chemical model \citep[see also][who come to a 
similar conclusion]{ms04}.

\subsection{Sensitivity to uncertainties in the chemical rate coefficients}
\label{res:chem}
\subsubsection{Reaction RA18}
\begin{figure}
\centering
\epsfig{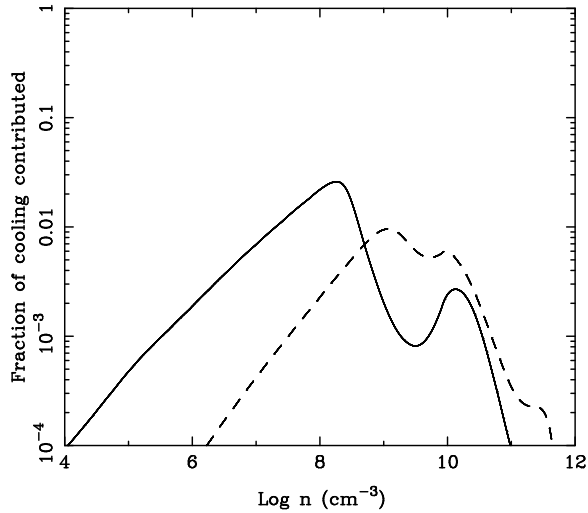} 
\caption{Ratio of the $\htp$ cooling rate to the total cooling rate in runs
REF (solid line) and RA (dashed line). These runs used values for 
$k_{\rm RA18}$ that differed by four orders of magnitude, and that 
were taken from \citet{gh92} and \citet{sld98}, respectively.
\label{h3pra_sense}}
\end{figure}

As we have already discussed in \S\ref{chem}, large uncertainties exist in the
rate coefficients of a number of the processes included in our chemical
model. Of particular relevance to this paper is the huge uncertainty that
appears to exist in the value of the rate coefficient for reaction RA18, the
formation of $\htp$ by the radiative association of $\mHt$ with $\Hp$.
In Figure~\ref{h3pra_sense}, we compare the contribution of $\htp$
cooling in our reference calculation (solid line), in which we adopt the 
large \citet{gh92} rate coefficient for reaction RA18, with the contribution of $\htp$
cooling in run RA, a similar calculation that adopts the smaller \citet{sld98}
rate coefficient (dashed line). We see that $\htp$ is less effective in the latter case,
and that its effectiveness also peaks at a later point in the simulation.

However, the reduction in the $\htp$ contribution is less than one might
expect given the very large difference in $k_{\rm RA18}$ between the
two runs. The reason for this is that although the rate of $\htp$ formation
by radiative association is strongly suppressed, other $\htp$ formation 
mechanisms remain unaffected. The $\htp$ formed in run RA is 
produced primarily by the familiar reaction
\begin{equation}
\mHtp + \mHt \rightarrow \htp + \mH,
\end{equation}
with the necessary $\mHtp$ coming mainly from reaction RA3, namely
\begin{equation}
\mH + \Hp \rightarrow \mHtp + \gamma.
\end{equation}
The persistence of a significant $\htp$ contribution at later times in
run RA than in run REF is a clear consequence of the fact that the rate 
at which $\Hp$ ions are removed from the gas by reaction RA18 is
smaller in the former run than in the latter. As a result, the familiar
rapid fall-off in the $\Hp$ abundance that occurs once the $\Hp$
removal timescale, $t_{\rm loss}$, becomes smaller than the free-fall
time (see \S\ref{res:basic}) takes place at a later time in run RA than in run REF, 
and hence the corresponding fall-off in the $\htp$ abundance also
occurs at a later point in the evolution of the gas.

\subsubsection{Reactions TB1 and TB2}
\begin{figure}
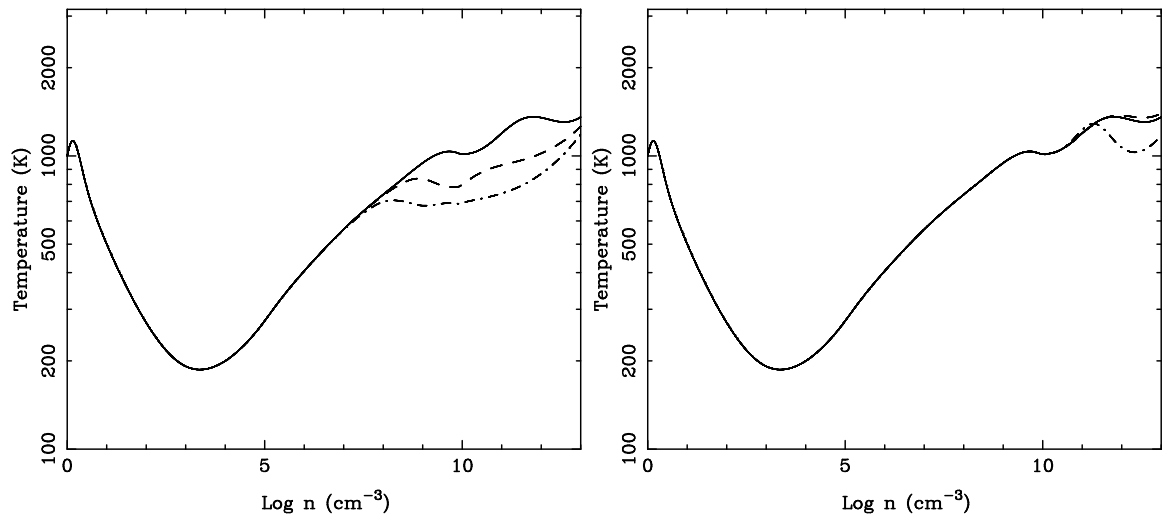

\centering
\epsfig{figure=f18a.eps,width=16pc,angle=270,clip=} 
\epsfig{figure=f18b.eps,width=16pc,angle=270,clip=} 
\caption{(a) Temperature evolution as a function of density, 
plotted for runs REF (solid line), 3B1 (dashed line) and
3B2 (dash-dotted line). We use different values for the rate
of reaction TB1 in these three runs. In run REF,
we use the rate coefficient from \citet{abn02}, in run 3B1 the rate coefficient
from \citet{pss83} and in run 3B2 the rate coefficient from \citet{fh07}. 
(b) As (a), but for runs REF (solid line), 3B3 (dashed line)
and 3B4 (dash-dotted line). In these runs, the rate of reaction 
TB2 is varied. In run REF, we use the rate coefficient from \citet{pss83},
while runs 3B3 and 3B4 use the rate coefficients from \citet{cw83}
and \citet{fh07}, respectively. \label{temp-TB}}
\end{figure}

\begin{figure}
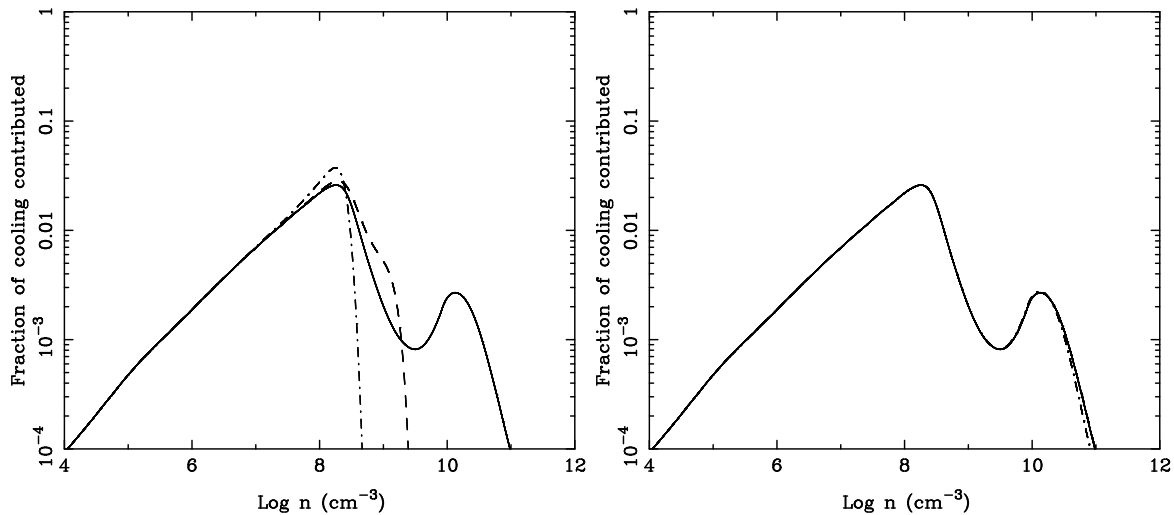

\centering
\epsfig{figure=f19a.eps,width=16pc,angle=270,clip=} 
\epsfig{figure=f19b.eps,width=16pc,angle=270,clip=} 
\caption{(a) Evolution of the ratio of the $\htp$ cooling rate to 
the total cooling rate in runs REF, 3B1 and 3B2.
(b) As (a), but for runs REF, 3B3 and 3B4. \label{cont-TB}}
\end{figure}

Large uncertainties also exist in the rate coefficients for three-body
$\mHt$ formation (reactions TB1 and TB2). As illustrated in 
Figure~\ref{temp-TB}, these uncertainties significantly affect the
temperature evolution of the gas at densities $n > 10^{8} \: {\rm cm^{-3}}$,
particularly the uncertainty in the rate of reaction TB1. The use of
larger values for the three-body reaction rates leads to faster
production of $\mHt$ at high densities, and hence a greater $\mHt$
cooling rate. This has the effect of slowing the rise in the gas
temperature at these densities, which may affect the ability of
the gas to fragment at late times \citep[see][]{cgk08}. However,
in the present context, the effect of the faster $\mHt$ formation rates
is to make cooling by $\htp$ even less effective at late times than
in our reference calculation, as demonstrated in Figure~\ref{cont-TB},
where we examine the effect that varying the rate of both reactions
has on the contribution that $\htp$ makes to the total cooling rate.

\subsubsection{Reaction AD1}
\label{res:chem:ad1}
The uncertainty in the rate of reaction AD1 discussed in \citet{gsj06} also 
affects the temperature evolution of the gas. We have examined two cases, runs
AR1 and AR2. In run AR1, we use a rate coefficient $k_{\rm AD1} = 0.65 \times 
10^{-9} \: {\rm cm^{3}} \: {\rm s^{-1}}$ for reaction AD1, taken from \citet{gsj06},
which is a plausible lower limit on the rate coefficient. In run AR2, on the other hand, we use 
a rate coefficient $k_{\rm AD1} = 5.0 \times 10^{-9} \: {\rm cm^{3}} \: {\rm s^{-1}}$ 
for reaction AD1, again taken from \citet{gsj06}, which is a plausible upper limit.

\begin{figure}
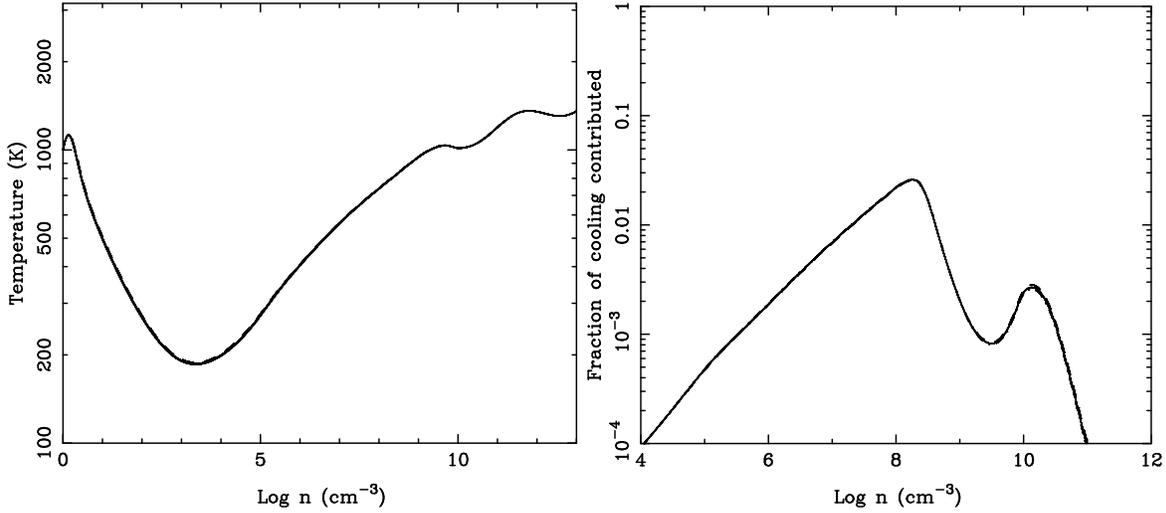

\centering
\epsfig{figure=f20a.eps,width=16pc,angle=270,clip=} 
\epsfig{figure=f20b.eps,width=16pc,angle=270,clip=} 
\caption{(a) Temperature evolution as a function of density in
runs REF (solid line), AR1 (dashed line) and AR2 (dash-dotted
line). In run AR1, the value for the rate coefficient of
reaction AD1 is chosen so as to minimize $\mHt$ production, while 
in run AR2, $\mHt$ production is maximized, as discussed in more detail 
in the text.
(b) As (a), but showing how the ratio of the $\htp$ cooling
rate to the total cooling rate varies in these three models.
\label{sense-MNAD}}
\end{figure}

In Figure~\ref{sense-MNAD}a, we show how the temperature 
of the gas evolves in these two runs, as well as in run REF 
for comparison. It is clear from the figure that the uncertainty in 
$k_{\rm AD1}$ has only a slight impact on the temperature evolution of 
the gas. In Figure~\ref{sense-MNAD}b we show a similar plot of the
ratio of the $\htp$ cooling rate to the total cooling rate. Again,
the rate coefficient uncertainty has only a small effect.
This result is in line with previous work showing that these
are unimportant when starting from cold, low ionization initial
conditions \citep{gsj06}. If, instead, we start with hot, ionized
gas, then the effect of the uncertainties on the temperature
evolution is much greater \citep{ga08}. However, even in
this case, the largest effects are seen at densities $n \simless 10^{4} \: 
{\rm cm^{-3}}$, far below the densities at which $\htp$ could
conceivably become important, and so our basic conclusion regarding the
unimportance of $\htp$ cooling remains unaffected.

\subsubsection{Reactions RA20 and CD26}
\label{res:chem:lih2p}
Finally, we have investigated the effects of varying the rates of two of
the key reactions involved in the $\lihhp$ chemistry: $\lihhp$ formation
by radiative association of $\lip$ and $\mHt$ (reaction RA20) and 
the collisional dissociation of $\lihhp$ by $\mHt$ (reaction CD26). 
As we have already discussed in section~\ref{lihhp_chem}, 
the rate coefficients for
both of these reactions are unknown. In our reference model, we
adopted values of $k_{\rm RA20} = 10^{-22} \: {\rm cm^{3}} \: {\rm s^{-1}}$
and $k_{\rm CD26} = 1.0 \times 10^{-9} \expf{-}{3250}{T} \: {\rm cm^{3}}
\: {\rm s^{-1}}$ for the rate coefficients; 
i.e., a small value for reaction RA20 and a large value for reaction CD26. 
These choices serve to minimize the role played by $\lihhp$ in the chemical 
evolution of the gas, and hence we considered them to be the most conservative
options in the circumstances. In run LP1, we adopted instead a much
larger value for the rate coefficient of reaction RA20, $k_{\rm RA20} =
10^{-17} \: {\rm cm^{3}} \: {\rm s^{-1}}$, but kept the same value for
$k_{\rm CD26}$ as in our reference model. In run LP2, we used our
reference value for $k_{\rm RA20}$, but adopted a much smaller value
for the rate coefficient of reaction CD26: $k_{\rm CD26} = 1.0 \times 10^{-13} 
\expf{-}{3250}{T}  \: {\rm cm^{3}} \: {\rm s^{-1}}$. Finally, in run LP3, we altered 
both rate coefficients, using the larger value for $k_{\rm RA20}$ and the smaller for
$k_{\rm CD26}$.

\begin{figure}
\centering
\epsfig{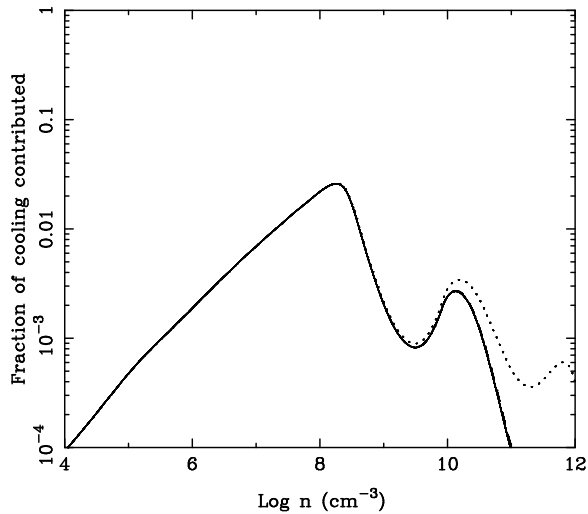} 
\caption{(a) Evolution of the ratio of the $\htp$ cooling rate to the
total cooling rate as a function of density
in runs REF (solid line), LP1 (dashed line), LP2 (dash-dotted line)
and LP3 (dotted line); note that the first three of these lines are
not distinguishable in the plot.  In run LP1, the rate coefficient for reaction 
RA20 was increased by a large factor compared to our reference value, while
in run LP2, the rate coefficient for reaction CD26 was decreased by a 
large factor. In run LP3, both changes were made.  \label{lp-sense}}
\end{figure}

In Figure~\ref{lp-sense}, we show how the contribution of $\htp$ cooling to
the total cooling rate varies with density in runs LP1, LP2 and LP3, along with 
run REF for comparison. It is clear that in runs LP1 and LP2, the behaviour is
essentially the same as in our reference run. This can be understood if we
consider the timescale on which $\lip$ ions are destroyed by the reaction 
sequence
\begin{equation}
\lip + \mHt \rightarrow \lihhp + \gamma,
\end{equation}
followed by
\begin{equation}
\lihhp + \me \rightarrow {\rm products}.
\end{equation}
This sequence of reactions removes $\lip$ on a timescale
\begin{equation}
t_{\rm loss} = \frac{1}{k_{\rm RA20} n_{\mHt} f_{\rm DR}}
\end{equation}
where here $f_{\rm DR}$ is given by
\begin{equation}
f_{\rm DR} = \frac{(k_{\rm DR19} + k_{\rm DR20} + k_{\rm DR21}) n_{\me}}{(k_{\rm DR19} 
+ k_{\rm DR20} + k_{\rm DR21}) n_{\me} +  k_{\rm CD26} n_{\mHt}},
\end{equation}
and represents the fraction of $\lihhp$ ions destroyed by dissociative recombination,
rather than by collisional dissociation. In our reference run, at a gas density $n = 10^{10}
\: {\rm cm^{-3}}$, 
the temperature $T \simeq 1000 \: {\rm K}$ and the electron abundance $x_{\me} \simeq 
5 \times 10^{-11}$, 
and hence $f_{\rm DR} \simeq 1.4 \times 10^{-7}$ and $t_{\rm loss} \simeq 1.4 \times 10^{19} 
\: {\rm s}$, many orders of magnitude longer than the dynamical timescale. Thus, in our reference
run, the $\lihhp$ chemistry has almost no effect on the $\lip$ abundance. In run LP1,
$k_{\rm RA20}$ is a factor of $10^{5} $ larger than in our reference run, and in run LP2,
$k_{\rm CD26}$ is a factor of $10^{4} $ smaller, and so in both runs, $t_{\rm loss}$ is
significantly reduced. However, it still remains far greater than the free-fall timescale, 
which is $\sim 10^{10} \: {\rm s}$ at this density. Thus, in these runs, the $\lihhp$ chemistry
still has almost no effect. 

In run LP3, however, where we both increase $k_{\rm RA20}$ and decrease $k_{\rm CD26}$,
$t_{\rm loss}$ is reduced by a factor of $10^{9}$, making it $t_{\rm loss} \sim 1.4 \times 10^{10}
\: {\rm s}$ at $n = 10^{10} \: {\rm cm^{-3}}$, of the same order of magnitude as the free-fall collapse 
time. Moreover, as $t_{\rm loss}$ scales with density as $t_{\rm loss}
\propto n^{-1}$, while the free-fall time scales as $t_{\rm ff} \propto n^{-1/2}$, $t_{\rm loss}$
becomes smaller than $t_{\rm ff}$ at densities not very much greater than $10^{10} \: {\rm cm^{-3}}$.
Therefore, in this run, the $\lihhp$ chemistry does have a noticeable effect on the $\lip$ 
abundance, reducing it by a factor of roughly fifty in comparison to the reference run by the
end of the simulation. This reduction in the $\lip$ abundance reduces the number of free electrons
available for destroying $\htp$, and so limits the rate at which its abundance declines at very
high densities, much as in runs performed without any lithium (c.f.\ section~\ref{res:elem}).
Nevertheless, it is clear from Figure~\ref{lp-sense} that this change in the lithium chemistry
does not change our basic results: $\htp$ cooling remains ineffective, albeit somewhat less
ineffective at high densities than in our reference run.

\subsection{Sensitivity to the details of the dynamical model}
\label{res:dyn}
A major limitation of our current study is the highly simplified dynamical
treatment that we use in our one-zone model. In all of the calculations
that we have presented so far, we have assumed that the gas is collapsing
gravitationally at the free-fall rate. However, it is well known from 
more detailed three-dimensional hydrodynamical models 
\citep[e.g.,][]{abn02,yoha06} that in realistic primordial clouds, the collapse
speed is significantly slower than the free-fall rate, owing to the
non-negligible gas pressure. As a result, the gas takes longer to
evolve than assumed here, and the impact of compressional
heating is also somewhat smaller. A crude way of taking this into account 
in a one-zone calculation is to artifically slow down the collapse of the 
gas. In other words, instead of assuming that the density evolves 
according to the standard free-fall relationship
\begin{equation}
 \frac{{\rm d} \rho}{{\rm d} t} =  \frac{\rho}{t_{\rm ff}} 
\end{equation}
where $t_{\rm ff}$ is the free-fall time, we can instead assume that
\begin{equation}
 \frac{{\rm d} \rho}{{\rm d} t} = \eta \frac{\rho}{t_{\rm ff}} 
\end{equation}
with $\eta < 1$. The effect of this change is to lengthen the time taken
for the gas to collapse to any given density by a factor $(1 / \eta)$.

\begin{figure}
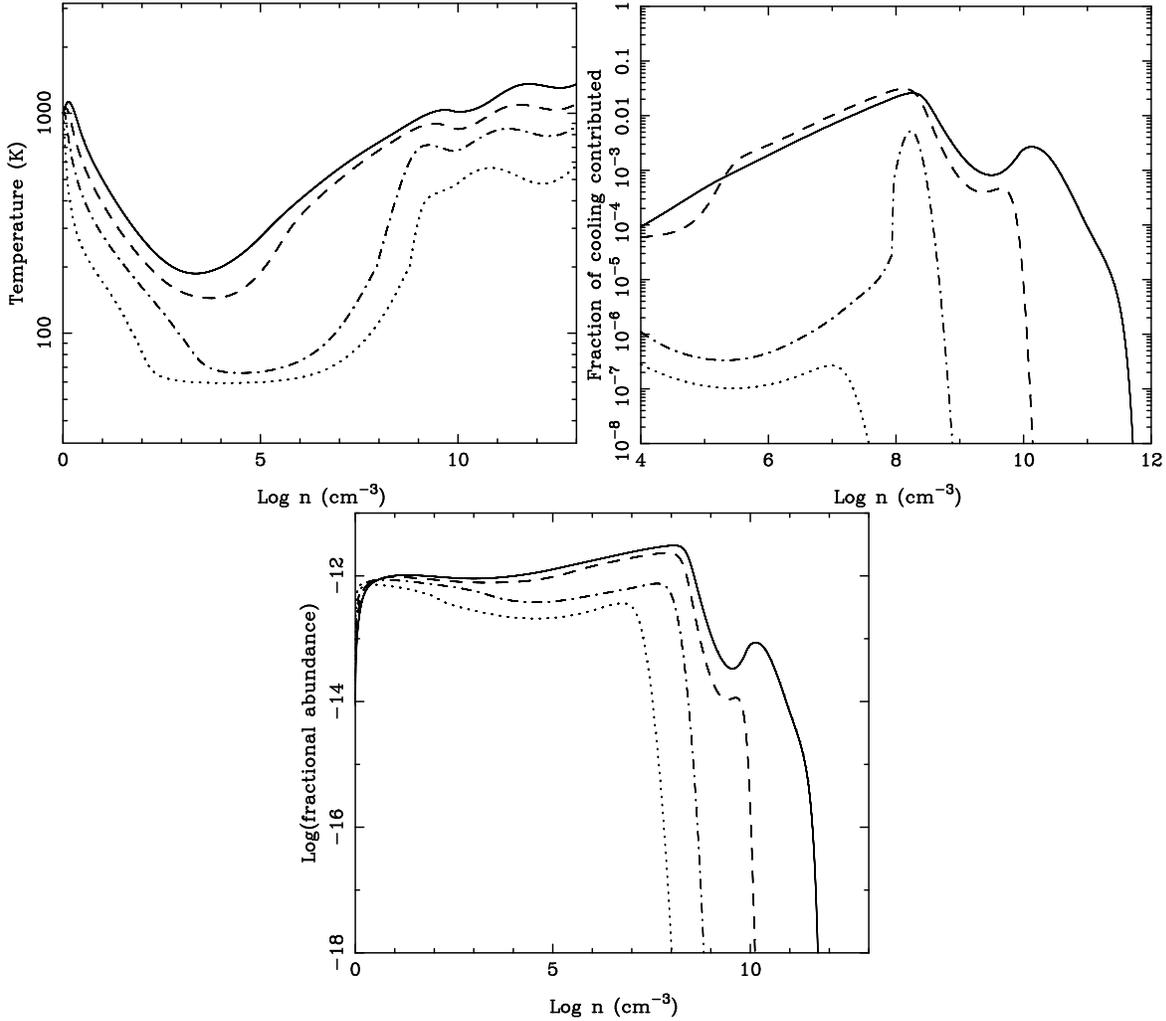

\centering
\epsfig{figure=f22a.eps,width=16pc,angle=270,clip=}  
\epsfig{figure=f22b.eps,width=16pc,angle=270,clip=}  
\epsfig{figure=f22c.eps,width=16pc,angle=270,clip=}  
\caption{(a) Temperature evolution as a function of density
in three runs in which the collapse parameter $\eta$ was
varied. Results are plotted for runs REF (solid line), 
DYN1 (dashed line),  DYN2 (dash-dotted line) and DYN3 
(dotted line), which had $\eta = 1.0$, 0.6, 0.3 and 0.1, respectively.
(b) As (a), but showing the ratio of the $\htp$ cooling rate to the
total cooling rate in the same three runs. 
(c) As (a), but showing the evolution of the $\htp$ abundance in
the three runs. \label{eta-sense}}
\end{figure}

In Figure~\ref{eta-sense}a we show the effect that 
slowing the collapse in this fashion has on the temperature evolution of the gas by 
comparing the results of three runs, DYN1, DYN2 and DYN3, with 
$\eta = 0.6$, $\eta = 0.3$ and $\eta = 0.1$, respectively, with the results of 
our reference run REF.
We see that reducing $\eta$ leads to a reduction in the temperature of
the gas throughout the run, a simple consequence of the reduction in the
compressional heating rate. Interestingly, in the $\eta = 0.3$ and $\eta = 0.1$
models, the reduced heating allows the gas to cool to 
temperatures low enough for chemical fractionation to strongly enhance
the HD fraction, allowing HD cooling to further cool the gas down to
temperatures close to $T_{\rm CMB}$. The fact that this effect is not 
seen in more realistic hydrodynamical models \citep[e.g.,][]{bcl02} 
suggests that in these models we are overestimating the extent to
which gas pressure slows the collapse, at least at early times.

In Figure~\ref{eta-sense}b, we show how reducing $\eta$ affects the
contribution that $\htp$ cooling makes to the total cooling rate. In the
$\eta = 0.6$ run, $\htp$ cooling is slightly more effective than in the
reference run at densities $10^{5} \simless n \simless 10^{8} \: 
{\rm cm^{-3}}$. This is a consequence of the slightly lower temperature
of the gas at these densities in run DYN1 compared to run REF, which
decreases the $\mHt$ cooling rate more than the $\htp$ cooling rate.
At lower densities, $\htp$ cooling in run DYN1 is slightly less effective
than in run REF, as the lower gas temperature makes HD cooling more
effective at these densities in the former run than in the latter. At higher
densities, $\htp$ cooling becomes far less effective in run DYN1 than
in run REF owing to a more rapid falloff in the $\htp$ abundance, as
illustrated in Figure~\ref{eta-sense}c. As in earlier runs, the reason for this
rapid falloff is that the timescale for the removal of $\Hp$ from
the gas by conversion to $\htp$ followed by destruction of
the $\htp$ by dissociative recombination, becomes shorter than 
the dynamical timescale of the gas at these densities. In most
of the runs that we have studied, the high gas temperature at these 
densities allows reaction TR17 to interfere with this process by 
converting most of the $\htp$ to $\mHtp$, following which 
reaction CT3 restores the original proton to the gas. As we saw 
in \S\ref{res:basic}, the effect of this is to lengthen the time 
required to remove all of the $\Hp$, which delays the precipitous
falloff until $n \sim 10^{11} \: {\rm cm^{-3}}$. However, in run DYN1,
the lower gas temperature means that reaction TR17 is less effective,
and so the delay is much shorter. Moreover, the dynamical timescale
itself is longer. Consequently, the rapid falloff occurs at a lower density.

A similar kind of behaviour is seen in runs DYN2 and DYN3. However,
in these runs the much lower temperature at low densities renders HD
cooling dominant for significantly longer, while the lower temperature
of the gas at high densities, plus the longer dynamical timescale, allow
the rapid falloff in the $\htp$ abundance to occur sooner. Consequently,
$\htp$ cooling never becomes significant in these runs.

\section{Conclusions}
\label{conc}
We have examined the contribution that $\htp$ cooling makes to 
the total cooling rate of gravitationally collapsing primordial gas 
in a wide range of different models using a newly-developed
$\htp$ cooling function along with the most detailed model of
primordial gas chemistry published to date. Our results demonstrate
that in general $\htp$ cooling is not important, although it comes
close to being so at densities $n = 10^{7}$--$10^{9} \: {\rm cm^{-3}}$,
contributing at its peak a few percent of the total amount of cooling.
We come to this conclusion despite making several assumptions
(regarding the collapse rate of the gas, the formation rate of $\htp$
by radiative association, and the collisional excitation rate of its
excited vibrational states) that favour $\htp$ cooling, and thus
we have confidence that our conclusion is robust.

As $\htp$ comes so close to being important, it is instructive to
examine why it ultimately fails to dominate. This can be ascribed
to a combination of two main effects. First, at high densities ($n \simgreat
10^{8} \: {\rm cm^{-3}}$), the gas temperature becomes high
enough to make the endothermic reaction TR17
\begin{equation}
\htp + \mH \rightarrow \mHtp + \mHt
\end{equation}
the most important $\htp$ destruction mechanism, which
significantly suppresses the $\htp$ abundance at these
densities. Second, the formation rate of $\htp$ is strongly
suppressed by the rapid removal of $\Hp$ from the gas
at densities $n \simgreat 10^{11} \: {\rm cm^{-3}}$. At these
densities, the fractional ionization of the gas is so low that
the main loss route for the $\Hp$ is conversion to $\htp$,
followed by $\htp$ dissociative recombination, and once
the timescale for $\Hp$ removal via this combination of
reactions becomes short compared to the free-fall time,
the $\Hp$ abundance decreases by orders of magnitude
within a short space of time, effectively switching off the 
formation of $\htp$. Moreover, destruction of $\htp$ by
dissociative recombination remains effective despite
the fall-off in $x_{\Hp}$ thanks to the contribution of
electrons from ionized lithium, $\lip$, which for $n > 3 \times 10^{8} \: 
{\rm cm^{-3}}$ is the most abundant positive ion in the gas.

In our study, the only situation in which we found $\htp$
cooling to be important is if the gas is illuminated by a strong flux of
cosmic rays or X-rays. If the incident flux is strong enough
to produce an ionization rate $\simgreat 10^{-18} \: {\rm s^{-1}}$
at densities $n \ge 10^{10} \: {\rm cm^{-3}}$, then the
high density $\htp$ abundance can be significantly
increased, and $\htp$ can even become the dominant
coolant. However, the necessary flux of cosmic rays or
hard X-rays is difficult to produce in the high-redshift
Universe. As the estimates in \S\S~\ref{res:cosmic}
and \ref{res:X} demonstrate,
the required flux is orders of magnitude greater than 
the size of any plausible extragalactic background, 
and will only be achieved within gas that is very close
to a local source (i.e., within 5--10~pc). However, gas that
is this close to a supernova remnant or miniquasar will
have been strongly affected by radiative feedback from
the progenitor star and so is not a promising place to
expect to find ongoing population III star formation.

Furthermore, even if the ionization rate is high enough
to make $\htp$  an important or dominant coolant  at
high densities, the effect of $\htp$ cooling on the
thermal evolution of the gas remains relatively small;
the difference it makes to the temperature evolution
at $n > 10^{8} \: {\rm cm^{-3}}$ is smaller than the error
introduced by the uncertainty in the three-body
$\mHt$ formation rate (reaction TB1).

Finally, our model has also allowed us to explore the effects of
cooling from the other minor ionic and molecular species present
in the gas (e.g., $\mHtp$, ${\rm H_{2}D^{+}}$, $\lih$, etc.). Despite
making rather optimistic assumptions regarding the cooling from
these species, we find that they are orders of magnitude less effective
than $\htp$ at cooling high density gas, and hence are never
significant.

\section*{Acknowledgments}
The authors would like to thank P.\ C.\ Stancil for useful discussions
regarding the rates of reactions involving deuterium, and C.\ Greene
for providing us with an estimate of the $\lihhp$ dissociative 
recombination rate. DWS was 
supported in part by the NASA Astronomy and Physics Research 
and Analysis program and the NSF Astronomy and Astrophysics 
Grants program.

\clearpage

\appendix
\section{Chemical network}
In Tables~\ref{chemtab_ci}--\ref{cr_pd_tab} we list the chemical reactions included
in our model of primordial gas, along with the rate coefficients adopted and the references 
from which these rate coefficients were taken. Some of these reactions are discussed in more detail
in \S\ref{react-discuss}. In these tables, $T$ is the gas temperature in K, $T_{3} = T / 300 \: {\rm K}$, and
$T_{\rm e}$ is the gas temperature in units of eV.

\begin{table*}
\begin{minipage}{126mm}
\caption{Chemical processes: collisional ionization (CI) \label{chemtab_ci}}
\begin{tabular}{lllc}
\hline
No.\ & Reaction & Rate coefficient (${\rm cm}^{3} \: {\rm s}^{-1}$) & Ref.\  \\
\hline
CI1  &  $\mH + \me  \rightarrow  \Hp + \me + \me $ &
$k_{\rm CI1} =  \exp[-3.271396786 \times 10^{1}$ & 1 \\
& & $\phantom{k_{\rm CI1}=} \mbox{}  + 1.35365560 \times 10^{1} \ln T_{\rm e}$ & \\
& & $\phantom{k_{\rm CI1}=} \mbox{}  - 5.73932875 \times 10^{0} (\ln T_{\rm e})^{2}$ & \\
& & $\phantom{k_{\rm CI1}=} \mbox{}  + 1.56315498 \times 10^{0} (\ln T_{\rm e})^{3}$ & \\
& & $\phantom{k_{\rm CI1}=} \mbox{}  -  2.87705600 \times 10^{-1} (\ln T_{\rm e})^{4}$ & \\
& & $\phantom{k_{\rm CI1}=} \mbox{}  + 3.48255977 \times 10^{-2} (\ln T_{\rm e})^{5}$ & \\
& & $\phantom{k_{\rm CI1}=} \mbox{}   - 2.63197617 \times 10^{-3} (\ln T_{\rm e})^{6}$ & \\ 
& & $\phantom{k_{\rm CI1}=} \mbox{}  + 1.11954395\times 10^{-4} (\ln T_{\rm e})^{7}$ & \\
& & $\phantom{k_{\rm CI1}=} \mbox{}   -  2.03914985 \times 10^{-6} (\ln T_{\rm e})^{8}]$ & \\
CI2  &  $\mD + \me  \rightarrow  \Dp + \me + \me $ & $k_{\rm CI2} = k_{\rm CI1}$ & 1  \\ 
CI3  &  $\He + \me  \rightarrow  \Hep + \me + \me $ &
$k_{\rm CI3} = \exp[-4.409864886 \times 10^{1}$ & 1 \\
& & $\phantom{k_{\rm CI3}=} \mbox{} + 2.391596563 \times 10^{1} \ln T_{\rm e}$ & \\
& & $\phantom{k_{\rm CI3}=} \mbox{} - 1.07532302 \times 10^{1} (\ln T_{\rm e})^{2}$ & \\
& & $\phantom{k_{\rm CI3}=} \mbox{} +3.05803875 \times 10^{0} (\ln T_{\rm e})^{3}$ & \\
& & $\phantom{k_{\rm CI3}=} \mbox{} - 5.68511890 \times 10^{-1} (\ln T_{\rm e})^{4}$ & \\
& & $\phantom{k_{\rm CI3}=} \mbox{} +6.79539123 \times 10^{-2} (\ln T_{\rm e})^{5}$ & \\
& & $\phantom{k_{\rm CI3}=} \mbox{} -5.00905610 \times 10^{-3} (\ln T_{\rm e})^{6}$ & \\
& & $\phantom{k_{\rm CI3}=} \mbox{} + 2.06723616\times 10^{-4}  (\ln T_{\rm e})^{7}$ & \\
& & $\phantom{k_{\rm CI3}=} \mbox{} - 3.64916141 \times 10^{-6} (\ln T_{\rm e})^{8}]$ & \\ 
CI4  &  $\li + \me  \rightarrow  \lip + \me + \me $ &
$k_{\rm CI4} = 3.11 \times 10^{-8} T_{3}^{0.163} \expf{-}{62700}{T}$ & 2 \\
\hline
\end{tabular}
\medskip
\\
{\bf Notes}: T is the gas temperature in K, $T_{3} = T / 300 \: {\rm K}$, and
$T_{\rm e}$ is the gas temperature in eV. \\
{\bf References}:  1 -- \citet{j87}; 2 -- \citet{v97}
\end{minipage}
\end{table*}

\begin{table*}
\begin{minipage}{126mm}
\caption{Chemical processes: photorecombination (PR)
\label{tab:PR}}
\begin{tabular}{llllc}
\hline
No.\ & Reaction & Rate coefficient (${\rm cm}^{3} \: {\rm s}^{-1}$) & Notes & Ref.\  \\
\hline
PR1 &  $\Hp + \me  \rightarrow  \mH + \gamma $ &
 $k_{\rm PR1} = 2.753 \times 10^{-14} \left(\frac{315614}{T}\right)^{1.500}
[1.0+ \left(\frac{115188}{T}\right)^{0.407}]^{-2.242} $ & & 1 \\ 
PR2  &  $\Dp + \me  \rightarrow \mD + \gamma $ & $k_{\rm PR2} = k_{\rm PR1}$ & & 1 \\ 
PR3 & $\Hep + \me \rightarrow \He + \gamma$ & 
$k_{\rm PR3,  rr, A} = 10^{-11} T^{-0.5} \left[12.72 - 1.615 \log{T} \right. $ & Case A & 2  \\
& & $\left. \phantom{k_{\rm PR3, rr, A} = } \mbox{} 
- 0.3162 (\log{T})^{2} + 0.0493 (\log{T})^{3}\right]$ & & \\
& & $k_{\rm PR3, rr, B} = 10^{-11} T^{-0.5} \left[11.19 - 1.676 \log{T} \right. $ & Case B & 2  \\
& & $\left. \phantom{k_{\rm PR3, rr, A} = } \mbox{} - 0.2852 (\log{T})^{2} 
+ 0.04433 (\log{T})^{3} \right]$ & & \\
& & $k_{\rm PR3,  di} = T^{-1.5} \left[5.966 \times 10^{-4} \exp \left(\frac{-455600}{T} \right) \right.$ & & 3 \\
& & $\phantom{k_{\rm PR3, di} =} \mbox{} + 1.613 \times 10^{-4}   \exp \left(\frac{-555200}{T} \right)$ 
& & \\
& & $\phantom{k_{\rm PR3, di} =} \left. \mbox{} - 2.223 \times 10^{-5}   \exp \left(\frac{-898200}{T} \right)
\right]$ & & \\
PR4 &  $\lip + \me  \rightarrow  \li + \gamma $ &
$k_{\rm PR4, rr} = 1.036 \times 10^{-11} \left(\frac{T}{107.7}\right)^{-0.5} 
\left[1.0 + \left(\frac{T}{107.7}\right)^{0.5}\right]^{-0.612}$ & & 4 \\ 
& & $\phantom{k_{\rm PR4} =}  \times \left[1.0 + \left(\frac{T}{1.177 
\times 10^{7}}\right)^{0.5}\right]^{-1.388} $ & & \\ 
& & $k_{\rm PR4, di} = T^{-1.5} \left[2.941 \times 10^{-5} \exp \left(\frac{-634500}{T} \right) \right.$ & & 5 \\
& & $\phantom{k_{\rm PR4, di} =} \mbox{} + 6.068 \times 10^{-5}   \exp \left(\frac{-702400}{T} \right)$ 
& & \\
& & $\phantom{k_{\rm PR4, di} =} \left. \mbox{} - 7.753 \times 10^{-7}   \exp \left(\frac{-827100}{T} \right)
\right]$ & & \\
\hline
\end{tabular}
\medskip
\\
{\bf Notes}: T is the gas temperature in K. Note that the recently revised values for PR1 and
for the radiative recombination portions of PR3 and PR4 presented by \citet{bad06rr} do 
not differ from the older rate coefficients quoted here by more than a couple of percent
at the temperatures of interest in this study.
\\
{\bf References}: 1 -- \citet{fer92}; 2 -- \citet{hs98}; 3 -- \citet{bad06}; 4 -- \citet{vf96}; 5 -- \citet{bb07}
\end{minipage}
\end{table*}

\begin{table*}
\begin{minipage}{126mm}
\caption{Chemical processes: dissociative recombination (DR)
\label{tab:DR}}
\begin{tabular}{llllc}
\hline
No.\ & Reaction & Rate coefficient (${\rm cm}^{3} \: {\rm s}^{-1}$)  & Notes & Ref.\  \\
\hline
DR1  &  $\mHtp + \me  \rightarrow  \mH + \mH $ &
$k_{\rm DR1} = 1.0 \times 10^{-8}$  & $T \le 617 \: {\rm K}$ & 1 \\
& & $\phantom{k_{\rm DR1}} = 1.32 \times 10^{-6} T^{-0.76}$ & $T > 617 \: {\rm K}$ & \\
DR2  &  $\hdp + \me  \rightarrow  \mH + \mD $ &
$k_{\rm DR2} = 7.2 \times 10^{-8} T^{-0.5}$ & & 2 \\ 
DR3 & $\ddp + \me \rightarrow \mD + \mD$  & $k_{\rm DR3} = 3.4 \times 10^{-9} T_{3}^{-0.4}$ &  & 3 \\
DR4  &  $\htp + \me  \rightarrow  \mHt + \mH $ &
$k_{\rm DR4} = 2.34 \times 10^{-8} T_{3}^{-0.52}$ & & 4 \\
DR5  &  $\htp + \me  \rightarrow  \mH + \mH + \mH $ &
$k_{\rm DR5} =  4.36 \times 10^{-8} T_{3}^{-0.52}$ & & 4 \\
DR6  &  $\hhdp + \me  \rightarrow  \mH + \mH + \mD $ &
$k_{\rm DR6} = 4.38 \times 10^{-8} T_{3}^{-0.5}$ & & 5 \\ 
DR7  &  $\hhdp + \me  \rightarrow  \mHt + \mD $ & 
$k_{\rm DR7} = 4.2 \times 10^{-9} T_{3}^{-0.5}$ & & 5 \\ 
DR8 &  $\hhdp + \me  \rightarrow  \mH + \hd $ & 
$k_{\rm DR8} = 1.2 \times 10^{-8} T_{3}^{-0.5}$ & & 5 \\ 
DR9 & $\hddp + \me \rightarrow \mD + \mD + \mH$ & 
$k_{\rm DR9} = 4.38 \times 10^{-8} T_{3}^{-0.5}$ & & 6 \\
DR10 & $\hddp + \me \rightarrow \DD + \mH $ & 
$k_{\rm DR10} = 4.2 \times 10^{-9} T_{3}^{-0.5}$ & & 6 \\
DR11 & $\hddp + \me \rightarrow \hd + \mD$ &
$k_{\rm DR11} = 1.2 \times 10^{-8} T_{3}^{-0.5}$ & & 6 \\
DR12 & $\dtp + \me \rightarrow \DD + \mD$ & 
$k_{\rm DR12} = 5.4 \times 10^{-9} T_{3}^{-0.5}$  & & 7 \\
DR13 &$\dtp + \me \rightarrow \mD + \mD + \mD$ & 
$k_{\rm DR13} = 2.16 \times 10^{-8} T_{3}^{-0.5}$ & & 7 \\
DR14  &  $\hehp + \me  \rightarrow  \He + \mH $ &
$k_{\rm DR14} = 3.0 \times 10^{-8} T_{3}^{-0.47}$ & & 8 \\ 
DR15  &  $\hedp + \me  \rightarrow  \He + \mD $ & 
$k_{\rm DR15} = 3.0 \times 10^{-8} T_{3}^{-0.47}$ & & 9 \\ 
DR16  &  $\hehep + \me  \rightarrow  \He + \He $ &
$k_{\rm DR16} = 6.1 \times 10^{-11} T_{3}^{-0.9}$ & & 10 \\ 
DR17 &  $\lihp + \me  \rightarrow  \li + \mH $ &
$k_{\rm DR17} = 3.8 \times 10^{-7} T_{3}^{-0.47}$ & & 11 \\ 
DR18 & $\lidp + \me \rightarrow \li + \mD$ & 
$k_{\rm DR18} = 3.8 \times 10^{-7} T_{3}^{-0.47}$ & & 12 \\
DR19 & $\lihhp + \me \rightarrow \li + \mHt$ & 
$k_{\rm DR19} = 1.6 \times 10^{-7} T_{3}^{-0.5}$ & & 13 \\
DR20 & $\lihhp + \me \rightarrow \lih + \mH$ & 
$k_{\rm DR20} = 2.0 \times 10^{-8} T_{3}^{-0.5}$ & & 13 \\
DR21 & $\lihhp + \me \rightarrow \li + \mH + \mH$ & 
$k_{\rm DR21} = 2.0 \times 10^{-8} T_{3}^{-0.5}$ & & 13 \\
\hline
\end{tabular}
\medskip
\\
{\bf Notes}: T is the gas temperature in K, and $T_{3} = T / 300 \: {\rm K}$. \\
{\bf References}: 1 -- \citet{sdgr94};  2 -- \citet{sss95};  
3 -- \citet{wfp04};  4 -- \citet{mac04}; 
5 -- \citet{l96};  6 -- \citet{rhm04}, based on \citet{l96};  7 -- \citet{l97};
8 -- \citet{g94}; 9 -- \citet{sld98}, based on \citet{g94}; 10 -- \citet{cos99}; 
11 -- \citet{k01}; 12 -- same as corresponding H reaction;
13 -- \citet{thomas06}, C. Greene (private communication) \\
\end{minipage}
\end{table*}

\begin{table*}
\begin{minipage}{126mm}
\caption{Chemical processes: charge transfer (CT)}
\begin{tabular}{llllc}
\hline
No.\ & Reaction & Rate coefficient (${\rm cm}^{3} \: {\rm s}^{-1}$) & Notes & Ref.\  \\
\hline
CT1 &  $\mH + \Dp  \rightarrow  \mD + \Hp $ & 
$k_{\rm CT1} = 2.06 \times 10^{-10} T^{0.396}  \exp \left(-\frac{33}{T} \right)$ & & 1 \\
& & $\phantom{k_{\rm CT1} =}\mbox{} + 2.03 \times 10^{-9} T^{-0.332}$ & & \\ 
CT2  &  $\mH + \Dm  \rightarrow  \mD + \Hm $ & $k_{\rm CT2} =  6.4 \times 10^{-9} T_{3}^{0.41}$ & & 2 \\ 
CT3  &  $\mH + \mHtp  \rightarrow  \mHt + \Hp $ & $k_{\rm CT3} = 6.4 \times 10^{-10}$ & & 3 \\ 
CT4  &  $\mH + \hdp  \rightarrow  \hd + \Hp $ & $k_{\rm CT4} = 6.4 \times 10^{-10}$ & & 4 \\ 
CT5  &  $\mH + \ddp  \rightarrow  \DD + \Hp $ & $k_{\rm CT5} = 6.4 \times 10^{-10}$ & & 4 \\ 
CT6  &  $\mH + \Hep  \rightarrow  \He + \Hp + \gamma $ & $k_{\rm CT6} = 1.25 \times 10^{-15}
T_{3}^{0.25}$ & & 5 \\ 
CT7  &  $\mH + \hehep  \rightarrow  \He + \He + \Hp $ &
$k_{\rm CT7} = 1.0 \times 10^{-9}$ & & 6 \\
CT8 &  $\mH + \lihp  \rightarrow  \lih + \Hp $ & 
$k_{\rm CT8} = 1.0 \times 10^{-11} \expf{-}{67900}{T}$ & & 7 \\ 
CT9 &  $\mH + \lidp  \rightarrow  \liD + \Hp $ & 
$k_{\rm CT9} = 1.0 \times 10^{-11} \expf{-}{67900}{T}$ & & 4 \\ 
CT10  &  $\mD + \Hp \rightarrow  \mH + \Dp $ & 
$k_{\rm CT10} = 2.0 \times 10^{-10} T^{0.402}  \exp \left(-\frac{37.1}{T} \right)$ & & 1 \\
& & $\phantom{k_{\rm CT10}=} \mbox{}- 3.31 \times 10^{-17} T^{1.48}$ & &  \\
CT11  &  $\mD + \Hm  \rightarrow  \mH + \Dm $ &
$k_{\rm CT11} = 6.4 \times 10^{-9} T_{3}^{0.41}$ & & 2 \\ 
CT12  &  $\mD + \mHtp  \rightarrow  \mHt + \Dp $ & $k_{\rm CT12} = 6.4 \times 10^{-10}$ & & 4 \\ 
CT13  &  $\mD + \hdp  \rightarrow  \hd + \Dp $ & $k_{\rm CT13} = 6.4 \times 10^{-10}$  & & 4 \\ 
CT14  &  $\mD + \ddp  \rightarrow  \DD + \Dp $ & $k_{\rm CT14} = 6.4 \times 10^{-10}$  & & 4 \\ 
CT15 &  $\mD + \Hep \rightarrow  \He + \Dp + \gamma $ & 
$k_{\rm CT15} = 1.1 \times 10^{-15} T_{3}^{0.25}$ & & 8 \\
CT16  &  $\mD + \hehep  \rightarrow  \He + \He + \Dp $ &
$k_{\rm CT16} = 7.5 \times 10^{-10}$ & & 9 \\
CT17  &  $\mD + \lihp  \rightarrow  \lih + \Dp $ & 
$k_{\rm CT17} = 1.0 \times 10^{-11} \expf{-}{67900}{T}$ & & 4 \\ 
CT18 &  $\mD + \lidp  \rightarrow  \liD + \Dp $ & 
$k_{\rm CT18} = 1.0 \times 10^{-11} \expf{-}{67900}{T} $ & & 4 \\
CT19  &  $\mHt + \Hp  \rightarrow  \mH + \mHtp $ &
$k_{\rm CT19} = [- 3.3232183 \times 10^{-7}$ & & 10 \\
 & & $\phantom{k_{\rm CT19} =}  \mbox{} + 3.3735382 \times 10^{-7}  \ln{T}$  & & \\
 & & $\phantom{k_{\rm CT19} =}  \mbox{} - 1.4491368 \times 10^{-7}  (\ln{T})^2$ & & \\
 & & $\phantom{k_{\rm CT19} =}  \mbox{} + 3.4172805 \times 10^{-8}  (\ln{T})^3$ & & \\
 & & $\phantom{k_{\rm CT19} =}  \mbox{} - 4.7813720 \times 10^{-9}  (\ln{T})^4$ & & \\
 & & $\phantom{k_{\rm CT19} =}  \mbox{} + 3.9731542 \times 10^{-10} (\ln{T})^5$ & & \\
 & & $\phantom{k_{\rm CT19} =}  \mbox{}  - 1.8171411 \times 10^{-11}  (\ln{T})^6$ & & \\
 & & $\phantom{k_{\rm CT19} =}  \mbox{}  + 3.5311932 \times 10^{-13} (\ln{T})^7 ]$ & & \\
 & & $\phantom{k_{\rm CT19} =} \mbox{} \times \exp \left(\frac{-21237.15}{T} \right)$ & & \\
CT20  &  $\mHt + \Dp  \rightarrow  \mD + \mHtp $ & $k_{\rm CT20} = k_{\rm CT19}$ & & 4 \\ 
CT21  &  $\mHt + \Hep  \rightarrow  \He + \mHtp $ & $k_{\rm CT21} = 7.2 \times 10^{-15}$ & & 11 \\ 
CT22  &  $\mHt + \Hep  \rightarrow  \He + \mH + \Hp $ &
$k_{\rm CT22} = 3.7 \times 10^{-14} \expf{}{35}{T}$ & & 11 \\ 
CT23 & $\mHt + \lip \rightarrow \li + \mHtp$ & $k_{\rm CT23} = 3.0 \times 10^{-10} T_{3}^{-1.5} 
\expf{-}{116000}{T}$ & & 12 \\
CT24 & $\hd + \Hp \rightarrow \mH + \hdp $ & $k_{\rm CT24} = k_{\rm CT19}$ & & 4 \\
CT25 & $\hd + \Dp \rightarrow \mD + \hdp $ & $k_{\rm CT25} = k_{\rm CT19}$ & & 4 \\
CT26 & $\hd + \Hep \rightarrow \He + \hdp $ & $k_{\rm CT26} = 7.2 \times 10^{-15}$ & & 4 \\
CT27  &  $\hd + \Hep  \rightarrow  \He + \Hp + \mD $ &
$k_{\rm CT27} = 1.85 \times 10^{-14} \expf{}{35}{T}$ & & 13 \\
CT28  &  $\hd + \Hep  \rightarrow  \He + \mH + \Dp $ & 
$k_{\rm CT28} = 1.85 \times 10^{-14} \expf{}{35}{T}$ & & 13 \\
CT29 & $\hd + \lip \rightarrow \li + \hdp$ & $k_{\rm CT29} = k_{\rm CT23}$ & & 4 \\
CT30 & $\DD + \Hp \rightarrow \mH + \ddp$ & $k_{\rm CT30} = k_{\rm CT19}$ &  & 4 \\
CT31 & $\DD + \Dp \rightarrow \mD + \ddp$ & $k_{\rm CT31} = k_{\rm CT19}$ & & 4 \\
CT32 & $\DD + \Hep \rightarrow \He + \ddp$ & $k_{\rm CT32} = 2.5 \times 10^{-14}$ & & 14 \\
CT33  &  $\DD + \Hep  \rightarrow  \He + \Dp + \mD $ & 
$k_{\rm CT33} = 1.1 \times 10^{-13} T_{3}^{-0.24}$ & & 14 \\
CT34 & $\DD + \lip \rightarrow \li + \ddp$ & $k_{\rm CT34} = k_{\rm CT23}$ & & 4 \\
CT35 & $\He + \Hp \rightarrow \mH + \Hep$ &
$k_{\rm CT35} = 1.26 \times 10^{-9} T^{-0.75} \expf{-}{127500}{T}$ & $ T \leq 10000 \: {\rm K}$ & 15 \\
& & $\phantom{k_{\rm CT35}} = 4.0 \times 10^{-37} T^{4.74}$ & $T > 10000 \: {\rm K}$ & \\
CT36 & $\He + \Dp \rightarrow \mD + \Hep$ & $k_{\rm CT36} = k_{\rm CT35}$ & & 4 \\
\hline
\end{tabular}
\end{minipage}
\end{table*}

\begin{table*}
\begin{minipage}{126mm}
\contcaption{}
\begin{tabular}{llllc}
\hline
No.\ & Reaction & Rate coefficient (${\rm cm}^{3} \: {\rm s}^{-1}$) & Notes & Ref.\  \\
\hline
CT37 &  $\li + \Hp  \rightarrow  \mH + \lip$ &
$k_{\rm CT37} = 2.5 \times 10^{-40} T^{7.9} \expf{-}{T}{1210}$ & & 16 \\ 
CT38 &  $\li + \Hp  \rightarrow  \mH + \lip + \gamma $ &
$k_{\rm CT38} = 1.7 \times 10^{-13} T^{-0.051} \expf{-}{T}{282000}$ & & 17 \\ 
CT39  &  $\li + \Dp  \rightarrow  \mD + \lip$ &
$k_{\rm CT39} = 8.0 \times 10^{-22} T_{3}^{6.8} \expf{-}{T}{1800}$ & & 18 \\
CT40  &  $\li + \Dp  \rightarrow  \mD + \lip + \gamma $ &
$k_{\rm CT40} = 1.1 \times 10^{-13} T_{3}^{-0.051} \expf{-}{T}{282000}$ & & 19 \\
CT41 & $\li + \mHtp \rightarrow \mHt + \lip$ & $k_{\rm CT41} = 3.0 \times 10^{-10} T_{3}^{-1.5}$ & & 20 \\
CT42 & $\li + \hdp \rightarrow \hd + \lip$ & $k_{\rm CT42} = k_{\rm CT41}$ & & 4 \\
CT43 & $\li + \ddp \rightarrow \DD + \lip$ & $k_{\rm CT43} = k_{\rm CT41}$ & & 4 \\
CT44 &  $\lih + \Hp  \rightarrow  \mH + \lihp$ & $k_{\rm CT44} = 2.0 \times 10^{-15}$ & & 21 \\
CT45  &  $\lih + \Dp  \rightarrow \mD + \lihp$ & $k_{\rm CT45} = 2.0 \times 10^{-15}$ & & 21 \\ 
CT46 &  $\liD + \Hp  \rightarrow  \mH + \lidp$ & $k_{\rm CT46} = 2.0 \times 10^{-15}$ & & 21 \\
CT47  &  $\liD + \Dp  \rightarrow \mD + \lidp$ & $k_{\rm CT47} = 2.0 \times 10^{-15}$ & & 21 \\ 
\hline
\end{tabular}
\medskip
\\
{\bf Notes}: T is the gas temperature in K, and $T_{3} = T / 300 \: {\rm K}$. \\
{\bf References}: 1 -- \citet{sav02}; 2 -- \citet{dm56}, scaled by D reduced mass;
3 -- \citet{kah79}; 4 -- same as corresponding H reaction;
5 -- \citet{z89}; 6 -- Estimate by \citet{sld98}, based on \citet{sbd93};  
7 -- \citet{sld96}; 8 -- \citet{z89}, scaled by D reduced mass;
9 -- As ref.\ 6, but scaled by D reduced mass; 10 -- \citet{skhs04}; 
11 -- \citet{b84}; 12 -- Estimate, based on low-energy extrapolation of cross-section
in \citet{wutte97}; 13 -- total rate coefficient from \citet{b84}, branching ratios from
\citet{prf89}; 14 -- \citet{wfp04}; 15 -- \citet{kldd93}; 
16 -- \citet{kds94}; 17 -- \citet{sz96}; 18 -- \citet{kds94}, scaled by D reduced mass;
19 -- \citet{sz96}, scaled by D reduced mass; 
20 -- From detailed balance applied to inverse reaction; 21 -- \citet{bgmr01} \\
\end{minipage}
\end{table*}

\begin{table*}
\begin{minipage}{126mm}
\caption{Chemical processes: radiative attachment and radiative association (RA)
\label{tab:RA}}
\begin{tabular}{llllc}
\hline
No.\ & Reaction & Rate coefficient (${\rm cm}^{3} \: {\rm s}^{-1}$) & Notes & Ref.\  \\
\hline
RA1  &  $\mH + \me  \rightarrow  \Hm + \gamma $ &
$k_{\rm RA1} = {\rm dex}[-17.845 + 0.762 \log{T}$ & $T \le 6000 \: {\rm K}$ & 1 \\
& & $\phantom{k_{\rm RA1}={\rm dex}[} \mbox{}+ 0.1523 (\log{T})^{2}$ & & \\
& & $\phantom{k_{\rm RA1}= {\rm dex}[} \mbox{} - 0.03274 (\log{T})^{3}] $ & & \\
& & $ \phantom{k_{\rm RA1}} = {\rm dex}[-16.420 + 0.1998 (\log{T})^{2}$ & $T > 6000 \: {\rm K}$ & \\ 
& & $ \phantom{k_{\rm RA1} = {\rm dex}[} \mbox{}-5.447  \times 10^{-3}  (\log{T})^{4}$ & & \\ 
& & $ \phantom{k_{\rm RA1} = {\rm dex}[} \mbox{}+ 4.0415 \times 10^{-5} (\log{T})^{6}]$  & & \\
RA2  &  $\mD + \me  \rightarrow  \Dm + \gamma $ & $k_{\rm RA2} = k_{\rm RA1}$ & & 1 \\
RA3  &  $\mH + \Hp  \rightarrow  \mHtp + \gamma $ &
$k_{\rm RA3} = {\rm dex}[-19.38 - 1.523 \log{T} $ & & 2 \\
& & $\phantom{k_{\rm RA3}=} \mbox{} + 1.118 (\log{T})^{2}  - 0.1269 (\log{T})^{3}]$ & & \\ 
RA4  &  $\mH + \Dp  \rightarrow  \hdp + \gamma $ &
$k_{\rm RA4} = 3.9 \times 10^{-19} T_{3}^{1.8} \expf{}{20}{T}$ & & 3 \\
RA5 & $\mH + \mD \rightarrow \hd + \gamma$ & 
$k_{\rm RA5} = 10^{-25} \left[2.80202 - 6.63697 \ln T 
 \right.$ & $10 < T \le 200 \: {\rm K}$ & 4 \\
& & $\phantom{k_{\rm RA5}=} \mbox{} + 4.75619 (\ln T)^{2} -1.39325 (\ln T)^{3}$ & & \\
& & $\left. \phantom{k_{\rm RA5}=} \mbox{} + 0.178259 (\ln T)^{4} - 0.00817097 (\ln T)^{5} \right]$ & & \\
& & $\phantom{k_{\rm RA5}} = 10^{-25} \exp \left[507.207 - 370.889 \ln T 
\right.$ & $T > 200 \: {\rm K}$ & \\
& & $\phantom{k_{\rm RA5}=}  \mbox{} + 104.854 (\ln T)^2  - 14.4192 (\ln T)^{3} $ & & \\
& & $\left. \phantom{k_{\rm RA5}=} \mbox{} + 0.971469 (\ln T)^{4} - 0.0258076 (\ln T)^{5} \right]$ & & \\
RA6  &  $\mH + \mHtp  \rightarrow  \htp + \gamma $ & 
$k_{\rm RA6} = 1.5 \times 10^{-17} T_{3}^{1.8} \expf{}{20}{T}$ & & 5 \\ 
RA7  &  $\mH + \hdp  \rightarrow  \hhdp + \gamma $ & 
$k_{\rm RA7} = 1.2 \times 10^{-17} T_{3}^{1.8} \expf{}{20}{T}$ & & 6 \\
RA8 & $\mH + \ddp \rightarrow \hddp + \gamma$ & 
$k_{\rm RA8} = 1.1 \times 10^{-17} T_{3}^{1.8} \expf{}{20}{T}$ & & 6 \\
RA9  &  $\mH + \Hep \rightarrow  \hehp + \gamma $ &
$k_{\rm RA9} = 4.16 \times 10^{-16} T_{3}^{-0.37} \expf{-}{T}{87600}$ & & 7 \\
RA10 &  $\mH + \lip  \rightarrow  \lihp + \gamma $ & 
$k_{\rm RA10} = {\rm dex} \left[-22.4 + 0.999 \log{T} \right.$ & & 8 \\
& & $\left. \phantom{k_{\rm RA10}=} \mbox{} - 0.351 (\log{T})^{2} \right]$ & & \\
RA11  &  $\mD + \Hp  \rightarrow  \hdp + \gamma $ &
$k_{\rm RA11} = 3.9 \times 10^{-19} T_{3}^{1.8} \expf{}{20}{T}$ & & 3 \\ 
RA12 & $\mD + \Dp \rightarrow \ddp + \gamma$ & 
$k_{\rm RA12} =  1.9 \times 10^{-19} T_{3}^{1.8} \expf{}{20}{T}$ & & 3 \\
RA13  &  $\mD + \mHtp  \rightarrow  \hhdp + \gamma $ & 
$k_{\rm RA13} = 7.0 \times 10^{-18} T_{3}^{1.8} \expf{}{20}{T}$ & & 6 \\ 
RA14 & $\mD + \hdp \rightarrow \hddp + \gamma$ & 
$k_{\rm RA14} = 5.2 \times 10^{-18} T_{3}^{1.8} \expf{}{20}{T}$ & & 6 \\
RA15 & $\mD + \ddp \rightarrow \dtp + \gamma$ & 
$k_{\rm RA15} = 4.3 \times 10^{-18} T_{3}^{1.8} \expf{}{20}{T}$  & & 6 \\
RA16  &  $\mD + \Hep  \rightarrow  \hedp + \gamma $ & 
$k_{\rm RA16} = 5.0 \times 10^{-16} T_{3}^{-0.37} \expf{-}{T}{87600}$ & & 6 \\ 
RA17 &  $\mD + \lip  \rightarrow  \lidp + \gamma $ & 
$k_{\rm RA17} = 1.5 \times 10^{-22} T_{3}^{-0.9} \expf{-}{T}{7000}$ & & 9 \\
RA18  &  $\mHt + \Hp  \rightarrow  \htp + \gamma $ &
$k_{\rm RA18} = 1.0 \times 10^{-16}$ & & 10 \\
RA19 &  $\mHt + \Dp  \rightarrow  \hhdp + \gamma $ & $k_{\rm RA19} = 1.0 \times 10^{-16}$  & & 11  \\ 
RA20 & $\lip + \mHt \rightarrow \lihhp + \gamma$ & $k_{\rm RA20} = 1.0 \times 10^{-22}$ & & 12 \\
RA21 &  $\hd + \Hp  \rightarrow  \hhdp + \gamma $ & $k_{\rm RA21} = 1.0 \times 10^{-16}$  & & 11 \\ 
RA22 & $\hd + \Dp \rightarrow \hddp + \gamma$ & $k_{\rm RA22} = 1.0 \times 10^{-16}$  & & 11 \\
RA23  &  $\DD + \Hp  \rightarrow  \hddp + \gamma $ & $k_{\rm RA23} = 1.0 \times 10^{-16}$  & & 11 \\ 
RA24 & $\DD + \Dp \rightarrow \dtp + \gamma$ & $k_{\rm RA24} = 1.0 \times 10^{-16}$ & & 11 \\
RA25  &  $\He + \Hp  \rightarrow  \hehp + \gamma $ &
$k_{\rm RA25} = 8.0 \times 10^{-20} T_{3}^{-0.24} \expf{-}{T}{4000}$ & & 13 \\ 
RA26 &  $\He + \Dp  \rightarrow  \hedp + \gamma $ & 
$k_{\rm RA26} = 1.0 \times 10^{-19} T_{3}^{-0.24} \expf{-}{T}{4000}$ & & 6 \\
RA27  &  $\He + \Hep  \rightarrow  \hehep + \gamma $ &
$k_{\rm RA27} = 4.76 \times 10^{-20} T_{3}^{1.82} \expf{}{29}{T}$ & & 14 \\ 
RA28 &  $\li + \me  \rightarrow  \Limi + \gamma $ & 
$k_{\rm RA28} = 6.1 \times 10^{-17} T^{0.58} \expf{-}{T}{17200}$ & & 15 \\ 
RA29  &  $\li + \Hp  \rightarrow  \lihp + \gamma $ &
$k_{\rm RA29} = 4.8 \times 10^{-14} T^{-0.49}$ & & 16 \\
RA30  &  $\li + \Dp  \rightarrow  \lidp + \gamma $ &
$k_{\rm RA30} = 6.4 \times 10^{-14} T^{-0.49}$ & & 6 \\
RA31 &  $\li + \mH  \rightarrow  \lih + \gamma $ & 
$k_{\rm RA31} = 10^{-20} \left\{3.22 + [0.0657 (T / 1000)^{-2.45} \right.$ & & 17 \\
& & $\phantom{k_{\rm RA31} =} \left. \mbox{} + 6.3 \times 10^{-3} (T / 1000)^{0.837} ] ^{-1} \right\}$ & & \\
RA32 &  $\li + \mD  \rightarrow  \liD + \gamma $ & 
$k_{\rm RA32} = 5.5 \times 10^{-20} T_{3}^{-0.28} \expf{-}{T}{3300}$ & & 9 \\
\hline
\end{tabular}
\medskip
\\
{\bf Notes}: T is the gas temperature in K, and $T_{3} = T / 300 \: {\rm K}$. \\
{\bf References}: 1 -- \citet{wish79}; 2 -- \citet{rp76}; 3 -- \citet{rp76} and \citet{fp78}, scaled by D reduced mass;
4 -- \citet{dic05}; 5 -- \citet{dm56}; 6 -- Same as corresponding H reaction, but scaled by D reduced
mass;  7 -- \citet{ksj95}; 8 -- \citet{dks96, ggg96};  9 --  \citet{sld96}, scaled by D reduced mass;
10 -- \citet{gh92}; 11 -- estimate, based on \citet{gh92}: highly uncertain;
12 -- estimate - see also \S\ref{lihhp_chem};
13 -- \citet{jsk95};  14 -- \citet{sbd93}; 15 -- \citet{rbb94}; 16 -- \citet{dks96}; 17 -- \citet{bdlg03} \\
\end{minipage}
\end{table*}

\begin{table*}
\begin{minipage}{126mm}
\caption{Chemical processes: associative detachment, dissociative attachment 
and associative ionization (AD)
\label{tab:AD}}
\begin{tabular}{llllc}
\hline
No.\ & Reaction & Rate coefficient (${\rm cm}^{3} \: {\rm s}^{-1}$) & Notes & Ref.\  \\
\hline
AD1  &  $\mH + \Hm  \rightarrow  \mHt + \me $ &
$k_{\rm AD1} = 1.5 \times 10^{-9} T_{3}^{-0.1}$ & & 1\\
AD2  &  $\mD + \Hm  \rightarrow  \hd + \me $ & $k_{\rm AD2} = 1.5 \times 10^{-9} T_{3}^{-0.1}$ 
& & 2 \\ 
AD3  &  $\mH + \Dm  \rightarrow  \hd + \me $ & $k_{\rm AD3} = 1.5 \times 10^{-9} T_{3}^{-0.1}$  & & 2 \\ 
AD4 & $\mD + \Dm \rightarrow \DD + \me$ & $k_{\rm AD4} = 1.6 \times 10^{-9} T_{3}^{-0.1}$  & & 2 \\
AD5  &  $\mHt + \me  \rightarrow  \mH + \Hm $ &
$k_{\rm AD5} = 2.7 \times 10^{-8} T^{-1.27} \expf{-}{43000}{T}$ & & 3 \\ 
AD6 &  $\hd + \me  \rightarrow  \mH + \Dm $ & 
$k_{\rm AD6} = 1.35 \times 10^{-9} T^{-1.27} \expf{-}{43000}{T}$ & & 4 \\
AD7 &  $\hd + \me  \rightarrow  \mD + \Hm $ & 
$k_{\rm AD7} = 1.35 \times 10^{-9} T^{-1.27} \expf{-}{43000}{T}$ & & 4 \\
AD8 &  $\DD + \me  \rightarrow  \mD + \Dm $ & 
$k_{\rm AD8} = 6.7 \times 10^{-11} T^{-1.27} \expf{-}{43000}{T}$ & & 4 \\
AD9  &  $\Hp + \Hm  \rightarrow  \mHtp + \me $ & 
$k_{\rm AD9} =  6.9 \times 10^{-9}  T^{-0.35}$  & $T \le 8000 \: {\rm K}$ & 5 \\
 & & $\phantom{k_{\rm AD9}} =  9.6 \times 10^{-7} T^{-0.90}$ & $T > 8000 \: {\rm K}$ & \\
AD10 & $\Hp + \Dm \rightarrow \hdp + \me$ & 
$k_{\rm AD10} = 1.1 \times 10^{-9} T_{3}^{-0.4}$ & & 2 \\
AD11 & $\Dp + \Hm \rightarrow \hdp + \me$ & 
$k_{\rm AD11} = 1.1 \times 10^{-9} T_{3}^{-0.4}$ & & 2 \\
AD12 & $\Dp + \Dm \rightarrow \ddp + \me$ & 
$k_{\rm AD12} = 1.3 \times 10^{-9} T_{3}^{-0.4}$ & & 2 \\
AD13  &  $\mHtp + \Hm  \rightarrow  \htp + \me $ &
$k_{\rm AD13} = 2.7 \times 10^{-10} T_{3}^{-0.485} \expf{}{T}{31200}$ & & 6 \\ 
AD14  &  $\mHtp + \Dm  \rightarrow  \hhdp + \me $ &
$k_{\rm AD14} = 2.24 \times 10^{-10} T_{3}^{-0.49} \expf{}{T}{43600}$ & & 6  \\
AD15 & $\hdp + \Hm \rightarrow \hhdp + \me$ & 
$k_{\rm AD15} = 2.9 \times 10^{-10} T_{3}^{-0.485} \expf{}{T}{31200}$ & & 2 \\
AD16 & $\hdp + \Dm \rightarrow \hddp + \me$ & 
$k_{\rm AD16} =  3.7 \times 10^{-10} T_{3}^{-0.485} \expf{}{T}{31200}$ & & 2 \\
AD17 & $\ddp + \Hm \rightarrow \hddp + \me$ & 
$k_{\rm AD17} = 3.0 \times 10^{-10} T_{3}^{-0.485} \expf{}{T}{31200}$ & & 2 \\
AD18 & $\ddp + \Dm \rightarrow \dtp + \me$ & 
$k_{\rm AD18} = 3.9 \times 10^{-10} T_{3}^{-0.485} \expf{}{T}{31200}$ & & 2 \\
AD19 &  $\li + \Hm  \rightarrow  \lih + \me $ & $k_{\rm AD19} = 4.0 \times 10^{-10}$  & & 7 \\ 
AD20 & $\li + \Dm  \rightarrow  \liD + \me $ & $k_{\rm AD20} =  4.0 \times 10^{-10}$ & & 2 \\
AD21 &  $\Limi + \mH  \rightarrow  \lih + \me $ & $k_{\rm AD21} =  4.0 \times 10^{-10}$ & & 7 \\ 
AD22 & $\Limi + \mD  \rightarrow  \liD + \me $ & $k_{\rm AD22} =  4.0 \times 10^{-10}$ & & 2 \\
\hline
\end{tabular}
\medskip
\\
{\bf Notes}: T is the gas temperature in K, and $T_{3} = T / 300 \: {\rm K}$. \\
{\bf References}: 1 -- \citet{llz91}; 2 -- Same as corresponding H reaction, but scaled 
by D reduced mass; 3 -- \citet{sa67}; 4 -- \citet{xf01}; 5 -- \citet{pbcmv78};  6 -- \citet{naj98}; 
7 -- \citet{sld96} \\
\end{minipage}
\end{table*}

\begin{table*}
\begin{minipage}{126mm}
\caption{Chemical processes: collisional detachment and collisional dissociation (CD)
\label{tab:CD}}
\begin{tabular}{llllc}
\hline
No.\ & Reaction & Rate coefficient (${\rm cm}^{3} \: {\rm s}^{-1}$) & Notes & Ref.\  \\
\hline
CD1  &  $\Hm + \me  \rightarrow  \mH + \me + \me $ &
$ k_{\rm CD1}  = \exp [-1.801849334 \times 10^{1}$ & & 1 \\
& & $\phantom{k_{\rm CD1}=} \mbox{} + 2.36085220 \times 10^{0} \ln T_{\rm e}$ & &  \\
& & $\phantom{k_{\rm CD1}=} \mbox{} - 2.82744300 \times 10^{-1} (\ln T_{\rm e})^{2}$ & & \\
& & $\phantom{k_{\rm CD1}=} \mbox{}  +1.62331664\times 10^{-2} (\ln T_{\rm e})^{3}$ & & \\
& & $\phantom{k_{\rm CD1}=} \mbox{} -3.36501203 \times 10^{-2} (\ln T_{\rm e})^{4}$ & & \\
& & $\phantom{k_{\rm CD1}=} \mbox{}   +1.17832978\times 10^{-2} (\ln T_{\rm e})^{5}$ & & \\ 
& & $\phantom{k_{\rm CD1}=} \mbox{}  -1.65619470\times 10^{-3} (\ln T_{\rm e})^{6}$ & & \\ 
& & $\phantom{k_{\rm CD1}=} \mbox{}   +1.06827520\times 10^{-4} (\ln T_{\rm e})^{7}$ & & \\
& & $\phantom{k_{\rm CD1}=} \mbox{}  -2.63128581\times 10^{-6} (\ln T_{\rm e})^{8} ]$ & & \\
CD2  &  $\Hm + \mH  \rightarrow  \mH + \mH + \me $ &
$k_{\rm CD2} = 2.5634 \times 10^{-9} T_{\rm e}^{1.78186}$ & $T_{\rm e} \leq 0.1 \: \rm{eV}$ & 1 \\
& & $\phantom{k_{\rm CD2}} = \exp[-2.0372609 \times 10^{1}$ & $T_{\rm e} > 0.1 \: \rm{eV}$  & \\
& & $\phantom{k_{\rm CD2} =} \mbox{}+1.13944933 \times 10^{0} \ln T_{\rm e}$ & & \\
& & $\phantom{k_{\rm CD2} =} \mbox{}-1.4210135 \times 10^{-1} (\ln T_{\rm e})^{2}$ & & \\
& & $\phantom{k_{\rm CD2} =} \mbox{}+8.4644554 \times 10^{-3} (\ln T_{\rm e})^{3}$ & & \\
& & $\phantom{k_{\rm CD2} =} \mbox{}-1.4327641 \times 10^{-3} (\ln T_{\rm e})^{4}$  & & \\
& & $\phantom{k_{\rm CD2} =} \mbox{}+2.0122503 \times 10^{-4} (\ln T_{\rm e})^{5}$ & & \\
& & $\phantom{k_{\rm CD2} =} \mbox{}+8.6639632 \times 10^{-5} (\ln T_{\rm e})^{6}$ & & \\
& & $\phantom{k_{\rm CD2} =} \mbox{}-2.5850097 \times 10^{-5} (\ln T_{\rm e})^{7}$ & & \\
& & $\phantom{k_{\rm CD2} =} \mbox{}+ 2.4555012\times 10^{-6} (\ln T_{\rm e})^{8}$ & & \\
& & $\phantom{k_{\rm CD2} =} \mbox{} -8.0683825\times 10^{-8} (\ln T_{\rm e})^{9}]$ & & \\
CD3 & $\Hm + \mD \rightarrow \mH + \mD + \me$ & $k_{\rm CD3} = k_{\rm CD2}$ & & 2 \\
CD4 &  $\Hm + \He  \rightarrow  \mH + \He + \me $ &
$k_{\rm CD4} = 4.1 \times 10^{-17} T^{2} \expf{-}{19870}{T}$ & & 3 \\ 
CD5 &  $\Dm + \me  \rightarrow  \mD + \me + \me $ & $k_{\rm CD5} = k_{\rm CD1}$ & & 2 \\
CD6  &  $\Dm + \mH  \rightarrow  \mD + \mH + \me $ & $k_{\rm CD6} = k_{\rm CD2}$ & & 2 \\
CD7  &  $\Dm + \mD  \rightarrow  \mD + \mD + \me $ & $k_{\rm CD7} = k_{\rm CD2}$ & & 2 \\
CD8  & $\Dm + \He  \rightarrow  \mD + \He + \me $ & 
$k_{\rm CD8} = 1.5 \times 10^{-17} T^{2} \expf{-}{19870}{T}$ & & 4 \\
CD9 & $\mHt + \mH  \rightarrow  \mH + \mH + \mH$  & 
$k_{\rm CD9} = 6.67 \times 10^{-12} T^{0.5} \exp \left[-(1+ \frac{63593}{T}) \right]$ & v=0 & 5 \\
& & $\phantom{k_{\rm CD9}} = k_{\rm TB1} / K$ & LTE & 6 \\
CD10 & $\mHt + \mHt \rightarrow \mH + \mH + \mHt$ & 
$k_{\rm CD10} = \frac{5.996 \times 10^{-30} T^{4.1881}}{(1.0 + 6.761 \times 10^{-6} T)^{5.6881}}
 \exp \left(-\frac{54657.4}{T} \right)$ & $v=0$ & 7 \\
& & $\phantom{\rm CD10} = k_{\rm TB2} / K$ & LTE & 6 \\
CD11 & $\mHt + \He \rightarrow \mH + \mH + \He$ & 
$k_{\rm CD11} = {\rm dex} \left[ -27.029 + 3.801 \log{T} - \frac{29487}{T} \right]$ & $v=0$ & 8 \\
& & $\phantom{k_{\rm CD11}} =  6.6 \times 10^{-10} T^{0.115} \expf{-}{52000}{T}$ & LTE & 9 \\
CD12 &  $\mHt + \me  \rightarrow  \mH + \mH + \me $ &
$k_{\rm CD12} = 4.49 \times 10^{-9} T^{0.11} \expf{-}{101858}{T}$ & $v=0$ & 10 \\
& & $\phantom{k_{\rm CD12}} = 1.91 \times 10^{-9} T^{0.136} \expf{-}{53407.1}{T}$ & LTE & 10 \\
CD13 & $\hd + \mH \rightarrow \mH + \mD + \mH$ &  $k_{\rm CD13} = k_{\rm CD9}$ 
& See \S\ref{cdhd} & 2 \\
CD14 & $\hd + \mHt \rightarrow \mH + \mD + \mHt$ & $k_{\rm CD14} = k_{\rm CD10}$ 
& See \S\ref{cdhd} & 2 \\
CD15 & $\hd + \He \rightarrow \mH + \mD + \He$ & $k_{\rm CD15} = k_{\rm CD11}$
& See \S\ref{cdhd} & 2 \\
CD16 & $\hd + \me \rightarrow \mH + \mD + \me$ & 
$k_{\rm CD16} = 5.09 \times 10^{-9} T^{0.128}  \expf{-}{103258}{T}$ & $v=0$ & 11 \\
& & $\phantom{k_{\rm CD16}} = 1.04 \times 10^{-9} T^{0.218}  \expf{-}{53070.7}{T}$  & LTE & 11 \\
CD17 & $\DD + \mH \rightarrow \mD + \mD + \mH$ & $k_{\rm CD17} = k_{\rm CD9}$ & & 2 \\
CD18 & $\DD + \mHt \rightarrow \mD + \mD + \mHt$ & $k_{\rm CD18} = k_{\rm CD10}$ & & 2 \\
CD19 & $\DD + \He \rightarrow \mD + \mD + \He$ & $k_{\rm CD19} = k_{\rm CD11}$ & & 2 \\
CD20 & $\DD + \me \rightarrow \mD + \mD + \me$ & 
$k_{\rm CD20} =  8.24 \times 10^{-9} T^{0.126} \expf{-}{105388}{T}$ & $v=0$ & 10 \\
& & $\phantom{k_{\rm CD20}} = 2.75 \times 10^{-9} T^{0.163} \expf{-}{53339.7}{T}$ & LTE & 10 \\
CD21 & $\lihp + \mD \rightarrow \lip + \mH + \mD$ & 
$k_{\rm CD21} = 1.0 \times 10^{-9} \expf{-}{1400}{T}$ & & 12 \\
CD22 & $\lihp + \mD \rightarrow \li + \Hp + \mD$ & 
$k_{\rm CD22} = 1.0 \times 10^{-9} \expf{-}{97500}{T}$ & & 12 \\
CD23 & $\lihp + \mD \rightarrow \li + \mH + \Dp$ & 
$k_{\rm CD23} = 1.0 \times 10^{-9} \expf{-}{97500}{T}$ & & 12 \\
CD24 & $\lidp + \mD \rightarrow \lip + \mD + \mD$ & 
$k_{\rm CD24} = 1.0 \times 10^{-9} \expf{-}{1400}{T}$ & & 12 \\
CD25 & $\lidp + \mD \rightarrow \li + \Dp + \mD$ & 
$k_{\rm CD25} = 1.0 \times 10^{-9} \expf{-}{97500}{T}$ & & 12 \\
CD26 & $\lihhp + \mHt \rightarrow \lip + \mHt + \mHt$ & $k_{\rm CD26} = 1.0 \times 10^{-9}
\expf{-}{3000}{T}$ & & 13 \\
\hline
\end{tabular}
\medskip
\\
{\bf Notes}: T is the gas temperature in K and $T_{\rm e}$ is the gas temperature in eV. 
$K$ is the equilibrium constant relating reactions TB1 and CD9, and reactions 
TB2 and CD10; its value is given in \S\ref{h2cd}. \\
{\bf References}: 1 -- \citet{j87}; 2 -- Assumed same as corresponding H reaction;
3 -- \citet{h82};  4 -- Same as corresponding H reaction, but scaled by D reduced mass;
5 -- \citet{ms86}; 6 -- determined from three-body rate coefficient by detailed balance (see \S\ref{tbhf});
7 -- \citet{mkm98};  8 -- \citet{drcm87}; 9 -- determined from the \citet{wk75} rate coefficient for
reaction TB3 by detailed balance; 10 -- \citet{tt02a};  11 -- \citet{tt02b}; 13 -- estimate - see
also \S\ref{lihhp_chem} \\
\end{minipage}
\end{table*}

\begin{table*}
\begin{minipage}{126mm}
\caption{Chemical processes: mutual neutralization (MN)
\label{tab:MN}}
\begin{tabular}{lllc}
\hline
No.\ & Reaction & Rate coefficient (${\rm cm}^{3} \: {\rm s}^{-1}$) & Ref.\  \\
\hline
MN1  &  $\Hp + \Hm  \rightarrow  \mH + \mH $ & 
$k_{\rm MN1} = 2.4 \times 10^{-6} T^{-1/2} \left(1.0 + 5.0 \times 10^{-5} T \right)$ & 1 \\
MN2  &  $\Dp + \Hm  \rightarrow  \mD + \mH $ & $k_{\rm MN2} = 1.1 \times k_{\rm MN1}$ & 2 \\
MN3  &  $\Hp + \Dm  \rightarrow  \mD + \mH $ & $k_{\rm MN3} = 1.1 \times k_{\rm MN1}$ & 2 \\
MN4  &  $\Dp + \Dm  \rightarrow  \mD + \mD $ & $k_{\rm MN4} = 1.3 \times k_{\rm MN1}$ & 2 \\
MN5  &  $\mHtp + \Hm  \rightarrow  \mHt + \mH $ & $k_{\rm MN5} = 1.4 \times 10^{-7} T_{3}^{-0.5}$
& 3 \\ 
MN6  &  $\mHtp + \Hm  \rightarrow  \mH + \mH + \mH $ & 
$k_{\rm MN6} = 1.4 \times 10^{-7} T_{3}^{-0.5}$ & 3 \\ 
MN7 & $\mHtp + \Dm \rightarrow \mHt  + \mD$ & 
$k_{\rm MN7} = 1.7 \times 10^{-7} T_{3}^{-0.5}$ & 2 \\
MN8 & $\mHtp + \Dm \rightarrow \mH + \mH  + \mD$ & 
$k_{\rm MN8} = 1.7 \times 10^{-7} T_{3}^{-0.5}$ & 2 \\
MN9 & $\hdp + \Hm \rightarrow \hd  + \mH$ & 
$k_{\rm MN9} = 1.5 \times 10^{-7} T_{3}^{-0.5}$ & 2 \\
MN10 & $\hdp + \Hm \rightarrow \mD + \mH + \mH$ & 
$k_{\rm MN10} = 1.5 \times 10^{-7} T_{3}^{-0.5}$ & 2 \\
MN11 & $\hdp + \Dm \rightarrow \hd  + \mD$ & 
$k_{\rm MN11} = 1.9 \times 10^{-7} T_{3}^{-0.5}$ & 2 \\
MN12 & $\hdp + \Dm \rightarrow \mD + \mH + \mD$ & 
$k_{\rm MN12} = 1.9 \times 10^{-7} T_{3}^{-0.5}$ & 2 \\
MN13 & $\ddp + \Hm \rightarrow \DD + \mH$ & 
$k_{\rm MN13} = 1.5 \times 10^{-7} T_{3}^{-0.5}$ & 2 \\
MN14 & $\ddp + \Hm \rightarrow \mD + \mD + \mH$ & 
$k_{\rm MN14} = 1.5 \times 10^{-7} T_{3}^{-0.5}$  & 2 \\
MN15 & $\ddp + \Dm \rightarrow \DD + \mD$ & 
$k_{\rm MN15} = 2.0 \times 10^{-7} T_{3}^{-0.5}$ & 2 \\
MN16 & $\ddp + \Dm \rightarrow \mD + \mD + \mD$ & 
$k_{\rm MN16} = 2.0 \times 10^{-7} T_{3}^{-0.5}$ & 2 \\
MN17  &  $\htp + \Hm  \rightarrow  \mHt + \mH + \mH $ &
$k_{\rm MN17} = 2.3 \times 10^{-7} T_{3}^{-0.5}$ & 4 \\
MN18  &  $\htp + \Hm  \rightarrow  \mHt + \mHt $ &  
$k_{\rm MN18} = 2.3 \times 10^{-7} T_{3}^{-0.5}$ & 5 \\ 
MN19 & $\htp + \Dm \rightarrow \mHt + \mH + \mD$ &
$k_{\rm MN19} = 2.9 \times 10^{-7} T_{3}^{-0.5}$ & 6 \\
MN20 & $\htp + \Dm \rightarrow \mHt + \hd$ &
$k_{\rm MN20} = 2.9 \times 10^{-7} T_{3}^{-0.5}$ & 6 \\
MN21 & $\hhdp + \Hm \rightarrow \mHt + \mH + \mD$ &
$k_{\rm MN21} = 1.6 \times 10^{-7} T_{3}^{-0.5}$ & 6 \\
MN22 & $\hhdp + \Hm \rightarrow \mHt + \hd$ &
$k_{\rm MN22} = 1.6 \times 10^{-7} T_{3}^{-0.5}$ & 6 \\
MN23 & $\hhdp + \Hm \rightarrow \hd + \mH + \mH$ &
$k_{\rm MN23} = 1.6 \times 10^{-7} T_{3}^{-0.5}$ & 6 \\
MN24 & $\hhdp + \Dm \rightarrow \mHt + \mD + \mD$ &
$k_{\rm MN24} = 1.5 \times 10^{-7} T_{3}^{-0.5}$ & 6 \\
MN25 & $\hhdp + \Dm \rightarrow \mHt + \DD$ &
$k_{\rm MN25} = 1.5 \times 10^{-7} T_{3}^{-0.5}$ & 6 \\
MN26 & $\hhdp + \Dm \rightarrow \hd + \mH + \mD$ &
$k_{\rm MN26} = 1.5 \times 10^{-7} T_{3}^{-0.5}$ & 6 \\
MN27 & $\hhdp + \Dm \rightarrow \hd + \hd$ &
$k_{\rm MN27} = 1.5 \times 10^{-7} T_{3}^{-0.5}$ & 6 \\
MN28 & $\hddp + \Hm \rightarrow \mHt + \DD$ &
$k_{\rm MN28} = 1.2 \times 10^{-7} T_{3}^{-0.5}$ & 6 \\
MN29 & $\hddp + \Hm \rightarrow \hd + \mH + \mD$ &
$k_{\rm MN29} = 1.2 \times 10^{-7} T_{3}^{-0.5}$ & 6 \\
MN30 & $\hddp + \Hm \rightarrow \hd + \hd$ & 
$k_{\rm MN30} = 1.2 \times 10^{-7} T_{3}^{-0.5}$ & 6 \\
MN31 & $\hddp + \Hm \rightarrow \DD + \mH + \mH$ &
$k_{\rm MN31} = 1.2 \times 10^{-7} T_{3}^{-0.5}$ & 6 \\
MN32 & $\hddp + \Dm \rightarrow \hd + \mD + \mD$ &
$k_{\rm MN32} = 2.1 \times 10^{-7} T_{3}^{-0.5}$ & 6 \\
MN33 & $\hddp + \Dm \rightarrow \hd + \DD$ &
$k_{\rm MN33} = 2.1 \times 10^{-7} T_{3}^{-0.5}$ & 6 \\
MN34 & $\hddp + \Dm \rightarrow \DD + \mH + \mD$ & 
$k_{\rm MN34} = 2.1 \times 10^{-7} T_{3}^{-0.5}$ & 6 \\
MN35 & $\dtp + \Hm \rightarrow \hd + \DD$ & 
$k_{\rm MN35} = 2.4 \times 10^{-7} T_{3}^{-0.5}$ & 6 \\
MN36 & $\dtp + \Hm \rightarrow \DD + \mH + \mD$ &
$k_{\rm MN36} = 2.4 \times 10^{-7} T_{3}^{-0.5}$ & 6 \\
MN37 & $\dtp + \Dm \rightarrow \DD + \mD + \mD$ &
$k_{\rm MN37} = 3.3 \times 10^{-7} T_{3}^{-0.5}$ & 2 \\
MN38 & $\dtp + \Dm \rightarrow \DD + \DD$ &
$k_{\rm MN38} = 3.3 \times 10^{-7} T_{3}^{-0.5}$ & 2 \\
MN39  &  $\Hep + \Hm  \rightarrow  \He + \mH $ &
$k_{\rm MN39} = 2.32 \times 10^{-7} T_{3}^{-0.52} \exp \left(\frac{T}{22400}\right)$ & 7 \\ 
MN40 &  $\Hep + \Dm  \rightarrow  \He + \mD $ & 
$k_{\rm MN40} =  3.03 \times 10^{-7} T_{3}^{-0.52} \exp \left(\frac{T}{22400}\right)$ & 2 \\ 
MN41 &  $\lip + \Hm  \rightarrow  \li + \mH $ &
$k_{\rm MN41} = 2.93 \times 10^{-7} T_{3}^{-0.477} \expf{}{T}{23200}$ & 1 \\ 
MN42  &  $\lip + \Dm  \rightarrow  \li + \mD $ &
$k_{\rm MN42} = 2.06 \times 10^{-7} T_{3}^{-0.5} \expf{}{T}{18300}$  & 7 \\ 
MN43 &  $\Limi + \Hp  \rightarrow  \li + \mH $ &
$k_{\rm MN43} = 1.8 \times 10^{-7} T_{3}^{-0.477} \expf{}{T}{23200}$ & 1 \\
MN44  &  $\Limi + \Dp  \rightarrow  \li + \mD $ & 
$k_{\rm MN44} = 2.06 \times 10^{-7} T_{3}^{-0.5} \expf{}{T}{18300}$  & 7 \\ 
\hline
\end{tabular}
\medskip
\\
{\bf Notes}: T is the gas temperature in K, and $T_{3} = T / 300 \: {\rm K}$. Some
of the mutual neutralization reactions listed here also include dissociation or
transfer in the process. \\
{\bf References}: 1 -- \citet{cdg99}; 2 -- Same as corresponding H reaction, but scaled by D reduced mass;
3 -- \citet{dl87}; 4 -- \citet{dm56}; 5 -- \citet{lmm00}; 6 -- As 2, with the additional assumption 
of equally probable outcomes; 7 -- \citet{ph94} \\
\end{minipage}
\end{table*}

\begin{table*}
\begin{minipage}{126mm}
\caption{Chemical processes: three-body association (TB)
\label{tab:TB}}
\begin{tabular}{llllc}
\hline
No.\ & Reaction & Rate coefficient (${\rm cm}^{6} \: {\rm s}^{-1}$) & Ref.\  \\
\hline
TB1  &  $\mH + \mH + \mH  \rightarrow  \mHt + \mH $ & See \S\ref{tbhf} & --- \\ 
TB2  &  $\mH + \mH + \mHt  \rightarrow  \mHt + \mHt $ & See \S\ref{tbhf} & --- \\ 
TB3  &  $\mH + \mH + \He  \rightarrow  \mHt + \He $ & 
$k_{\rm TB3} = 6.9 \times 10^{-32} T^{-0.4}$ & 1 \\ 
TB4 & $\mH + \mD + \mH \rightarrow \hd + \mH$ & See \S\ref{tbhf} & --- \\
TB5 & $\mH + \mD + \mHt \rightarrow \hd + \mHt$ & See \S\ref{tbhf}  & --- \\
TB6 & $\mH + \mD + \He \rightarrow \hd + \He$ & 
$k_{\rm TB6} = 6.9 \times 10^{-32} T^{-0.4}$ & 2 \\
TB7 & $\mD + \mD + \mH \rightarrow \DD + \mH$ & See \S\ref{tbhf}  & --- \\
TB8 & $\mD + \mD + \mHt \rightarrow \DD + \mHt$ & See \S\ref{tbhf} & --- \\
TB9 & $\mD + \mD + \He \rightarrow \DD + \He$ & 
$k_{\rm TB9} = 6.9 \times 10^{-32} T^{-0.4}$ & 2 \\
TB10 &  $\Hp + \mH + \mH  \rightarrow  \mHtp + \mH $ & 
$k_{\rm TB10} = 1.203 \times 10^{-29} T^{-1.041}$ & 3 \\ 
TB11 & $\Dp + \mH + \mH  \rightarrow  \hdp + \mH $ & 
$k_{\rm TB11} = 1.203 \times 10^{-29} T^{-1.041}$ & 2 \\
TB12 & $\Hp + \mD + \mH  \rightarrow  \hdp + \mH $ & 
$k_{\rm TB12} = 1.203 \times 10^{-29} T^{-1.041}$ & 2 \\
TB13 & $\Dp + \mD + \mH  \rightarrow  \ddp + \mH $ & 
$k_{\rm TB13} = 1.203 \times 10^{-29} T^{-1.041}$ & 2 \\
TB14 & $\Hp + \mHt + \mH  \rightarrow  \htp + \mH $ & $k_{\rm TB14} = 1.0 \times 10^{-28}$ & 4 \\
TB15 &  $\Hp + \mHt + \mHt  \rightarrow  \htp + \mHt $ &
$k_{\rm TB15} = 5.4 \times 10^{-29}$ & 5 \\ 
TB16  &  $\Hp + \mHt + \He  \rightarrow  \htp + \He $ & $k_{\rm TB16} = 1.07 \times 10^{-28}$ & 5  \\ 
TB17 & $\Dp + \mHt + \mH  \rightarrow  \hhdp + \mH $ & $k_{\rm TB17} = 1.0 \times 10^{-28}$ & 4 \\
TB18  &  $\Dp + \mHt + \mHt  \rightarrow  \hhdp + \mHt $ & $k_{\rm TB18} = 5.4 \times 10^{-29}$ & 2 \\
TB19  &  $\Dp + \mHt + \He  \rightarrow  \hhdp + \He $ & $k_{\rm TB19} = 1.07 \times 10^{-28}$ & 2 \\
TB20 & $\Hp + \hd + \mH  \rightarrow  \hhdp + \mH $ & $k_{\rm TB20} = 1.0 \times 10^{-28}$ & 4 \\
TB21  &  $\Hp + \hd + \mHt  \rightarrow  \hhdp + \mHt $ & $k_{\rm TB21} = 5.4 \times 10^{-29}$ & 2 \\
TB22  &  $\Hp + \hd + \He  \rightarrow  \hhdp + \He $ & $k_{\rm TB22} = 1.07 \times 10^{-28}$ & 2 \\
TB23 & $\Dp + \hd + \mH  \rightarrow  \hddp + \mH $ & $k_{\rm TB23} = 1.0 \times 10^{-28}$ & 4 \\
TB24  &  $\Dp + \hd + \mHt  \rightarrow  \hddp + \mHt $ & $k_{\rm TB24} = 5.4 \times 10^{-29}$ & 2 \\
TB25  &  $\Dp + \hd + \He  \rightarrow  \hddp + \He $ & $k_{\rm TB25} = 1.07 \times 10^{-28}$ & 2 \\
TB26 & $\Hp + \DD + \mH  \rightarrow  \hddp + \mH $ & $k_{\rm TB26} = 1.0 \times 10^{-28}$ & 4 \\
TB27  &  $\Hp + \DD + \mHt  \rightarrow  \hddp + \mHt $ & $k_{\rm TB27} = 5.4 \times 10^{-29}$ & 2 \\
TB28  &  $\Hp + \DD + \He  \rightarrow  \hddp + \He $ & $k_{\rm TB28} = 1.07 \times 10^{-28}$ & 2 \\
TB29 & $\Dp + \DD + \mH  \rightarrow  \dtp + \mH $ & $k_{\rm TB29} = 1.0 \times 10^{-28}$ & 4 \\
TB30  &  $\Dp + \DD + \mHt  \rightarrow  \dtp + \mHt $ & $k_{\rm TB30} = 5.4 \times 10^{-29}$ & 2 \\
TB31  &  $\Dp + \DD + \He  \rightarrow  \dtp + \He $ & $k_{\rm TB31} = 1.07 \times 10^{-28}$ & 2 \\
TB32  &  $\li + \mH + \mH  \rightarrow  \lih + \mH $ & 
$k_{\rm TB32} = 2.5 \times 10^{-29} T^{-1}$ &  6 \\ 
TB33  &  $\li + \mH + \mHt  \rightarrow  \lih + \mHt $ & 
$k_{\rm TB33} = 4.1 \times 10^{-30} T^{-1}$ & 6 \\ 
TB34 &  $\li + \mD + \mH  \rightarrow  {\rm LiD} + \mH $ &
$k_{\rm TB34} = 2.5 \times 10^{-29} T^{-1}$ & 2 \\
TB35  & $\li + \mD + \mHt  \rightarrow  {\rm LiD} + \mHt $ & 
$k_{\rm TB35} = 4.1 \times 10^{-30} T^{-1}$ & 2 \\
\hline
\end{tabular}
\medskip
\\
{\bf Notes}: T is the gas temperature in K. \\
{\bf References}:  1 -- \citet{wk75}; 
2 -- Same as corresponding H reaction; 
3 -- \citet{kjs03}; 
4 -- Estimate;
5 -- \citet{gh92};  
6 -- \citet{mon05} \\
\end{minipage}
\end{table*}

\begin{table*}
\begin{minipage}{126mm}
\caption{Chemical processes: isotopic exchange (IX)
\label{tab:IX}}
\begin{tabular}{llllc}
\hline
No.\ & Reaction & Rate coefficient (${\rm cm}^{3} \: {\rm s}^{-1}$) & Notes & Ref.\  \\
\hline
IX1  &  $\mHtp + \mD  \rightarrow  \hdp + \mH $ &
$k_{\rm IX1} = 1.07 \times 10^{-9} T_{3}^{0.062} \expf{-}{T}{41400}$ & & 1 \\ 
IX2 & $\mHtp + \mD \rightarrow \hd + \Hp$ & $k_{\rm IX2} = 1.0 \times 10^{-9}$ & & 2 \\
IX3 & $\hdp + \mH  \rightarrow  \mHtp + \mD $ & 
$k_{\rm IX3} = 1.0 \times 10^{-9} \expf{-}{154}{T}$ & & 3 \\ 
IX4 & $\hdp + \mH \rightarrow \mHt + \Dp$ & $k_{\rm IX4} = 1.0 \times 10^{-9}$ & & 2 \\
IX5 & $\hdp + \mD \rightarrow \ddp + \mH$ & $k_{\rm IX5} = 1.0 \times 10^{-9}$ & &  4 \\
IX6 & $\hdp + \mD \rightarrow \DD + \Hp$ & $k_{\rm IX6} = 1.0 \times 10^{-9}$ & & 2 \\
IX7 & $\ddp + \mH \rightarrow \hdp + \mD$ & $k_{\rm IX7} = 1.0 \times 10^{-9} \expf{-}{472}{T}$ & &  4 \\
IX8 & $\ddp + \mH \rightarrow \hd + \Dp$ & $k_{\rm IX8} = 1.0 \times 10^{-9}$  & & 2 \\
IX9  &  $\mHt + \Dp  \rightarrow  \hd + \Hp $ &
$k_{\rm IX9} = 4.17 \times 10^{-10} + 8.46 \times 10^{-10} \log{T}$ & & 5 \\
& & $\phantom{k_{\rm IX9}=} \mbox{} - 1.37 \times 10^{-10} (\log{T})^{2}$ & & \\
IX10 & $\mHt + \Dp \rightarrow \hdp + \mH$ & 
$k_{\rm IX10} = \left[ 1.04 \times 10^{-9} + 9.52 \times 10^{-9} \left(\frac{T}{10000}\right) \right.$ & & 6 \\
& & $\phantom{k_{\rm IX10}=} \left. \mbox{} -1.81 \times 10^{-9} \left(\frac{T}{10000}\right)^{2} 
\right]\expf{-}{21000}{T}$ & &  \\
IX11  &  $\hd + \Hp  \rightarrow  \mHt + \Dp $ & 
$k_{\rm IX11} = 1.1 \times 10^{-9} \expf{-}{488}{T}$ & & 5 \\ 
IX12 & $\hd + \Hp  \rightarrow \mHtp + \mD$ &
$k_{\rm IX12} = 1.0 \times 10^{-9} \expf{-}{21600}{T}$ & & 2 \\
IX13 & $\hd + \Dp \rightarrow \DD + \Hp$ & $k_{\rm IX13} = 1.0 \times 10^{-9}$ & &  4 \\
IX14 & $\hd + \Dp \rightarrow \ddp + \mH$ & 
$k_{\rm IX14} = \left[ 3.54 \times 10^{-9} + 7.50 \times 10^{-10} \left(\frac{T}{10000}\right) \right.$ & & 6 \\
& & $\phantom{k_{\rm IX14}=} \left. \mbox{} -2.92 \times 10^{-10} \left(\frac{T}{10000}\right)^{2} 
\right]\expf{-}{21100}{T}$ & & \\
IX15 & $\DD + \Hp \rightarrow \hd + \Dp$ & $k_{\rm IX15} = 2.1 \times 10^{-9} \expf{-}{491}{T}$ & &  4 \\
IX16 & $\DD + \Hp \rightarrow \hdp + \mD$ & 
$k_{\rm IX16} =  \left[ 5.18 \times 10^{-11} + 3.05 \times 10^{-9} \left(\frac{T}{10000}\right) \right.$ & 
& 6 \\
& & $\phantom{k_{\rm IX16}=} \left. \mbox{} -5.42 \times 10^{-10} \left(\frac{T}{10000}\right)^{2} 
\right]\expf{-}{20100}{T}$ & & \\
IX17  &  $\mHt + \mD  \rightarrow  \hd + \mH $ &
$k_{\rm IX17} = {\rm dex}\left[-56.4737 + 5.88886\log{T} \right.$ & $T \le 2000 \: {\rm K}$ & 7 \\
& & $\phantom{k_{\rm IX17} = {\rm dex}[} \mbox{} + 7.19692 (\log{T})^{2}$ & & \\
& & $\phantom{k_{\rm IX17} = {\rm dex}[} \mbox{} + 2.25069 (\log{T})^{3}$ & & \\
& & $\phantom{k_{\rm IX17} = {\rm dex}[} \mbox{} - 2.16903  (\log{T})^{4}$ & & \\
& & $\left. \phantom{k_{\rm IX17} = {\rm dex}[} \mbox{} + 0.317887 (\log{T})^{5} \right]$ & & \\
& & $\phantom{k_{\rm IX17}} = 3.17 \times 10^{-10} \expf{-}{5207}{T}$ & $T > 2000 \: {\rm K}$ & \\
IX18  &  $\hd + \mH  \rightarrow  \mHt + \mD $ &
$k_{\rm IX18} = 5.25 \times 10^{-11} \expf{-}{4430}{T}$ & $T \le 200 \: {\rm K}$ & 8 \\
& & $\phantom{k_{\rm IX18}} = 5.25 \times 10^{-11} \exp \left(-\frac{4430}{T} +
 \frac{173900}{T^{2}}\right)$ & $T > 200 \: {\rm K}$ & \\
IX19 & $\hd + \mD \rightarrow \DD + \mH$ & 
$k_{\rm IX19} = 1.15 \times 10^{-11} \expf{-}{3220}{T}$ & & 8 \\
IX20 & $\DD + \mH \rightarrow \hd + \mD$ & 
$k_{\rm IX20} = {\rm dex}\left[-86.1558 + 4.53978 \log{T} \right.$ & $T \le 2200 \: {\rm K}$ & 7 \\
& & $\phantom{k_{\rm IX20} = {\rm dex}[} \mbox{} + 33.5707 (\log{T})^{2}$ & & \\
& & $\phantom{k_{\rm IX20} = {\rm dex}[} \mbox{} - 13.0449 (\log{T})^{3}$ & & \\
& & $\phantom{k_{\rm IX20} = {\rm dex}[} \mbox{} + 1.22017  (\log{T})^{4}$ & & \\
& & $\left. \phantom{k_{\rm IX20} = {\rm dex}[} \mbox{} + 0.0482453 (\log{T})^{5} \right]$ & & \\
& & $\phantom{k_{\rm IX20}} = 2.67 \times 10^{-10} \expf{-}{5945}{T}$ & $T > 2200 \: {\rm K}$ & \\
IX21  &  $\htp + \mD  \rightarrow  \hhdp + \mH $ & $k_{\rm IX21} = 1.0 \times 10^{-9}$ & & 9 \\ 
IX22  &  $\hhdp + \mH  \rightarrow  \htp + \mD $ & 
$k_{\rm IX22} = 1.0 \times 10^{-9} \expf{-}{632}{T}$ & & 10 \\ 
IX23 & $\hhdp + \mD \rightarrow \hddp + \mH$ & $k_{\rm IX23} = 1.0 \times 10^{-9}$ & &  4 \\
IX24 & $\hddp + \mH \rightarrow \hhdp + \mD$ &  
$k_{\rm IX24} = 1.0 \times 10^{-9} \expf{-}{600}{T} $ & &  4 \\
IX25 & $\hddp + \mD \rightarrow \dtp + \mH$ &  $k_{\rm IX25} = 1.0 \times 10^{-9}$ & &  4 \\
IX26 & $\dtp + \mH \rightarrow \hddp + \mD$ &  
$k_{\rm IX26} = 1.0 \times 10^{-9} \expf{-}{655}{T}$ & &  4 \\
IX27  &  $\htp + \hd  \rightarrow  \hhdp + \mHt $ & $k_{\rm IX27} = 3.5 \times 10^{-10}$ & & 4 \\ 
IX28 & $\htp + \DD \rightarrow \hhdp + \hd$ & 
$k_{\rm IX28} = 3.5 \times 10^{-11} T_{3}^{-0.19}$ & & 11 \\
IX29 & $\htp + \DD \rightarrow \hddp + \mHt$ & 
$k_{\rm IX29} = 9.64 \times 10^{-10} T_{3}^{-0.024}$ & & 11 \\
IX30  &  $\hhdp + \mHt  \rightarrow  \htp + \hd $ &
$k_{\rm IX30} = 1.4 \times 10^{-10} \expf{-}{232}{T}$ & & 4 \\ 
IX31 & $\hhdp + \hd \rightarrow \htp + \DD$ & 
$k_{\rm IX31} = 1.75 \times 10^{-11} T_{3}^{-0.19} \expf{-}{153}{T}$ & & 12 \\
IX32 & $\hhdp + \hd \rightarrow \hddp + \mHt$ & $k_{\rm IX32} = 2.6 \times 10^{-10}$ & & 4 \\
IX33 & $\hhdp + \DD \rightarrow \hddp + \hd$ & $k_{\rm IX33} = 8.5 \times 10^{-10}$ & & 4 \\
IX34 & $\hhdp + \DD \rightarrow \dtp + \mHt$ & $k_{\rm IX34} = 8.5 \times 10^{-10}$  & & 4 \\
IX35 & $\hddp + \mHt \rightarrow \htp + \DD$ & 
$k_{\rm IX35} = 2.0 \times 10^{-10} \expf{-}{340.2}{T}$ & & 13 \\
\hline
\end{tabular}
\end{minipage}
\end{table*}

\begin{table*}
\begin{minipage}{126mm}
\contcaption{}
\begin{tabular}{llllc}
\hline
No.\ & Reaction & Rate coefficient (${\rm cm}^{3} \: {\rm s}^{-1}$) & Notes &  Ref.\  \\
\hline
IX36 & $\hddp + \mHt \rightarrow \hhdp + \hd$ & 
$k_{\rm IX36} = 1.0 \times 10^{-10} \expf{-}{187.2}{T}$ & & 13 \\
& & $\phantom{k_{\rm IX36}=} \mbox{} \times \left[1.0 + \expf{-}{87}{T}\right]$ & & \\
IX37 & $\hddp + \hd \rightarrow \hhdp + \DD$ & 
$k_{\rm IX37} = 1.0 \times 10^{-10} \expf{-}{108.4}{T} $ & & 13 \\
& & $\phantom{k_{\rm IX37}=} \mbox{} \times \left[1.0 + \expf{-}{86.5}{T} \right]$ & & \\
IX38 & $\hddp + \hd \rightarrow \dtp + \mHt$ & $k_{\rm IX38} = 2.0 \times 10^{-10}$ & & 4 \\
IX39 & $\hddp + \DD \rightarrow \dtp + \hd$ & $k_{\rm IX39} = 8.7 \times 10^{-10}$ & &14 \\
IX40 & $\dtp + \mHt \rightarrow \hhdp + \DD$ & 
$k_{\rm IX40} = 1.5 \times 10^{-9} \expf{-}{342.2}{T}$ & & 13 \\
IX41 & $\dtp + \mHt \rightarrow \hddp + \hd$ & 
$k_{\rm IX41} = 1.5 \times 10^{-9} \expf{-}{233.8}{T}$ & & 13 \\
IX42 & $\dtp + \hd \rightarrow \hddp + \DD$ & 
$k_{\rm IX42} = 3.75 \times 10^{-10} \expf{-}{155}{T}$ & & 13 \\
& & $\phantom{k_{\rm IX42} =} \mbox{} \times
\left[1.0 + 2.0 \expf{-}{50.4}{T} + \expf{-}{86}{T}\right]$ & &  \\
IX43  &  $\hehp + \mD  \rightarrow  \hedp + \mH $ &
$k_{\rm IX43} = 1.0 \times 10^{-9}$ & & 3 \\ 
IX44 &  $\hedp + \mH  \rightarrow  \hehp + \mD $ & 
$k_{\rm IX44} = 8.0 \times 10^{-10} \expf{-}{468}{T}$ & & 3 \\ 
IX45 & $\lihp + \mD \rightarrow \lidp + \mH$ & $k_{\rm IX45} = 1.0 \times 10^{-9}$ & & 2 \\
IX46 & $\lidp + \mH \rightarrow \lihp + \mD$ & $k_{\rm IX46} = 1.0 \times 10^{-9} \expf{-}{64}{T}$ & & 2 \\
\hline
\end{tabular}
\medskip
\\
{\bf Notes}: T is the gas temperature in K, and $T_{3} = T / 300 \: {\rm K}$. \\
{\bf References}: 1 -- \citet{ljb95}; 2 -- estimate;
3 -- \citet{dm56}, scaled as in \citet{sld98}; 
4 -- \citet{wfp04}; 5 -- \citet{ger82}; 
6 -- Our fits to cross-sections from \citet{ws02};
7 -- Our fits to \citet{mie03};
8 -- \citet{s59}; 9 -- \citet{mbh89};
10 -- \citet{prf89}; 11 -- \citet{mc03}; 
12 -- Derived from forward reaction, using equilibrium 
constant from \citet{rt04}; 13 -- \citet{fpw04};
14 -- Derived from inverse reaction in \citet{wfp04} \\
\end{minipage}
\end{table*}

\clearpage

\begin{table*}
\begin{minipage}{126mm}
\caption{Chemical processes: transfer reactions (TR)}
\begin{tabular}{lllc}
\hline
No.\ & Reaction & Rate coefficient (${\rm cm}^{3} \: {\rm s}^{-1}$) & Ref.\  \\
\hline
TR1  &  $\mHtp + \mHt  \rightarrow  \htp + \mH $ &
$k_{\rm TR1} = 2.24 \times 10^{-9} T_{3}^{0.042} \expf{-}{T}{46600}$ & 1 \\ 
TR2  &  $\mHtp + \hd  \rightarrow  \htp + \mD $ &$k_{\rm TR2} = 1.05 \times 10^{-9}$  & 2 \\ 
TR3  &  $\mHtp + \hd  \rightarrow  \hhdp + \mH $ & $k_{\rm TR3} = 1.05 \times 10^{-9}$ & 2 \\ 
TR4 & $\mHtp + \DD \rightarrow \hhdp + \mD$ & $k_{\rm TR4} = 1.05 \times 10^{-9}$ & 3 \\
TR5 & $\mHtp + \DD \rightarrow \hddp + \mH$ & $k_{\rm TR5} = 1.05 \times 10^{-9}$ & 3 \\
TR6  &  $\hdp + \mHt  \rightarrow  \htp + \mD $ & $k_{\rm TR6} = 0.5 \times k_{\rm TR1}$ & 1 \\ 
TR7 &  $\hdp + \mHt  \rightarrow  \hhdp + \mH $ & $k_{\rm TR7} = 0.5 \times k_{\rm TR1}$ & 1 \\ 
TR8 & $\hdp + \hd \rightarrow \hhdp + \mD$ & $k_{\rm TR8} = 1.05 \times 10^{-9}$ & 3 \\
TR9 & $\hdp + \hd \rightarrow \hddp + \mH$ & $k_{\rm TR9} = 1.05 \times 10^{-9}$ & 3 \\
TR10 & $\hdp + \DD \rightarrow \hddp + \mD$ & $k_{\rm TR10} = 1.05 \times 10^{-9}$ & 3 \\
TR11 & $\hdp + \DD \rightarrow \dtp + \mH$ & $k_{\rm TR11} = 1.05 \times 10^{-9}$ & 3 \\
TR12 & $\ddp + \mHt \rightarrow \hhdp + \mD$ & $k_{\rm TR12} = 1.05 \times 10^{-9}$ & 3 \\
TR13 & $\ddp + \mHt \rightarrow \hddp + \mH$ & $k_{\rm TR13} = 1.05 \times 10^{-9}$ & 3 \\
TR14 & $\ddp + \hd \rightarrow \hddp + \mD$ & $k_{\rm TR14} = 1.05 \times 10^{-9}$ & 3 \\
TR15 & $\ddp + \hd \rightarrow \dtp + \mH$ & $k_{\rm TR15} = 1.05 \times 10^{-9}$ & 3 \\
TR16 & $\ddp + \DD \rightarrow \dtp + \mD$ & $k_{\rm TR16} = 2.1 \times 10^{-9}$ & 3 \\
TR17  &  $\htp + \mH  \rightarrow  \mHtp + \mHt $ & 
$k_{\rm TR17} = 7.7 \times 10^{-9} \expf{-}{17560}{T}$ & 4 \\ 
TR18 & $\htp + \mD \rightarrow \mHtp + \hd$ & 
$k_{\rm TR18} = 0.5 \times k_{\rm TR17}$ & 5 \\
TR19 & $\htp + \mD \rightarrow \hdp + \mHt$ & 
$k_{\rm TR19} = 0.5 \times k_{\rm TR17}$ & 5 \\
TR20 & $\hhdp + \mH \rightarrow \mHtp + \hd$ & $k_{\rm TR20} = 0.5 \times k_{\rm TR17}$ & 5 \\
TR21 & $\hhdp + \mH \rightarrow \hdp + \mHt$ & $k_{\rm TR21} = 0.5 \times k_{\rm TR17}$ & 5 \\
TR22 & $\hhdp + \mD \rightarrow \mHtp + \DD$ & $k_{\rm TR22} = 0.333 \times k_{\rm TR17} $ & 5 \\
TR23 & $\hhdp + \mD \rightarrow \hdp + \hd$ & $k_{\rm TR23} = 0.333 \times k_{\rm TR17} $ & 5 \\
TR24 & $\hhdp + \mD \rightarrow \ddp + \mHt$ & $k_{\rm TR24} = 0.333 \times k_{\rm TR17} $ & 5 \\
TR25 & $\hddp + \mH \rightarrow \mHtp + \DD$ & $k_{\rm TR25} = 0.333 \times k_{\rm TR17} $ & 5 \\
TR26 & $\hddp + \mH \rightarrow \hdp + \hd$ & $k_{\rm TR26} = 0.333 \times k_{\rm TR17} $ & 5 \\
TR27 & $\hddp + \mH \rightarrow \ddp + \mHt$ & $k_{\rm TR27} = 0.333 \times k_{\rm TR17} $ & 5 \\
TR28 & $\hddp + \mD \rightarrow \hdp + \DD$ & $k_{\rm TR28} = 0.5 \times k_{\rm TR17} $ & 5 \\
TR29 & $\hddp + \mD \rightarrow \ddp + \hd$ & $k_{\rm TR29} = 0.5 \times k_{\rm TR17} $ & 5 \\
TR30 & $\dtp + \mH \rightarrow \hdp + \DD$ & $k_{\rm TR30} = 0.5 \times k_{\rm TR17} $ & 5 \\
TR31 & $\dtp + \mH \rightarrow \ddp + \hd$ & $k_{\rm TR31} = 0.5 \times k_{\rm TR17} $ & 5 \\
TR32 & $\dtp + \mD \rightarrow \ddp + \DD$ & $k_{\rm TR32} = k_{\rm TR17} $ & 5 \\
TR33  &  $\He + \mHtp  \rightarrow  \hehp + \mH $ &
$k_{\rm TR33} = 3.0 \times 10^{-10} \expf{-}{6717}{T}$ & 6 \\
TR34  &  $\He + \hdp  \rightarrow  \hehp + \mD $ & $k_{\rm TR34} = k_{\rm TR33}$ & 7 \\ 
TR35  &  $\He + \hdp  \rightarrow  \hedp + \mH $ & $k_{\rm TR35} = k_{\rm TR33}$ & 7 \\ 
TR36 & $\He + \ddp \rightarrow \hedp + \mD$ & $k_{\rm TR36} = k_{\rm TR33} $ & 8 \\
TR37  &  $\hehp + \mH  \rightarrow  \mHtp + \He $ & 
$k_{\rm TR37} = 1.04 \times 10^{-9} T_{3}^{0.13} \expf{-}{T}{33100}$ & 1 \\ 
TR38  &  $\hehp + \mD  \rightarrow  \hdp + \He $ & 
$k_{\rm TR38} = 8.5  \times 10^{-10} T_{3}^{0.13} \expf{-}{T}{33100}$  & 9 \\ 
TR39  &  $\hehp + \mHt  \rightarrow  \htp + \He $ &
$k_{\rm TR39} = 1.53 \times 10^{-9} T_{3}^{0.24} \expf{-}{T}{14800}$ & 1 \\ 
TR40  &  $\hehp + \hd  \rightarrow  \hhdp + \He $ & 
$k_{\rm TR40} = 1.20 \times 10^{-9} T_{3}^{0.24} \expf{-}{T}{14800}$ & 2 \\
TR41 & $\hehp + \DD \rightarrow \hddp + \He$ & 
$k_{\rm TR41} = 1.1 \times 10^{-9} T_{3}^{0.24} \expf{-}{T}{14800}$ & 10 \\
TR42  &  $\hedp + \mH  \rightarrow  \hdp + \He $ & 
$k_{\rm TR42} =  9.1 \times 10^{-10} T_{3}^{0.13} \expf{-}{T}{33100}$ & 9 \\
TR43 & $\hedp + \mD \rightarrow \ddp + \He$ & 
$k_{\rm TR43} = 8.5  \times 10^{-10} T_{3}^{0.13} \expf{-}{T}{33100}$ & 11 \\
TR44  &  $\hedp + \mHt  \rightarrow  \hhdp + \He $ &
$k_{\rm TR44} = 1.24 \times 10^{-9} T_{3}^{0.24} \expf{-}{T}{14800}$ & 2 \\
TR45 & $\hedp + \hd \rightarrow \hddp + \He$ & 
$k_{\rm TR45} = 1.2 \times 10^{-9} T_{3}^{0.24} \expf{-}{T}{14800}$ & 10 \\
TR46 & $\hedp + \DD \rightarrow \dtp + \He$ & 
$k_{\rm TR46} = 1.1\times 10^{-9} T_{3}^{0.24} \expf{-}{T}{14800}$ & 10 \\
TR47 &  $\lihp + \mH  \rightarrow  \lip + \mHt $ & $k_{\rm TR47} = 3.0 \times 10^{-10}$ & 12 \\
TR48 &  $\lihp + \mD  \rightarrow  \lip + \hd $ & $k_{\rm TR48} = 3.0 \times 10^{-10}$ & 13 \\
TR49 & $\lidp + \mH \rightarrow \lip + \hd$ & $k_{\rm TR49} = 3.0 \times 10^{-10}$ & 14 \\
TR50 & $\lidp + \mD \rightarrow \lip + \DD$ & $k_{\rm TR50} = 3.0 \times 10^{-10}$ & 14 \\
TR51 &  $\lihp + \mH  \rightarrow  \li + \mHtp $ &
$k_{\rm TR51} = 9.0 \times 10^{-10} \expf{-}{66400}{T}$ & 12 \\ 
TR52  &  $\lihp + \mD  \rightarrow  \li + \hdp $ & $k_{\rm TR52} = k_{\rm TR51}$ & 13 \\ 
TR53 & $\lidp + \mH \rightarrow \li + \hdp$ & $k_{\rm TR53} = k_{\rm TR51}$ & 14 \\
TR54 & $\lidp + \mD \rightarrow \li + \ddp$ & $k_{\rm TR54} = k_{\rm TR51}$ & 14 \\
\hline
\end{tabular}
\end{minipage}
\end{table*}

\begin{table*}
\begin{minipage}{126mm}
\contcaption{}
\begin{tabular}{lllc}
\hline
No.\ & Reaction & Rate coefficient (${\rm cm}^{3} \: {\rm s}^{-1}$) & Ref.\  \\
\hline
TR55 &  $\lih + \Hp  \rightarrow  \lip + \mHt $ & $k_{\rm TR55} = 2.0 \times 10^{-15}$ & 15 \\ 
TR56  &  $\lih + \Dp  \rightarrow  \lip + \hd $ & $k_{\rm TR56} = 2.0 \times 10^{-15}$  & 16 \\ 
TR57 & $\liD + \Hp \rightarrow \lip + \hd$ & $k_{\rm TR57} = 2.0 \times 10^{-15}$ & 16 \\
TR58 & $\liD + \Dp \rightarrow \lip + \DD$ &  $k_{\rm TR58} = 2.0 \times 10^{-15}$ & 16 \\
TR59  &  $\lih + \Hp  \rightarrow  \li + \mHtp $ & $k_{\rm TR59} = 1.0 \times 10^{-9}$ & 12 \\
TR60  &  $\lih + \Dp  \rightarrow  \li + \hdp $ & $k_{\rm TR60} = 1.0 \times 10^{-9}$  & 16 \\
TR61 & $\liD + \Hp \rightarrow \li + \hdp$ & $k_{\rm TR61} = 1.0 \times 10^{-9}$ & 16 \\
TR62 & $\liD + \Dp \rightarrow \li + \ddp$ & $k_{\rm TR62} = 1.0 \times 10^{-9}$ & 16 \\ 
TR63 &  $\lih + \mH  \rightarrow  \li + \mHt $ &
$k_{\rm TR63} = 1.55 \times 10^{-11} T^{0.4247}$ & 17 \\ 
TR64 &  $\lih + \mD  \rightarrow  \li + \hd $ & 
$k_{\rm TR64} = 1.2 \times 10^{-11} T^{0.4247}$ & 11 \\ 
TR65 & $\liD + \mH \rightarrow \li + \hd$ & 
$k_{\rm TR65} = 1.54 \times 10^{-11} T^{0.4247}$ & 11 \\
TR66 & $\liD + \mD \rightarrow \li + \DD$ & 
$k_{\rm TR66} = 1.2 \times 10^{-11} T^{0.4247}$ & 11 \\ 
\hline
\end{tabular}
\medskip
\\
{\bf Notes}: T is the gas temperature in K, and $T_{3} = T / 300 \: {\rm K}$.  \\
{\bf References}: 1 -- \citet{ljb95}; 2 -- \citet{sld98};
3 -- \citet{wfp04}; 4 -- \citet{smt92}; 
5 -- estimate, based on \citet{smt92}; 6 -- \citet{b78}; 
7 -- \citet{sld98}, based on \citet{b78}; 8 -- estimate, based on
\citet{b78};  9 -- \citet{ljb95}, scaled as in \citet{sld98};
10 -- Estimate, based on \citet{sld98};
11 -- Same as corresponding H reaction, but scaled by D reduced mass;
12 -- \citet{sld96};  13 -- \citet{sld98}, based on corresponding 
H reaction in \citet{sld96};
14 -- estimate, based on \citet{sld96}; 15 -- \citet{bgmr01}; 
16 -- same as corresponding H reaction;
17 -- \citet{dpgg05} \\
\end{minipage}
\end{table*}

\begin{table*}
\begin{minipage}{126mm}
\caption{Chemical processes: background radiation induced photodetachment, 
photodissociation and photoionization (BP) \label{phototab}}
\begin{tabular}{lllc}
\hline
No.\ & Reaction & Rate ($J_{21}^{-1} \: {\rm s}^{-1}$) & Ref.\ \\
\hline
BP1  &  $\Hm + \gamma  \rightarrow  \mH + \me $ & $R_{\rm BP1} = 1.36 \times 10^{-11}$ & 1 \\ 
BP2  &  $\Dm + \gamma  \rightarrow  \mD + \me $ & $R_{\rm BP2} = 1.36 \times 10^{-11}$ & 2 \\
BP3  &  $\mHtp + \gamma  \rightarrow  \mH + \Hp $ & $R_{\rm BP3} = 4.11 \times 10^{-12}$ & 3 \\ 
BP4  &  $\hdp + \gamma  \rightarrow  \mH + \Dp $ & $R_{\rm BP4} = 2.05 \times 10^{-12}$ & 2 \\ 
BP5 & $\hdp + \gamma  \rightarrow  \mD + \Hp $ & $R_{\rm BP5} = 2.05 \times 10^{-12}$ & 2 \\
BP6 & $\ddp + \gamma \rightarrow \mD + \Dp$ & $R_{\rm BP6} = 4.11 \times 10^{-12}$ & 2 \\
BP7  &  $\mHt + \gamma  \rightarrow  \mH + \mH $ & $R_{\rm BP7} = 1.3 \times 10^{-12} 
f_{\rm sh, \mHt}$ & 5 \\ 
BP8  &  $\hd + \gamma  \rightarrow  \mH + \mD $ & $R_{\rm BP8} = 1.45 \times 10^{-12}  
f_{\rm sh, \hd}$ & 6 \\ 
BP9 & $\DD + \gamma \rightarrow \mD + \mD$ & $R_{\rm BP9} = 1.3 \times 10^{-12}$ & 7 \\
BP10  &  $\htp + \gamma  \rightarrow  \mHtp + \mH $ & $R_{\rm BP10} = 2.4 \times 10^{-16}$ & 8 \\ 
BP11  &  $\htp + \gamma  \rightarrow  \mHt + \Hp $ & $R_{\rm BP11} = 2.4 \times 10^{-16}$ & 8 \\ 
BP12 & $\hhdp + \gamma \rightarrow \mHtp + \mD$ & $R_{\rm BP12} = 1.2 \times 10^{-16}$ & 9 \\
BP13 & $\hhdp + \gamma \rightarrow \mHt + \Dp$ &  $R_{\rm BP13} = 1.2 \times 10^{-16}$  & 9 \\
BP14 & $\hhdp + \gamma \rightarrow \hdp + \mH$ &  $R_{\rm BP14} = 1.2 \times 10^{-16}$  & 9 \\
BP15 & $\hhdp + \gamma \rightarrow \hd + \Hp$ &  $R_{\rm BP15} = 1.2 \times 10^{-16}$  & 9 \\
BP16 & $\hddp + \gamma \rightarrow \hdp + \mD$ & $R_{\rm BP16} = 1.2 \times 10^{-16}$ & 9 \\
BP17 & $\hddp + \gamma \rightarrow \hd + \Dp$ &  $R_{\rm BP17} = 1.2 \times 10^{-16}$  & 9 \\
BP18 & $\hddp + \gamma \rightarrow \ddp + \mH$ &  $R_{\rm BP18} = 1.2 \times 10^{-16}$  & 9 \\
BP19 & $\hddp + \gamma \rightarrow \DD + \Hp$ &  $R_{\rm BP19} = 1.2 \times 10^{-16}$  & 9 \\
BP20 & $\dtp + \gamma \rightarrow \ddp + \mD$ & $R_{\rm BP20} =  2.4 \times 10^{-16}$ & 9 \\ 
BP21 & $\dtp + \gamma \rightarrow \DD + \Dp$ & $R_{\rm BP21} =  2.4 \times 10^{-16}$ & 9 \\ 
BP22  &  $\hehp + \gamma  \rightarrow  \He + \Hp $ & $R_{\rm BP22} = 1.0 \times 10^{-17}$ & 10 \\ 
BP23 &  $\hedp + \gamma  \rightarrow  \He + \Dp $ & $R_{\rm BP23} = 1.0 \times 10^{-17}$ & 10 \\
BP24  &  $\hehep + \gamma  \rightarrow  \He + \Hep $ & $R_{\rm BP24} = 1.0 \times 10^{-12}$ & 11 \\ 
BP25  &  $\li + \gamma  \rightarrow  \lip + \me $ & $R_{\rm BP25} = 1.4 \times 10^{-12}$ & 12 \\ 
BP26  &  $\Limi + \gamma  \rightarrow  \li + \me $ & $R_{\rm BP26} = 1.2 \times 10^{-11}$ & 13 \\ 
BP27  &  $\lihp + \gamma  \rightarrow  \lip + \mH $ & $R_{\rm BP27} = 5.0 \times 10^{-18}$ & 13 \\ 
BP28  &  $\lihp + \gamma  \rightarrow  \li + \Hp $ & $R_{\rm BP28} = 9.3 \times 10^{-9}$ & 13 \\ 
BP29  &  $\lidp + \gamma  \rightarrow  \lip + \mD $ & $R_{\rm BP29} =  5.0 \times 10^{-18}$ & 2 \\
BP30  &  $\lidp + \gamma  \rightarrow  \li + \Dp $ & $R_{\rm BP30} = 9.3 \times 10^{-9}$  & 2 \\
BP31  &  $\lih + \gamma  \rightarrow  \li + \mH $ & $R_{\rm BP31} = 4.4 \times 10^{-14}$ & 14 \\ 
BP32  &  $\liD + \gamma  \rightarrow  \li + \mD $ & $R_{\rm BP32} = 4.4 \times 10^{-14}$ & 2 \\ 
\hline
\end{tabular}
\medskip
\\
{\bf Notes}: $\gamma$ represents a photon from the external background radiation
field. The listed reaction rates were computed assuming that this background has the
spectrum of a $10^{5} \: {\rm K}$ diluted black-body, cut-off above $h\nu = 13.6 \: {\rm eV}$,
as described in \S\ref{chem}. With this spectrum, reactions with threshold energies greater
than $13.6 \: {\rm eV}$ do not occur and are not listed in the table. 
$f_{\rm sh, \mHt}$ and $f_{\rm sh, HD}$ are the self-shielding factors for $\mHt$ and HD
photodissociation, respectively \citep[see e.g.,][]{gj07}. Note that in this paper, we consider
only the limiting cases $f_{\rm sh, \mHt} = f_{\rm sh, HD} = 0$ and
 $f_{\rm sh, \mHt} = f_{\rm sh, HD} = 1$. \\
{\bf References}: 1 -- \citet{wish79}; 2 -- assumed same as for corresponding H reaction;
3 -- \citet{d68};  4 -- total rate assumed same as for corresponding H reaction, individual 
outcomes assumed equally probable;
5 -- \citet{db96}; 6 -- \citet{ar06}; 7 -- estimate;
8 -- \citet{vd88}; 9 -- estimate, based on \citet{vd88};
10 -- \citet{rd82}; 11 -- \citet{s94}; 12 -- \citet{vf96};
13 -- \citet{gp98}; 14 -- \citet{kd78} \\
\end{minipage}
\end{table*}

\begin{table*}
\begin{minipage}{126mm}
\caption{Chemical processes: cosmic ray ionization (CR) \label{cr_ion_tab}}
\begin{tabular}{llcc}
\hline
No.\ & Process & Rate ($\zeta_{\rm i}/\zeta_{\mH}$) & Reference \\
\hline
CR1  &  $\mH + {\rm C.R.}  \rightarrow  \Hp + \me $ & 1.0 & --- \\ 
CR2  &  $\mHt + {\rm C.R.}  \rightarrow  \mHtp + \me $ & 2.09 &  1 \\ 
CR3 &  $\mHt + {\rm C.R.}  \rightarrow  \mH + \Hp + \me $ & 0.09 &  1 \\ 
CR4  &  $\mHt + {\rm C.R.}  \rightarrow  \mH+ \mH $ & 3.26 &  1 \\ 
CR5  &  $\He + {\rm C.R.}  \rightarrow  \Hep + \me $ & 1.09 & 1 \\ 
CR6  &  $\mD + {\rm C.R.}  \rightarrow  \Dp + \me $ & 1.0 & 2 \\ 
CR7  &  $\hd + {\rm C.R.}  \rightarrow  \hdp + \me $ & 2.09 & 2 \\ 
CR8  &  $\hd + {\rm C.R.}  \rightarrow  \mH + \Dp + \me $ & 0.04 & 2 \\ 
CR9  &  $\hd + {\rm C.R.}  \rightarrow  \Hp + \mD + \me $ & 0.04 & 2 \\ 
CR10  & $\hd + {\rm C.R.}  \rightarrow  \mH+ \mD $ & 3.26 & 2 \\ 
CR11  & $\DD + {\rm C.R.}  \rightarrow  \ddp + \me $ & 2.09 & 2 \\ 
CR12 &  $\DD + {\rm C.R.}  \rightarrow  \mD + \Dp + \me $ & 0.09 & 2 \\ 
CR13  & $\DD + {\rm C.R.}  \rightarrow  \mD+ \mD $ & 3.26 & 2 \\ 
\hline
\end{tabular}
\medskip
\\
{\bf Notes}: C.R. represents a cosmic ray.
$\zeta_{\mH}$, the cosmic ray ionization rate of atomic hydrogen, 
is an adjustable parameter in our models. \\
{\bf References}: 1 -- \citet{wfp04}; 2 -- assumed same as corresponding H process \\
\end{minipage}
\end{table*}

\begin{table*}
\begin{minipage}{126mm}
\caption{Chemical processes: cosmic ray induced photodetachment,
photodissociation and photoionization (CP) \label{cr_pd_tab}}
\begin{tabular}{llccc}
\hline
No.\ & Reaction & $\sigma_{\rm X, eff, \mHt}$ (Mb)
& $\sigma_{\rm X, eff, \mH}$ (Mb) &  Ref.\  \\
\hline
CP1 & $\Hm + \gamma_{\rm cr} \rightarrow \mH + \me $ & 5.0 & 5.8 & 1 \\
CP2 & $\Dm + \gamma_{\rm cr} \rightarrow \mD + \me $ & 5.0 & 5.8 & 2 \\
CP3  &  $\mHtp + \gamma_{\rm cr}   \rightarrow  \mH + \Hp $ & 5.0 & 6.6& 3 \\ 
CP4  &  $\hdp + \gamma_{\rm cr}   \rightarrow  \mH + \Dp $ & 2.5 & 3.3 & 2 \\
CP5 & $\hdp + \gamma_{\rm cr}   \rightarrow  \mD + \Hp $ & 2.5 & 3.3 & 2 \\
CP6 & $\ddp + \gamma_{\rm cr}  \rightarrow \mD + \Dp$ & 5.0 & 6.6 & 2 \\
CP7 & $\li + \gamma_{\rm cr}   \rightarrow  \lip + \me $ & 1.0 & 1.3  & 4 \\
CP8 &  $\Limi + \gamma_{\rm cr}   \rightarrow  \li + \me $ & 1.0 & 1.0 & 5 \\
CP9 & $\hehep + \gamma_{\rm cr}   \rightarrow  \He + \Hep $ & 5.0 & 5.0 & 6 \\
CP10  &  $\lihp + \gamma_{\rm cr}   \rightarrow  \li + \Hp $ & 100 & 100 & 7 \\
CP11  &  $\lidp + \gamma _{\rm cr}  \rightarrow  \li + \Dp $ & 100 & 100 & 2 \\
\hline
\end{tabular}
\medskip
\\
{\bf Notes}: $\gamma_{\rm cr}$ represents a secondary photon, produced
by cosmic-ray induced excitation of $\mH$ or $\mHt$, as described in \S\ref{cosmic_rays}. 
The references listed are the sources from which we have taken our photodissociation or
photoionization cross-sections. The emission probabilities $P_{\mHt}(\nu)$
used to calculate $\sigma_{\rm X, eff, \mHt}$ are rough estimates based on 
the emission spectra given in \citet{sdl87} and are likely accurate only to
within a factor of a few. \\
{\bf References}: 1 -- \citet{wish79}; 2 -- assumed same as for corresponding H reaction;
3 -- \citet{d68}; 4 -- \citet{vf96}; 5 -- order of magnitude estimate; 6 -- estimate, based on \citet{s94};
7 -- rough estimate, based on thermal rate in \citet{gp98} \\
\end{minipage}
\end{table*}

\end{document}